\documentclass{jfm}
\pdfoutput=1

\usepackage{graphicx}
\usepackage{natbib}
\usepackage{journals}
\usepackage{amsmath}
\usepackage{upmath}
\usepackage{amssymb}
\usepackage{longtable}
\usepackage{amsbsy}
\usepackage[colorlinks=True,citecolor=blue,linkcolor=blue,%
			urlcolor=blue,pdftex]{hyperref}

\newcommand{\tbli}{\lambda_T^{i}}
\newcommand{\tblo}{\lambda_T^{o}}
\newcommand{\ubli}{\lambda_U^{i}}
\newcommand{\ublo}{\lambda_U^{o}}
\newcommand{\umbli}{\lambda_{U,m}^{i}}
\newcommand{\umblo}{\lambda_{U,m}^{o}}

\usepackage{xcolor}

\def\vec#1{\ensuremath{\mathchoice{\mbox{\boldmath$\displaystyle#1$}}
{\mbox{\boldmath$\textstyle#1$}}
{\mbox{\boldmath$\scriptstyle#1$}}
{\mbox{\boldmath$\scriptscriptstyle#1$}}}}

\newsavebox{\astrutbox}
\sbox{\astrutbox}{\rule[-5pt]{0pt}{20pt}}

\title[Turbulent Rayleigh-B\'enard convection in spherical 
shells]{Turbulent Rayleigh-B\'enard convection in spherical shells}

\author[T. Gastine, J. Wicht and J.~M. Aurnou]%
{Thomas Gastine$^1$%
  \thanks{Email address for correspondence: gastine@mps.mpg.de},\ns
 Johannes Wicht$^1$,\ns Jonathan M. Aurnou$^2$}

\affiliation{$^1$Max Planck Institut f\"ur Sonnensystemforschung, 
Justus-von-Liebig-Weg 3, 37077 G\"ottingen, Germany, \\ $^2$Department of 
Earth, Planetary and Space Sciences, University of California, Los Angeles, CA, 
90095-1567 USA}

\pubyear{2015}
\volume{650}
\pagerange{119--126}
\date{?; revised ?; accepted ?. - To be entered by editorial office}
\begin{document}

\maketitle

%
%
\begin{abstract}
We simulate numerically Boussinesq convection in non-rotating spherical 
shells for a 
fluid with a unity Prandtl number and Rayleigh numbers up to $10^9$. In this 
geometry, curvature and radial variations of the gravitational acceleration 
yield asymmetric boundary layers. A systematic parameter study for various
radius ratios (from $\eta=r_i/r_o=0.2$ to $\eta=0.95$) and gravity profiles 
allows us to explore the dependence of the asymmetry on these parameters. We 
find that the average plume spacing is comparable between the 
spherical inner and outer bounding surfaces. An estimate of the average plume 
separation allows us to accurately predict the boundary layer asymmetry for the 
various spherical shell configurations explored here. The mean temperature and 
horizontal velocity profiles are in good agreement with 
classical Prandtl-Blasius laminar boundary layer profiles, provided the 
boundary layers are analysed in a dynamical frame that fluctuates with the 
local and instantaneous boundary layer thicknesses. The scaling properties of 
the Nusselt and Reynolds numbers are investigated by separating the 
bulk and boundary layer contributions to the thermal and viscous 
dissipation rates using numerical models with $\eta=0.6$ and a gravity 
proportional to $1/r^2$. We show that our spherical models are consistent with 
the predictions of Grossmann \& Lohse's (2000) theory and that $Nu(Ra)$ and 
$Re(Ra)$ scalings are in good agreement with plane layer results. 
\end{abstract}

\begin{keywords}
B\'enard convection, boundary layers, geophysical and geological flows
\end{keywords}

\section{Introduction}

\begin{figure}
 \centering
 \includegraphics[width=\textwidth]{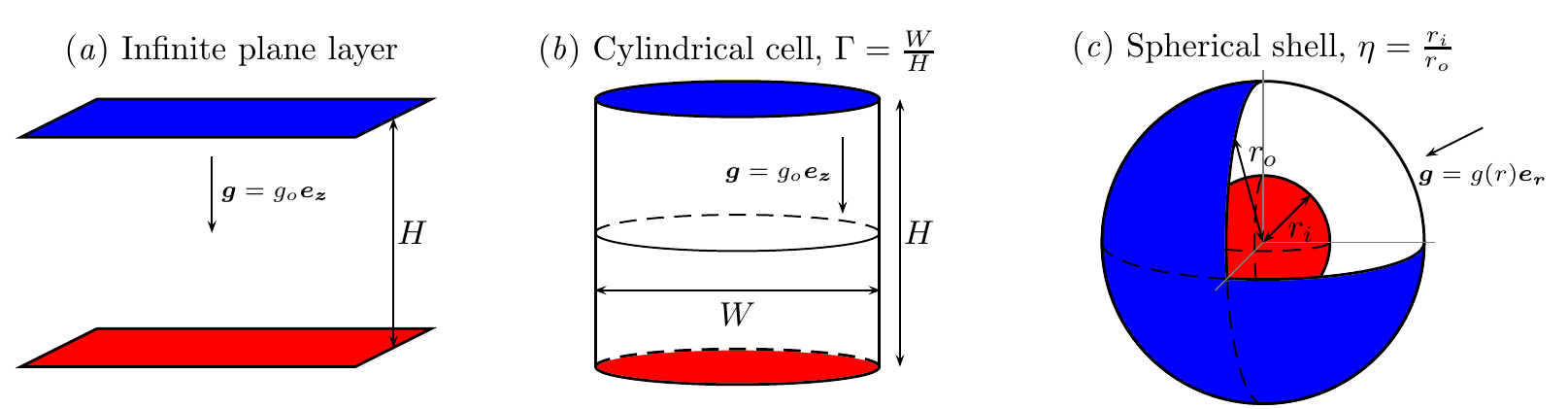}
 \caption{Schematic showing different Rayleigh-B\'enard convection setups. 
(\textit{a}) An infinitely extended fluid layer of height $H$ heated from below 
and cooled from above. (\textit{b}) The typical RB convection setup 
in a cylindrical cell of aspect ratio $\Gamma=W/H$. (\textit{c}) Convection in 
a spherical shell with a radius ratio $\eta=r_i/r_o$ with a radially inward 
gravity. In the three panels, the red and the blue surfaces correspond to the 
hot and cold boundaries held at constant temperatures.}
 \label{fig:intro}
\end{figure}

Thermal convection is ubiquitous in geophysical and astrophysical fluid 
dynamics and rules, for example, turbulent flows in the interiors of planets 
and stars. The so-called Rayleigh-B\'enard (hereafter RB) 
convection is probably the simplest paradigm to study heat 
transport phenomena in these natural systems. In this configuration, 
convection is driven in a planar fluid layer cooled from above and heated from 
below (figure~\ref{fig:intro}\textit{a}). The fluid is confined between two 
rigid impenetrable walls maintained at 
constant temperatures. The key issue in RB convection is to understand
the turbulent transport mechanisms of heat and momentum across the layer.
In particular, how does the heat transport, characterised by the Nusselt number 
$Nu$, and the flow amplitude, characterised by the Reynolds number $Re$,
depend on the various control parameters of the system, namely the 
Rayleigh number $Ra$, the Prandtl number $Pr$ and the cartesian aspect ratio 
$\Gamma$? In general, $\Gamma=W/H$ quantifies the fluid layer width 
$W$ over its height $H$ in classical planar or cylindrical RB cells. In 
spherical shells, we rather employ the ratio of the inner to the outer radius 
$\eta=r_i/r_o$ to characterise the geometry of the fluid layer.

Laboratory experiments of RB convection are classically performed in 
rectangular or in cylindrical tanks with planar upper and lower bounding 
surfaces where the temperature contrast is imposed 
(see figure~\ref{fig:intro}\textit{b}).
In such a system, the global dynamics are strongly influenced by the flow 
properties in the
thermal and kinematic boundary layers that form in the vicinity of the walls.
The characterisation of the structure of these boundary layers is crucial for a 
better 
understanding of the transport processes. The marginal stability theory by 
\cite{Malkus54} is the earliest boundary layer model and relies on the 
assumption that the 
thermal boundary layers adapt their thicknesses to maintain a critical boundary 
layer Rayleigh number, which implies $Nu\sim Ra^{1/3}$. Assuming 
that the boundary layers are sheared, \cite{Shraiman90} later derived a 
theoretical model that yields scalings of the form $Nu\sim 
Ra^{2/7}\,Pr^{-1/7}$ and $Re\sim Ra^{3/7}\,Pr^{-5/7}$ 
\citep[see also][]{Siggia94}. These asymptotic laws were generally consistent 
with most of the experimental results obtained in the 1990s up to $Ra 
\lesssim 10^{11}$. Within the 
typical experimental resolution of one percent, simple power laws of the form 
$Nu \sim Ra^\alpha Pr^\beta$ were found to provide an adequate representation 
with $\alpha$ exponents ranging from 0.28 to 0.31, in relatively good agreement 
with the Shraiman \& Siggia model 
\citep[e.g.][]{Castaing89,Chavanne97,Niemela00}. However, later high-precision 
experiments by \cite{Xu00} revealed that the dependence of $Nu$ upon $Ra$ cannot 
be accurately described by such simple power laws. In particular, the local 
slope of the function $Nu(Ra)$ has been found to increase slowly with 
$Ra$. The effective exponent $\alpha_{\text{eff}}$ of $Nu \sim 
Ra^{\alpha_{\text{eff}}}$ roughly ranges from values close to $0.28$ near 
$Ra\sim 10^7-10^8$ to $0.33$ when $Ra \sim 10^{11}-10^{12}$   
\citep[e.g.][]{Funfschilling05,Cheng15}. 

\cite{Grossmann00,Grossmann04} derived a competing theory capable of capturing 
this complex dynamics (hereafter GL). This scaling theory is built on the 
assumption of laminar boundary layers of Prandtl-Blasius (PB) type 
\citep{Prandtl04,Blasius08}. According to the GL theory, the flows are 
classified in four different regimes in the $Ra-Pr$ phase space according to 
the relative contribution of the bulk and boundary layer viscous and thermal 
dissipation rates. The theory predicts non-power-law behaviours for $Nu$ and 
$Ra$ in good agreement with the dependence $Nu=f(Ra,Pr,\Gamma)$ and 
$Re=f(Ra,Pr,\Gamma)$ observed in recent experiments and numerical simulations 
of RB convection in planar or cylindrical geometry \citep[see for recent 
reviews][]{Ahlers09,Chilla12}.

Benefiting from the interplay between experiments and direct numerical 
simulations (DNS), turbulent RB convection in planar and cylindrical cells has 
received a lot of interest in the past two 
decades. However, the actual geometry of several fundamental astrophysical 
and geophysical flows is essentially three-dimensional within concentric 
spherical upper and lower 
bounding surfaces under the influence of a radial buoyancy force that strongly 
depends on radius. The direct applicability of the results 
derived in the planar geometry to spherical shell convection is thus 
questionable. 

As shown in figure~\ref{fig:intro}(\textit{c}), convection in spherical shells 
mainly differs from the traditional plane layer configuration 
because of the introduction of curvature and the absence of side walls. These 
specific features of thermal convection in spherical shells yield 
significant dynamical 
differences with plane layers. For instance, the heat flux conservation through 
spherical surfaces implies that the temperature gradient is larger at the lower 
boundary than at the upper one to compensate for the smaller area of the bottom 
surface. This yields a much larger temperature drop at the inner 
boundary than at the outer one. In addition, this pronounced asymmetry in the 
temperature profile is accompanied by a difference between the thicknesses of 
the inner and the outer thermal boundary layers. Following Malkus's 
marginal stability arguments, \cite{Jarvis93} and \cite{Vangelov94} 
hypothesised 
that the thermal boundary layers in curvilinear geometries adjust their 
thickness to maintain the same critical boundary layer Rayleigh number at both 
boundaries. This criterion is however in poor agreement with the 
results from numerical models \citep[e.g.][]{Deschamps10}. The exact 
dependence of the boundary layer asymmetry on the radius ratio and the gravity 
distribution thus remains an open question in thermal convection in spherical 
shells \citep{Bercovici89,Jarvis95,Sotin99,Shahnas08,OFarrell13}. This open 
issue sheds some light on the possible dynamical influence of asymmetries 
between the hot and cold surfaces that originate due to both the boundary 
curvature and the radial dependence of buoyancy in spherical shells.

Ground-based laboratory experiments involving spherical geometry and a radial 
buoyancy forcing are limited by the fact that gravity is vertically downwards 
instead of radially inwards \citep{Scanlan70,Feldman13}. A possible way to 
circumvent this limitation is to conduct experiments under microgravity to 
suppress the vertically downward buoyancy force. Such an experiment was 
realised 
by \cite{Hart86} who designed a hemispherical shell that flew on board of the 
space shuttle Challenger in May 1985. The radial buoyancy force was modelled by 
imposing an electric field across the shell. The temperature dependence of the 
fluid's dielectric properties then produced an effective radial gravity that 
decreases with the fifth power of the radius (i.e. $g \sim 1/r^{5}$). More 
recently, a similar experiment named ``GeoFlow'' was run on the International 
Space Station, where much longer flight times are possible 
\citep{Futterer10,Futterer13}. This later experiment was designed to mimic the 
physical conditions in the Earth mantle. It was therefore mainly dedicated to 
the observation of plume-like structures in a high Prandtl number regime ($Pr 
> 40$) for $Ra \leq 10^6$. Unfortunately, this limitation to relatively 
small Rayleigh numbers makes the GeoFlow experiment quite restricted regarding 
asymptotic scaling behaviours in spherical shells.

To compensate the lack of laboratory experiments, three dimensional numerical 
models of convection in spherical shells have been developed since the 1980s 
\citep[e.g.][]{Zebib80,Bercovici89,Bercovici92,Jarvis95,Tilgner96,Tilgner97,
King10,Choblet12}. The vast majority of the numerical models of non-rotating 
convection in spherical shells has been developed with Earth's mantle in 
mind. 
These models therefore assume an infinite Prandtl number and most of them 
further include a strong dependence of viscosity on temperature to mimic the 
complex rheology of the mantle. Several recent studies of isoviscous 
convection with infinite Prandtl number in spherical shells have nevertheless 
been dedicated to the analysis of the scaling properties of the  
Nusselt number. For instance, \cite{Deschamps10} measured convective heat 
transfer in various radius ratios ranging from $\eta=0.3$ to $\eta=0.8$ and 
reported $Nu\sim Ra^{0.273}$ for 
$10^4\leq Ra \leq 10^7$, while \cite{Wolstencroft09} computed numerical models 
with Earth's mantle geometry ($\eta=0.55$) up to $Ra=10^8$ and found $Nu\sim 
Ra^{0.294}$. These studies also checked the possible influence of internal 
heating and reported quite similar scalings.

Most of the numerical models of convection in spherical shells have thus 
focused on the very specific dynamical regime of the infinite Prandtl number.
The most recent attempt to derive the scaling properties of $Nu$ and $Re$ in 
non-rotating spherical shells with finite Prandtl numbers is 
the study of \cite{Tilgner96}. He studied convection in self-graviting 
spherical shells (i.e. $g \sim r$) with $\eta=0.4$ spanning the range 
$0.06\leq Pr \leq 10$ and $4\times 10^3 \leq Ra \leq 8\times 10^5$. This study 
was thus restricted to low Rayleigh numbers, relatively close to the onset of 
convection, which prevents the derivation of asymptotic scalings for 
$Nu(Ra,Pr)$ and $Re(Ra,Pr)$ in spherical shells.

The objectives of the present work are twofold: (\textit{i}) to study the 
scaling properties of $Nu$ and $Re$ in spherical shells with finite Prandtl 
number; (\textit{ii}) to better characterise the inherent asymmetric boundary 
layers in thermal convection in spherical shells.  We therefore conduct two 
systematic parameter studies of turbulent RB convection in spherical shells with 
$Pr=1$ by means of three dimensional DNS. In the first set of models, we vary 
both the radius ratio (from $\eta=0.2$ to $\eta=0.95$) and the radial gravity 
profile (considering $g(r)\in [r/r_o,1,(r_o/r)^2,(r_o/r)^5]$) in a moderate 
parameter regime (i.e. $5 \leq Nu \leq 15$ for the majority of the 
cases) to study the influence of these 
properties on the boundary layer asymmetry. We then consider a second set of 
models with $\eta=0.6$ and $g\sim 1/r^2$ up to $Ra=10^9$. These DNS are used 
to check the applicability of the GL theory to thermal convection in spherical 
shells. We therefore numerically test the different basic prerequisites of the 
GL theory: we first analyse the nature of the boundary layers before deriving 
the individual scaling properties for the different contributions to the 
viscous and thermal dissipation rates.

The paper is organised as follows. In \S~\ref{sec:model}, we present the 
governing equations and the numerical models. We then focus on the asymmetry of 
the thermal boundary layers in \S~\ref{sec:thbl}. In \S~\ref{sec:viscBl}, we 
analyse the nature of the boundary layers and show that the boundary layer 
profiles are in agreement with the Prandtl-Blasius theory 
\citep{Prandtl04,Blasius08}. In \S~\ref{sec:dissip}, we investigate the scaling 
properties of the viscous and thermal dissipation rates before calculating the 
$Nu(Ra)$ and $Re(Ra)$ scalings in \S~\ref{sec:nurera}. We conclude
with a summary of our findings in \S~\ref{sec:conclu}.

\section{Model formulation} 
\label{sec:model}

\subsection{Governing hydrodynamical equations} 
 
We consider RB convection of a Boussinesq fluid contained in a 
spherical shell of outer radius $r_o$ and inner radius $r_i$. The boundaries are 
impermeable, no slip and at constant temperatures $T_{bot}$ and 
$T_{top}$. We adopt a 
dimensionless formulation using the shell gap $d=r_o-r_i$ as the reference 
lengthscale and the viscous dissipation time $d^2 /\nu$ as the reference 
timescale. Temperature is given in units of $\Delta T=T_{top}-T_{bot}$, the 
imposed temperature contrast over the shell. Velocity and pressure are 
expressed in units of $\nu/d$ and $\rho_o \nu^2/d^2$, respectively. Gravity is 
non-dimensionalised using its reference value at the outer boundary $g_o$. The 
dimensionless equations for the velocity $\vec{u}$, the pressure $p$ and the 
temperature $T$ are given by
\begin{equation}
  \vec{\nabla}\cdot\vec{u} = 0,
  \label{eq:divu}
\end{equation}
\begin{equation}
 \frac{\partial \vec{u}}{\partial t} + \vec{u}\cdot\vec{\nabla} \vec{u} = 
-\vec{\nabla} p + \frac{Ra}{Pr}\,g\,T\,\vec{e_r}+ 
\vec{\Delta}\vec{u},
 \label{eq:navier}
\end{equation} 
\begin{equation}
 \frac{\partial T}{\partial t} + \vec{u}\cdot\vec{\nabla} T = 
\frac{1}{Pr}\Delta T,
\label{eq:temp}
\end{equation}
where $e_r$ is the unit vector in the radial direction and $g$ is the gravity.  
Several gravity profiles have been classically considered to model convection 
in spherical shells. For instance, self-graviting spherical 
shells with a constant density correspond 
to $g\sim r$ \citep[e.g][]{Tilgner96}, while RB convection models with 
infinite Prandtl number usually assume a constant gravity in the perspective of 
modelling Earth's mantle \citep[e.g.][]{Bercovici89}. The assumption of a 
centrally-condensed mass has also been frequently assumed when modelling 
rotating convection \citep[e.g.][]{Glatz1,Jones11} and yields $g\sim 1/r^2$.  
Finally, the artificial central force field of the microgravity experiments 
takes effectively the form of $g \sim 1/r^5$
\citep[][]{Hart86,Feudel11,Futterer13}. To explore the possible impact of 
these various radial distribution of buoyancy on RB convection in spherical 
shells, we consider different models with the four 
following gravity profiles: $g\in[r/r_o,\,1,\,(r_o/r)^2,\,(r_o/r)^5]$. 
Particular attention will be paid to the cases with $g=(r_o/r)^2$, 
which is the only radial function compatible with an exact  analysis of the
dissipation rates (see below, \S~\ref{sect:dissipRel}).

The dimensionless set of equations (\ref{eq:divu}-\ref{eq:temp}) is governed by 
the Rayleigh number $Ra$, the Prandtl number $Pr$ and the radius ratio of the 
spherical shell $\eta$ defined by
\begin{equation}
 Ra = \frac{\alpha g_o \Delta T d^3}{\nu\kappa},\quad 
Pr=\frac{\nu}{\kappa},\quad 
\eta=\frac{r_i}{r_o},
\end{equation}
where $\nu$ and $\kappa$ are the viscous and thermal diffusivities and $\alpha$ 
is the thermal expansivity.

\subsection{Diagnostic parameters}

To quantify the impact of the different control parameters on 
the transport of heat and momentum, we analyse several diagnostic properties. 
We adopt the following notations regarding different averaging 
procedures. Overbars $\overline{\cdots}$ correspond to a time average
\[
 \overline{f} = \dfrac{1}{\tau}\int_{t_0}^{t_0+\tau} f\,{\rm d} t,
\]
where $\tau$ is the time averaging interval. Spatial averaging over the whole 
volume of the spherical shell are denoted by triangular brackets $\langle \cdots
\rangle$, while $\langle \cdots \rangle_s$ correspond to an average 
over a spherical surface:
\[
 \left\langle f \right\rangle = \frac{1}{V} \int_V 
f(r,\theta,\phi)\,{\rm d} V; \quad
\left\langle f \right\rangle_s = \frac{1}{4 \pi} \int_0^\pi \int_0^{2\pi} 
f(r,\theta,\phi)\sin\theta\,{\rm d}\theta\,{\rm d}\phi\,,
\]
where $V$ is the volume of the spherical shell, $r$ is the radius, $\theta$ 
the colatitude and $\phi$ the longitude. 

The convective heat transport is characterised by the Nusselt number $Nu$, the 
ratio of the total heat flux to the heat carried by conduction. In spherical 
shells with isothermal boundaries, the conductive temperature profile $T_c$ is 
the solution of 
\[
 \frac{\rm d }{{\rm d} r}\left( r^2 \frac{{\rm d} T_c}{{\rm d} r} \right) 
= 0, \quad T_c(r_i)=1,  \quad T_c(r_o)=0,
\]
which yields
\begin{equation}
 T_c(r)= \frac{\eta}{(1-\eta)^2}\frac{1}{r}-\frac{\eta}{1-\eta}\,.
 \label{eq:conducTemp}
\end{equation}
For the sake of clarity, we adopt in the following the notation 
$\vartheta$ for the time-averaged and horizontally-averaged radial temperature 
profile:
\[
 \vartheta(r) = \overline{\left\langle T \right\rangle_s}\, .
\]
The Nusselt number then reads
\begin{equation}
 Nu =  \frac{\overline{\left\langle u_r T \right\rangle_s} - 
\frac{1}{Pr}\frac{{\rm d} \vartheta}{{\rm d} r}}{-\frac{1}{Pr}\frac{{\rm d} 
T_c}{\rm dr}} = 
-\eta\frac{{\rm d} \vartheta}{{\rm d} 
r}(r=r_i) =-\frac{1}{\eta}\frac{{\rm d} 
 \vartheta}{{\rm d} r}(r=r_o)\,.
\label{eq:nudef}
\end{equation}
The typical rms flow velocity is given by the Reynolds number
\begin{equation}
 Re = \overline{\sqrt{\langle u^2\rangle}}=\overline{\sqrt{\langle 
u_r^2+u_\theta^2+u_\phi^2\rangle}}\,,
\end{equation}
while the radial profile for the time and horizontally-averaged horizontal 
velocity is defined by
\begin{equation}
 Re_h(r) = \overline{\left \langle \sqrt{u_\theta^2+u_\phi^2} \right\rangle_s}.
\end{equation}

\subsection{Exact dissipation relationships in spherical shells}
\label{sect:dissipRel}

The mean buoyancy power averaged over the whole 
volume of a spherical shell is expressed by
\[
 P = \frac{Ra}{Pr}\, \overline{\left \langle  g\, u_r T\right\rangle} = 
\frac{4\pi}{V}\,\frac{Ra}{Pr}
\int_{r_i}^{r_o} g\,r^2\,\overline{\left\langle u_r T\right\rangle_s}\,{\rm 
d}r\,, 
\]
Using the Nusselt number definition (\ref{eq:nudef}) and the conductive
temperature profile (\ref{eq:conducTemp}) then leads to
\[
 P = \frac{4\pi}{V}\,\frac{Ra}{Pr^2} \left( \int_{r_i}^{r_o} 
g\,r^2\,\dfrac{{\rm d}\vartheta}{{\rm d} r}\,{\rm d}r -
 Nu\,\frac{\eta}{(1-\eta)^2} \int_{r_i}^{r_o} g\, {\rm d}r \right)\,.
\]
The first term in the parentheses becomes identical to the imposed temperature 
drop $\Delta T$ for a gravity $g \sim 1/r^2$:
\[
 \int_{r_i}^{r_o} g\, r^2\,\dfrac{{\rm d}\vartheta}{{\rm d} r}\,{\rm d}r = 
r_o^2\left[\vartheta(r_o)-\vartheta(r_i)\right]=-r_o^2\,,
\]
and thus yields an analytical relation between $P$ and $Nu$.
For any other gravity model, we have to consider the actual 
spherically-symmetric radial temperature profile $\vartheta(r)$. 
\cite{Christensen06} solve this problem by approximating $\vartheta(r)$ by the 
diffusive solution (\ref{eq:conducTemp}) and obtain an approximate relation  
between $P$ and $\frac{Ra}{Pr^2}(Nu-1)$. This motivates our particular focus on 
the $g=(r_o/r)^2$ cases which allows us to conduct an exact analysis of the 
dissipation rates and therefore check the applicability of the GL 
theory to convection in spherical shells. 

Noting that $\frac{\eta}{(1-\eta)^2} 
\int_{r_i}^{r_o} g\, {\rm d} r = -\frac{1}{(1-\eta)^2}$ for 
$g=(r_o/r)^2$, one finally obtains 
the exact relation for the viscous dissipation rate $\epsilon_U$:
\begin{equation}
  \epsilon_U = \overline{\left\langle 
\left(\vec{\nabla}\times\vec{u}\right)^2\right\rangle}  = P = 
\dfrac{3}{1+\eta+\eta^2}\,\frac{Ra}{Pr^2}\,(Nu-1)\,.
  \label{eq:viscDiss}
 \end{equation}
The thermal dissipation rate can be obtained by multiplying the temperature 
equation (\ref{eq:temp}) by $T$ and integrate it over the whole volume of the 
spherical shell. This yields
\begin{equation}
 \epsilon_T= \overline{\left\langle (\nabla T)^2 \right\rangle} = 
\dfrac{3\eta}{1+\eta+\eta^2}\, 
Nu\,.
 \label{eq:tDiss}
\end{equation}
These two exact relations (\ref{eq:viscDiss}-\ref{eq:tDiss}) can be
used to validate the spatial resolutions of the numerical models with $g = 
(r_o/r)^2$. To do so, we introduce $\chi_{\epsilon_U}$ and $\chi_{\epsilon_T}$, 
the ratios of the two sides of Eqs~(\ref{eq:viscDiss}-\ref{eq:tDiss}):

\begin{equation}
\begin{aligned}
 \chi_{\epsilon_U} & = 
\frac{(1+\eta+\eta^2)\,Pr^2}{3\,Ra\,(Nu-1)}\,\epsilon_U\,, \\
   \chi_{\epsilon_T} & = 
\frac{(1+\eta+\eta^2)}{3\eta\,Nu}\,\epsilon_T\,. \\
 \end{aligned}
\label{eq:chis}
\end{equation}


\subsection{Setting up a parameter study}

\subsubsection{Numerical technique}

The numerical simulations have been carried out with the magnetohydrodynamics 
code MagIC \citep{Wicht02}. MagIC has been validated via several benchmark 
tests for convection and dynamo action \citep{Christensen01,Jones11}. To solve 
the system of equations (\ref{eq:divu}-\ref{eq:temp}), the solenoidal velocity 
field is decomposed into a poloidal and a toroidal contribution
\[
 \vec{u} = \vec{\nabla}\times\left(\vec{\nabla}\times W \vec{e_r}\right) +
\vec{\nabla}\times Z \vec{e_r},
\]
where $W$ and $Z$ are the poloidal and toroidal potentials. $W$, $Z$, $p$ and
$T$ are then expanded in spherical harmonic functions up to degree
$\ell_{\text{max}}$ in the angular variables $\theta$ and $\phi$ and in
Chebyshev polynomials up to degree $N_r$ in the radial direction. The combined 
equations governing $W$ and $p$ are obtained by taking the radial component and 
the horizontal part of the divergence of (\ref{eq:navier}). The equation 
for $Z$ is obtained by taking the radial component of the curl of 
(\ref{eq:navier}). The equations are time-stepped by advancing the 
nonlinear terms using an explicit second-order Adams-Bashforth
scheme, while the linear terms are time-advanced using an
implicit Crank-Nicolson algorithm. At each time step, all 
the nonlinear products are calculated in the physical space ($r$, $\theta$, 
$\phi$) and transformed back into the spectral space ($r$, $\ell$, $m$).
For more detailed descriptions of the numerical method and 
the associated spectral transforms, the reader is referred to
\citep{Glatz1,Tilgner97,Christensen07}.

\subsubsection{Parameter choices}

\begin{table}
 \begin{center}
 \def~{\hphantom{0}}
 \begin{tabular}{cccccccccc}
$Ra$ & $Nu$ & $Re$ & $\lambda_T^{i}/\lambda_T^{o}$ & 
$\lambda_U^{i}/\lambda_U^{o}$ & $\epsilon_T^{bu} (\%)$ & $\epsilon_U^{bu} (\%)$ 
& $\chi_{\epsilon_T}$ & $\chi_{\epsilon_U}$ & $N_r\times \ell_{max}$ \\
\hline
$1.5\times10^{3}$ & 1.33 & 4.4 & - & - & - & - & 1.000 & 1.000 & $49 \times 85$ \\
$2\times10^{3}$ & 1.59 & 6.7 & - & - & - & - & 1.000 & 1.000 & $49 \times 85$ \\
$3\times10^{3}$ & 1.80 & 9.6 & - & - & - & - & 1.000 & 1.000 & $49 \times 85$ \\
$5\times10^{3}$ & 2.13 & 14.4 & - & - & - & - & 1.000 & 1.000 & $49 \times 85$ \\
$7\times10^{3}$ & 2.20 & 17.5 & $0.186/0.251$ & $0.076/0.104$ & 0.11 & 0.57 & 1.000 & 1.000 & $49 \times 85$ \\
$9\times10^{3}$ & 2.43 & 21.7 & $0.168/0.223$ & $0.070/0.094$ & 0.14 & 0.60 & 1.000 & 1.000 & $49 \times 85$ \\
$1\times10^{4}$ & 2.51 & 23.3 & $0.162/0.217$ & $0.069/0.092$ & 0.15 & 0.61 & 1.000 & 1.000 & $49 \times 85$ \\
$1.5\times10^{4}$ & 2.81 & 29.8 & $0.143/0.196$ & $0.062/0.086$ & 0.17 & 0.62 & 1.000 & 1.000 & $49 \times 85$ \\
$2\times10^{4}$ & 3.05 & 35.0 & $0.130/0.185$ & $0.059/0.082$ & 0.15 & 0.64 & 1.000 & 1.000 & $49 \times 85$ \\
$3\times10^{4}$ & 3.40 & 44.0 & $0.116/0.167$ & $0.054/0.077$ & 0.17 & 0.64 & 1.000 & 1.000 & $49 \times 85$ \\
$5\times10^{4}$ & 3.89 & 57.5 & $0.102/0.147$ & $0.049/0.069$ & 0.18 & 0.67 & 1.000 & 1.000 & $49 \times 85$ \\
$7\times10^{4}$ & 4.27 & 68.5 & $0.093/0.133$ & $0.046/0.062$ & 0.18 & 0.69 & 1.000 & 1.000 & $49 \times 85$ \\
$1\times10^{5}$ & 4.71 & 82.3 & $0.085/0.120$ & $0.043/0.058$ & 0.18 & 0.70 & 1.000 & 1.000 & $49 \times 85$ \\
$1.5\times10^{5}$ & 5.28 & 101.2 & $0.076/0.107$ & $0.039/0.053$ & 0.19 & 0.71 & 1.000 & 1.000 & $49 \times 85$ \\
$2\times10^{5}$ & 5.72 & 117.0 & $0.070/0.099$ & $0.037/0.050$ & 0.20 & 0.74 & 1.000 & 1.000 & $49 \times 85$ \\
$3\times10^{5}$ & 6.40 & 143.3 & $0.062/0.088$ & $0.033/0.046$ & 0.22 & 0.74 & 1.000 & 1.000 & $61 \times 106$ \\
$5\times10^{5}$ & 7.37 & 185.1 & $0.054/0.077$ & $0.030/0.042$ & 0.24 & 0.77 & 1.000 & 1.000 & $61 \times 106$ \\
$7\times10^{5}$ & 8.10 & 218.6 & $0.049/0.070$ & $0.028/0.040$ & 0.22 & 0.77 & 1.000 & 1.000 & $61 \times 106$ \\
$1\times10^{6}$ & 8.90 & 259.2 & $0.045/0.064$ & $0.026/0.037$ & 0.24 & 0.79 & 1.000 & 1.000 & $81 \times 170$ \\
$1.5\times10^{6}$ & 9.97 & 315.1 & $0.040/0.057$ & $0.024/0.034$ & 0.23 & 0.81 & 1.000 & 1.000 & $81 \times 170$ \\
$2\times10^{6}$ & 10.79 & 362.8 & $0.037/0.053$ & $0.023/0.032$ & 0.26 & 0.81 & 1.000 & 1.000 & $81 \times 170$ \\
$3\times10^{6}$ & 12.11 & 443.5 & $0.033/0.048$ & $0.020/0.030$ & 0.24 & 0.82 & 0.999 & 1.003 & $81 \times 170$ \\
$5\times10^{6}$ & 13.97 & 565.6 & $0.029/0.041$ & $0.018/0.027$ & 0.25 & 0.83 & 1.000 & 1.001 & $97 \times 266$ \\
$7\times10^{6}$ & 15.39 & 666.4 & $0.026/0.037$ & $0.017/0.025$ & 0.27 & 0.83 & 1.000 & 1.005 & $97 \times 266$ \\
$1\times10^{7}$ & 17.07 & 790.4 & $0.023/0.034$ & $0.016/0.024$ & 0.25 & 0.84 & 1.000 & 1.005 & $97 \times 266$ \\
$1.5\times10^{7}$ & 19.17 & 960.1 & $0.021/0.030$ & $0.015/0.021$ & 0.28 & 0.84 & 1.000 & 1.009 & $97 \times 341$ \\
$2\times10^{7}$ & 20.87 & 1099.7 & $0.019/0.028$ & $0.013/0.020$ & 0.27 & 0.85 & 1.000 & 1.005 & $121 \times 426$ \\
$3\times10^{7}$ & 23.50 & 1335.5 & $0.017/0.025$ & $0.012/0.018$ & 0.28 & 0.85 & 1.000 & 1.012 & $121 \times 426$ \\
$5\times10^{7}$ & 27.35 & 1690.2 & $0.014/0.021$ & $0.011/0.016$ & 0.27 & 0.86 & 1.000 & 1.010 & $161 \times 512$ \\
$7\times10^{7}$ & 30.21 & 1999.1 & $0.013/0.019$ & $0.010/0.015$ & 0.28 & 0.86 & 1.000 & 1.005 & $201 \times 576$ \\
$1\times10^{8}$ & 33.54 & 2329.9 & $0.012/0.017$ & $0.009/0.013$ & 0.29 & 0.87 
& 1.000 & 1.011 & $201 \times 682$ \\
$2\times10^{8}$ & 41.49 & 3239.3 & $0.010/0.014$ & $0.008/0.012$ & 0.29 & 0.88 & 
1.000 & 1.006 & $321 \times 682$ \\
$3\times10^{8}$ & 47.22 & 3882.5 & $0.008/0.012$ & $0.007/0.010$ & 0.29 & 0.87 & 
1.001 & 1.015 & $321 \times 768$ \\
$\mathit{5\times10^{8}}$ & \textit{56.67} & \textit{5040.0} & 
$\mathit{0.007/0.009}$ & $\mathit{0.005/0.008}$ & \textit{0.31} & \textit{0.85} 
& \textit{1.001} & \textit{1.050} & $\mathit{321 \times 682^\star}$ \\
$5\times10^{8}$ & 55.07 & 4944.1 & $0.007/0.011$ & $0.006/0.009$ & 0.29 & 0.88 
& 0.999 & 1.003 & $401 \times 853^\star$ \\
 $\mathit{1\times10^{9}}$ & \textit{73.50} & \textit{7039.4} & 
$\mathit{0.005/0.007}$ & $\mathit{0.004/0.006}$ & \textit{0.32}
& \textit{0.79} & \textit{1.002} & \textit{1.148} & $\mathit{401 \times 
682^\star}$ \\
$1\times10^{9}$ & 68.48 & 6802.5 & $0.006/0.009$ & $0.005/0.007$ & 0.30 & 0.89 
& 0.996 & 1.006 & $513 \times 1066^\star$ \\
\end{tabular}

 \caption{Summary table of $Pr=1$ numerical simulations with $\eta=0.6$ and 
$g=(r_o/r)^2$. The boundary layer thicknesses and the viscous and thermal 
dissipations are only given for the cases with $Ra\geq 7\times 10^{3}$ when 
boundary layers can be clearly identified. 
The italic lines indicate simulations with smaller truncations to highlight the 
possible resolution problems. The cases with $Ra=5\times 10^8$ and $Ra=10^{9}$ 
(highlighted with a star in the last column) have been computed assuming a 
two-fold symmetry, i.e. effectively simulating only half of the spherical shell 
in longitude.}
   \label{tab:results}
 \end{center}
\end{table}

One of the main focuses of this study is to investigate the global scaling 
properties of $Pr=1$ RB convection in spherical shell geometries. This is 
achieved via measurements of the Nusselt and Reynolds numbers. In particular, 
we aim to test the 
applicability of the GL theory to spherical shells. As demonstrated before, 
only the particular choice of a gravity profile of the form $g\sim 1/r^2$ 
allows the exactness of the relation (\ref{eq:viscDiss}). Our main set of 
simulations is thus built assuming $g=(r_o/r)^2$. The radius ratio is kept to 
$\eta=0.6$ and the Prandtl number to $Pr=1$ to allow future comparisons 
with the rotating convection models by \cite{Gastine12} and \cite{Gastine13} 
who adopted the same configuration. We consider 35 numerical cases
spanning the range $1.5\times 10^3 \leq Ra \leq 10^9$. 
Table~\ref{tab:results} summarises the main diagnostic quantities for this 
dataset of numerical simulations and shows that our models basically lie within 
the ranges $1 < Re < 7000$ and $1 < Nu <70$.

Another important issue in convection in spherical shells concerns the 
determination of the average bulk temperature and the possible boundary 
layer asymmetry between the inner and the outer boundaries 
\citep[e.g.][]{Jarvis93,Tilgner96}. To better understand the influence of 
curvature and the radial distribution of buoyancy, we thus compute 
a second set of numerical models. This additional dataset consists of 113 
additional simulations with various radius ratios and gravity profiles, 
spanning the range $0.2 \leq \eta \leq 0.95$ with 
$g\in[r/r_o,\,1,\,(r_o/r)^2,\,(r_o/r)^5]$.
To limit the numerical cost of this second dataset, these cases have 
been run at moderate Rayleigh number and typically span the range 
$5<Nu<15$ for the majority of the cases. Table~\ref{tab:etas}, given in 
the Appendix, summarises the main diagnostic quantities for this second dataset 
of numerical simulations.

\subsubsection{Resolution checks}

Attention must be paid to the numerical resolutions of the DNS
of RB convection \citep[e.g.][]{Shishkina10}. Especially, underresolving the 
fine structure of the turbulent flow leads to an overestimate of the Nusselt 
number, which then falsifies the possible scaling analysis \citep{Amati05}. One 
of the most reliable ways to validate the truncations employed in our numerical 
models consists of comparing the obtained viscous and thermal dissipation rates 
with the average Nusselt number \citep{Stevens10,Lakkaraju12,King12}. The 
ratios $\chi_{\epsilon_U}$ and $\chi_{\epsilon_T}$, defined in 
(\ref{eq:chis}), are found to be very 
close to unity for all the cases of Table~\ref{tab:results}, which supports the 
adequacy of the employed numerical resolutions. To further highlight the 
possible impact of inadequate spatial resolutions, two underresolved numerical 
models for the two highest Rayleigh numbers have also been included in 
Table~\ref{tab:results} (lines in italics). Because of the insufficient number 
of grid points in the boundary layers, the viscous dissipation rates are 
significantly higher than expected in the statistically stationary state. This 
leads to overestimated Nusselt numbers by similar percent differences 
($2-10\%$).

Table~\ref{tab:results} shows that the typical resolutions span the range from
($N_r=49,\,\ell_{\text{max}}=85$) to 
($N_r=513,\,\ell_{\text{max}}=1066$).
The two highest Rayleigh numbers have been computed assuming a two-fold 
azimuthal symmetry to ease the numerical computations. A comparison of 
test runs with or without the two-fold azimuthal symmetry at lower Rayleigh 
numbers ($ 5\times 10^7 \leq Ra \leq 3\times 10^8$) showed no significant 
statistical differences. This enforced symmetry is thus not considered to be 
influential. The total computational time for the two datasets of numerical 
models represents roughly 5 million Intel Ivy Bridge CPU hours.

\section{Asymmetric boundary layers in spherical shells}
\label{sec:thbl}

\subsection{Definitions}
\label{subsec:defs}

\begin{figure}
 \centering
 \includegraphics[width=8.5cm]{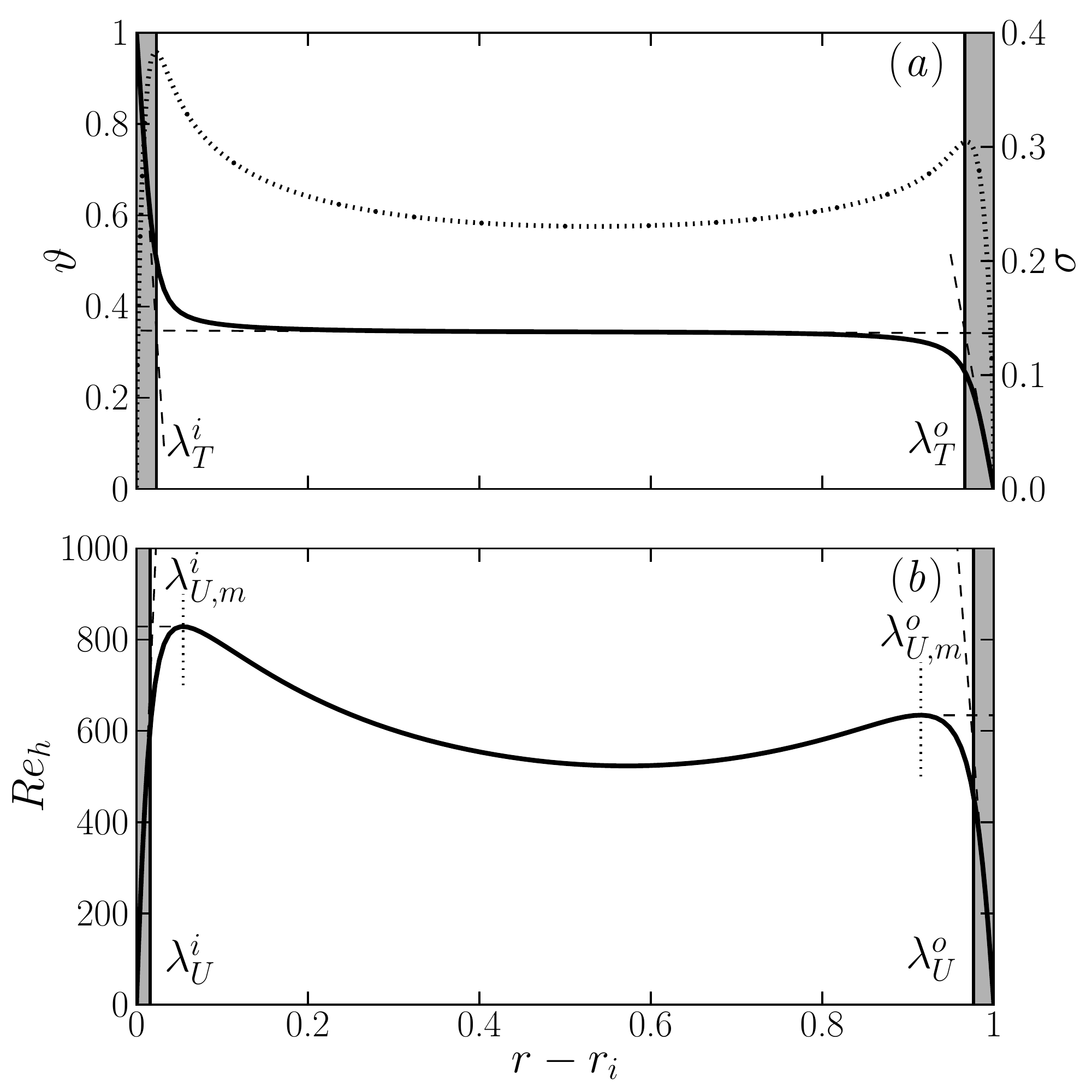}
 \caption{(\textit{a}) Radial profiles of the time and 
horizontally-averaged mean  temperature $\vartheta(r)$ (solid 
black line) and the temperature variance $\sigma$ (dotted black line).
The thermal boundary layers $\tbli$ and $\tblo$ are highlighted by the gray 
shaded area. They are defined as the 
depths where the linear fit to $\vartheta(r)$ near 
the top (bottom) crosses the linear fit to the temperature profile 
at mid-depth (dashed black lines).
(\textit{b}) Radial profiles of the time and horizontally-averaged horizontal 
velocity $Re_h(r)$. The viscous boundary layers are either 
defined by the local maxima of $Re_h$ (black dotted lines, $\umbli$, $\umblo$) 
or by the intersection of the linear fit to $Re_h$ near the inner (outer) 
boundary with the tangent to the local maxima (dashed black lines). This second 
definition is highlighted by a gray shaded area ($\ubli$, $\ublo$). These 
profiles have been obtained from a numerical model with $Ra=10^{7}$, 
$\eta=0.6$ and $g=(r_o/r)^2$.}
\label{fig:tempLayers}
\end{figure} 

Several different approaches are traditionally considered to define the 
thermal boundary layer thickness $\lambda_T$. They either rely on the 
horizontally-averaged mean radial temperature profile $\vartheta(r)$ or on the 
temperature fluctuation $\sigma$ defined as
\begin{equation}
 \sigma = \sqrt{\overline{\left\langle\left(T-\overline{\left\langle T
\right\rangle_s}\right)^2\right \rangle}}\,.
\end{equation}
Among the possible estimates 
based on $\vartheta(r)$, the slope method 
\citep[e.g.][]{Verzicco99,Breuer04,Liu11} defines $\lambda_T$ as the depth 
where the linear fit to $\vartheta(r)$ near the boundaries 
intersects the linear fit to the temperature profile 
at mid-depth. Alternatively, $\sigma$ exhibits sharp local maxima close to the 
walls. The radial distance separating those peaks from the corresponding 
nearest boundary can be used to define the thermal boundary layer thicknesses 
\citep[e.g.][]{Tilgner96,King13}.
Figure~\ref{fig:tempLayers}(\textit{a}) shows that both definitions of 
$\lambda_T$ actually yield nearly indistinguishable boundary layer thicknesses.  
We therefore adopt the slope method to define the thermal 
boundary layers. 

There are also several ways to define the viscous boundary layers.
Figure~\ref{fig:tempLayers}(\textit{b}) shows the vertical profile of the 
root-mean-square horizontal velocity $Re_h$. This profile exhibits strong 
increases close to the boundaries that are accompanied by well-defined peaks. 
Following \cite{Tilgner96} and \cite{Kerr00}, the first way to define the 
kinematic boundary layer is thus to measure the distance between the walls and 
these local maxima. This commonly-used definition gives $\umbli$ ($\umblo$) for 
the inner (outer) spherical boundary. Another possible method to estimate the 
viscous boundary layer follows a similar strategy as the slope method that we 
adopted for the thermal boundary layers \citep{Breuer04}. $\ubli$ ($\ublo$) is 
defined as the distance from the inner (outer) wall where the linear 
fit to $Re_h$  near the inner (outer) boundary intersects the horizontal line 
passing through the maximum horizontal velocity.

Figure~\ref{fig:tempLayers}(\textit{b}) reveals that these two definitions lead 
to very distinct viscous boundary layer thicknesses. In particular, the 
definition based on the position of the local maxima of $Re_h$ yields much 
thicker boundary layers than the tangent intersection method, i.e. 
$\lambda_{U,m}^{i,o} >\lambda_{U}^{i,o}$. The discrepancies of these two 
definitions are further discussed in \S~\ref{sec:viscBl}.

\subsection{Asymmetric thermal boundary layers and mean bulk temperature}

\begin{figure}
 \centering
 \includegraphics[width=\textwidth]{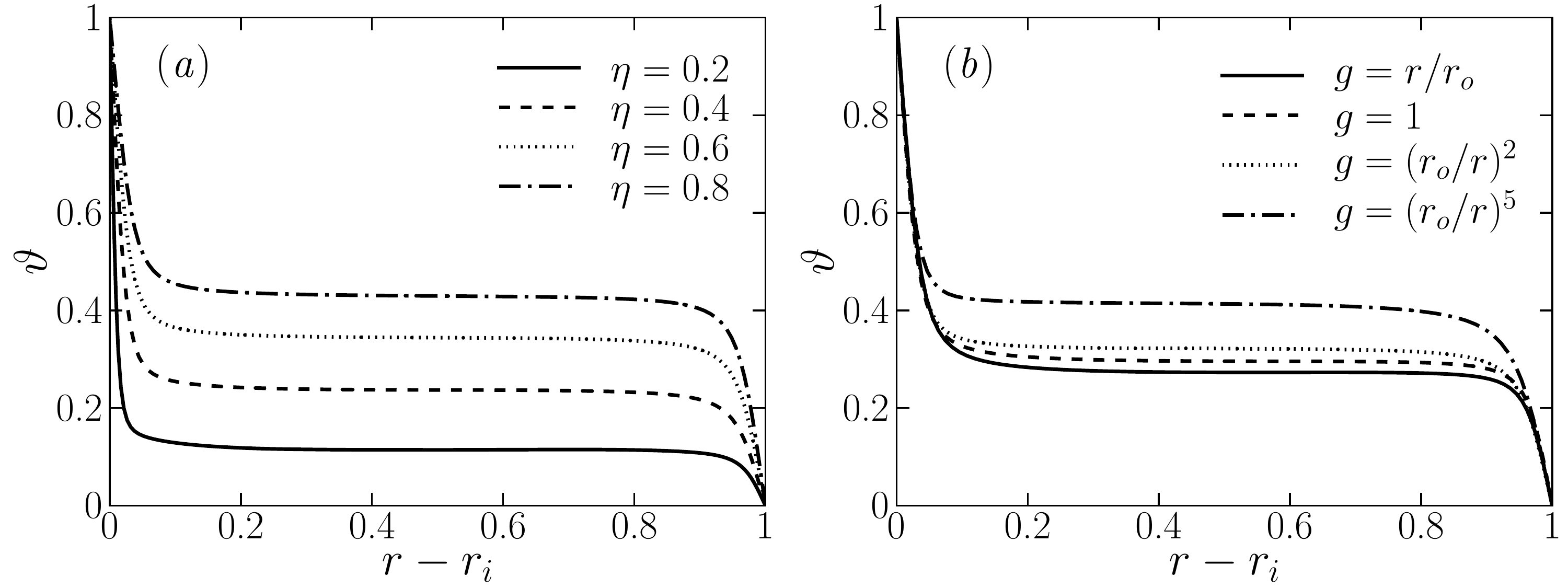}
 \caption{(\textit{a}) $\vartheta(r)$ for different
radius ratios $\eta$ with a gravity $g=(r_o/r)^2$. These models have 
approximately the same Nusselt number $12<Nu<14$. (\textit{b}) $\vartheta(r)$ 
for different gravity profiles with a fixed radius ratio $\eta=0.6$. 
These models have approximately the same Nusselt number $10<Nu<11$.}
 \label{fig:bLayers} 
\end{figure}

Figure~\ref{fig:tempLayers} also reveals a pronounced asymmetry 
in the mean temperature profiles with a much larger temperature drop at the 
inner boundary than at the outer boundary. As a consequence, the mean 
temperature of the spherical shell $T_m = \frac{1}{V} \int_V T\,{\rm d} 
V$ is much below $\Delta T/2$. Determining how the mean temperature 
depends on the radius ratio $\eta$ has been an ongoing open question in mantle 
convection studies with infinite Prandtl number 
\citep[e.g.][]{Bercovici89,Jarvis93,Vangelov94,Jarvis95,Sotin99,
Shahnas08,Deschamps10,OFarrell13}.
To analyse this issue in numerical models with $Pr=1$, we have 
performed a systematic parameter study varying both the radius ratio 
of the spherical shell $\eta$ and the gravity profile $g(r)$ (see 
Table~\ref{tab:etas}). Figure~\ref{fig:bLayers} shows some selected radial 
profiles of the mean temperature $\vartheta$ for various radius
ratios (panel \textit{a}) and gravity profiles (panel \textit{b}) for cases 
with similar $Nu$. For small values of $\eta$, the large difference between the 
inner and the outer surfaces lead to a 
strong asymmetry in the temperature distribution: nearly 90\% of the total 
temperature drop occurs at the inner boundary when $\eta=0.2$.  In thinner 
spherical shells, the mean temperature gradually approaches a more symmetric 
distribution to 
finally reach $T_m = 0.5$ when $\eta\rightarrow 1$ (no curvature). 
Figure~\ref{fig:bLayers}(\textit{b}) also reveals that a change in the gravity 
profile has a direct impact on the mean temperature profile. This shows that 
both the shell geometry and the radial distribution of buoyancy affect
the temperature of the fluid bulk in RB convection in spherical 
shells.

To analytically access the asymmetries in thickness and temperature drop
observed in figure~\ref{fig:bLayers}, we 
first assume that the heat is purely transported by conduction in the thin 
thermal boundary layers. The heat flux conservation through spherical surfaces 
(\ref{eq:nudef}) then yields
\begin{equation}
\frac{\Delta T^{o}}{\tblo}=\eta^2 \frac{\Delta T^{i}}{\tbli},
\label{eq:flux}
\end{equation}
where the thermal boundary layers are assumed to correspond to a linear 
conduction profile with a temperature drop $\Delta T^i$ ($\Delta T^o$) over a 
thickness $\tbli$ ($\tblo$). As shown in 
Figs.~\ref{fig:tempLayers}-\ref{fig:bLayers}, the fluid 
bulk is isothermal and forms the majority of the fluid by volume. We can thus 
further assume that the temperature drops occur only in the thin boundary 
layers, which leads to 
\begin{equation}
\Delta T^{o}+\Delta T^{i}=1.
\label{eq:jump}
\end{equation}
Equations~(\ref{eq:flux}) and (\ref{eq:jump}) are nevertheless not sufficient 
to determine the three unknowns $\Delta T^i$, $\Delta T^o$, $\tblo/\tbli$ and 
an additional physical assumption is required.

\begin{figure}
 \centering
 \includegraphics[width=\textwidth]{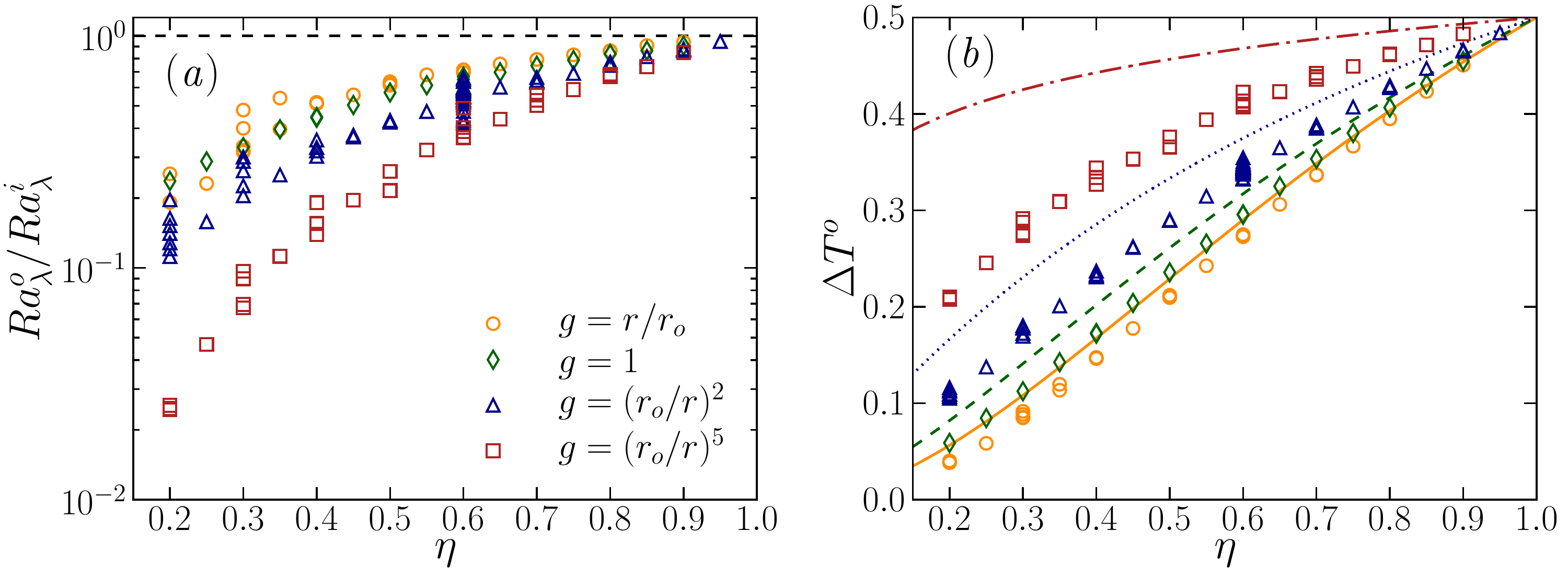}
 \caption{(\textit{a}) Ratio of boundary layer Rayleigh 
numbers (\ref{eq:malkus}) for various radius ratios and gravity profiles. 
The horizontal dashed line
corresponds to the hypothetical identity $Ra_\lambda^i = Ra_\lambda^o$.
(\textit{b}) Temperature drop at the outer boundary layer. The lines 
correspond to the theoretical prediction given in (\ref{eq:jarvis}).}
 \label{fig:ra} 
\end{figure}

A hypothesis frequently used in mantle convection models with infinite Prandtl 
number in spherical geometry \citep{Jarvis93,Vangelov94} is to 
further assume that both thermal boundary layers are marginally stable such that 
the local boundary layer Rayleigh numbers $Ra_\lambda^i$ and $Ra_\lambda^o$ are 
equal: 
\begin{equation}
Ra_\lambda^i=Ra_\lambda^o \,\rightarrow\, \frac{\alpha g_i\Delta T^i 
{\tbli}^3}{\nu\kappa} = \frac{\alpha g_o\Delta T^o {\tblo}^3}{\nu\kappa}\,.
\label{eq:malkus}
\end{equation}
This means that both thermal boundary layers adjust their 
thickness and temperature drop to yield $Ra_\lambda^i\sim Ra_\lambda^o \sim 
Ra_c \simeq 1000$ 
\citep[e.g.,][]{Malkus54}. The temperature drops at both boundaries and the 
ratio of the thermal boundary layer thicknesses can then be derived using 
Eqs.~(\ref{eq:flux}-\ref{eq:jump})
\begin{equation}
\Delta T^i=\frac{1}{1+\eta^{3/2}\,\chi_g^{1/4}},\quad
\Delta T^o \simeq T_m 
=\frac{\eta^{3/2}\,\chi_g^{1/4}}{1+\eta^{3/2}\,\chi_g^{1/4}}, 
\quad
 \frac{\tblo}{\tbli} =\frac{\chi_g^{1/4}}{\eta^{1/2}},
 \label{eq:jarvis}
\end{equation}
where 
\begin{equation}
\chi_g=\frac{g(r_i)}{g(r_o)}\,,
\end{equation}
is the ratio of the gravitational acceleration  between the inner and the outer 
boundaries. 
Figure~\ref{fig:ra}(\textit{a}) reveals that the marginal stability 
hypothesis is not fulfilled when different radius ratios and gravity profiles 
are considered. This is particularly obvious for 
small radius ratios where $Ra_\lambda^o$ is more than 10 times larger than 
$Ra_\lambda^i$. This discrepancy tends to vanish when 
$\eta\rightarrow 1$, when curvature and gravity variations become unimportant.
As a consequence, there is a significant mismatch between the predicted 
mean bulk temperature from (\ref{eq:jarvis}) and the actual values 
(figure~\ref{fig:ra}\textit{b}).
\cite{Deschamps10} also reported a similar deviation from 
(\ref{eq:jarvis}) in their spherical shell models with infinite Prandtl 
number. They suggest that assuming instead $Ra_\lambda^o/Ra_\lambda^i \sim 
\eta^2$ might help to improve the agreement with the data.
This however cannot account for the additional dependence on the gravity 
profile visible in figure~\ref{fig:ra}.
We finally note that $Ra_\lambda<400$ for the database of numerical 
simulations explored here, which suggests that the thermal boundary layers are 
stable in all our simulations.

\begin{figure}
 \centering
 \includegraphics[width=\textwidth]{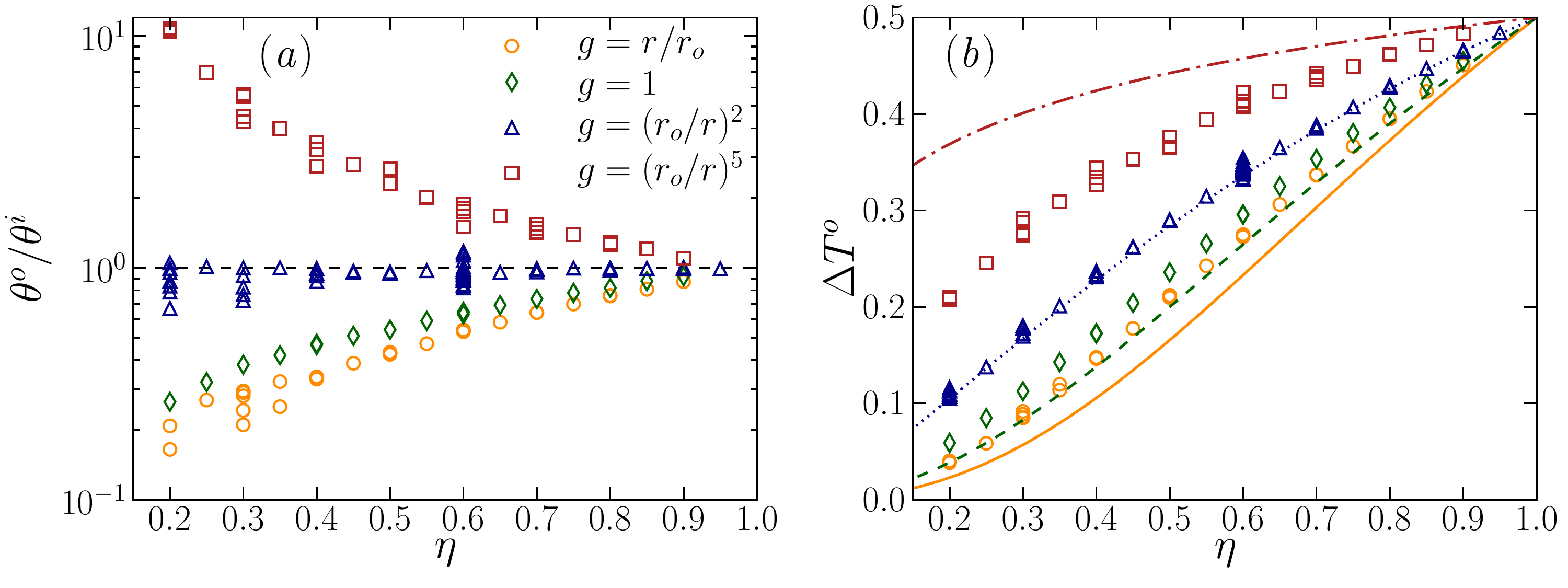}
 \caption{(\textit{a}) Ratio of boundary layer temperature scales 
(\ref{eq:wl91}) for various radius ratios and gravity profiles. 
The horizontal dashed line corresponds to the hypothetical identity 
$\theta^i = \theta^o$. (\textit{b}) Temperature drop 
at the outer boundary layer. The lines correspond to the theoretical prediction 
given in (\ref{eq:predwl91}).}
 \label{fig:wl} 
\end{figure}

Alternatively \citet{Wu91} and \cite{Zhang97} proposed that the thermal 
boundary layers adapt their thicknesses such that the mean hot and cold 
temperature fluctuations at mid-depth are equal. Their experiments with 
Helium indeed revealed that the statistical distribution of the temperature 
at mid-depth was symmetrical. They further assumed that the 
thermal fluctuations in the center can be identified with the boundary layer 
temperature scales  $\theta^i=\frac{\nu\kappa}{\alpha g_i {\tbli}^3}$ and 
$\theta^o=\frac{\nu\kappa}{\alpha g_o {\tblo}^3}$, which characterise the 
temperature scale of the thermal boundary layers in a different way 
than the relative temperature drops $\Delta T^i$ and $\Delta T^o$. This second 
hypothesis yields
 \begin{equation}
 \theta^i = \theta^o \,\rightarrow\, \frac{\nu\kappa}{\alpha g_i {\tbli}^3} 
=\frac{\nu\kappa}{\alpha g_o {\tblo}^3},
\label{eq:wl91}
\end{equation}
and the corresponding temperature drops and boundary layer thicknesses ratio
\begin{equation}
 \Delta T^i = \frac{1}{1+\eta^{2}\,\chi_g^{1/3}},\quad \Delta T^o = T_m =
\frac{\eta^2\,\chi_g^{1/3}}{1+\eta^{2}\,\chi_g^{1/3}}, \quad  
\frac{\tblo}{\tbli} = \chi_g^{1/3}.
\label{eq:predwl91}
\end{equation}
Figure~\ref{fig:wl}(\textit{a}) shows $\theta^o/\theta^i$ for different 
radius ratios and gravity profiles, while 
figure~\ref{fig:wl}(\textit{b}) shows a comparison between the predicted mean 
bulk temperature and the actual values. Besides the cases with $g=(r_o/r)^2$ 
which are in relatively good agreement with the predicted scalings, the 
identity of the boundary layer temperature scales is in general not fulfilled 
for the other gravity profiles. The actual mean bulk temperature is thus
poorly described by (\ref{eq:predwl91}). We note that previous findings by 
\citet{Ahlers06} already reported that the theory by Wu \& Libchaber's does also 
not hold when the transport properties depend on temperature (i.e. 
non-Oberbeck-Boussinesq convection).

\subsection{Conservation of the average plume density in spherical shells}

\begin{figure}
 \centering
 \includegraphics[width=\textwidth]{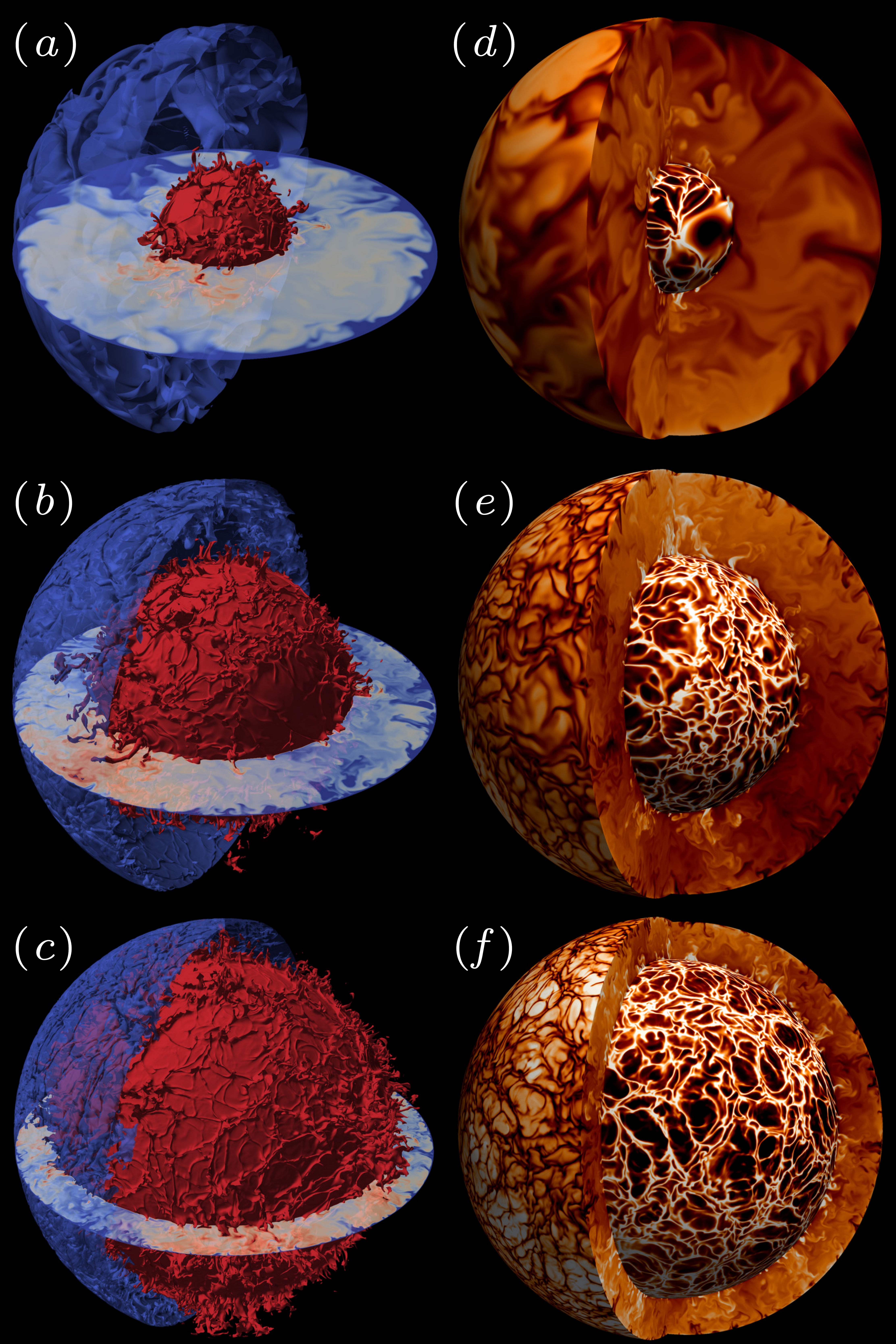}
 \caption{(\textit{a-c}) Isosurfaces and equatorial cut of the temperature for 
three numerical models: hot at the inner thermal boundary layer 
$T(r=r_i+\tbli)$ in red, cold at the outer thermal boundary layer 
$T(r=r_o-\tblo)$ in blue. (\textit{d-f}) Meridional cuts and surfaces of the 
temperature fluctuations $T'$. The inner (outer) surface corresponds to the 
location of the inner (outer) thermal boundary layers.  Color levels range from 
-0.2 (black) to 0.2 (white). Panels (\textit{a}) and (\textit{d}) correspond to 
a model with $Ra=3\times 10^{6},\,\eta=0.3$ and $g=(r_o/r)^5$. Panels 
(\textit{b}) and (\textit{e}) correspond to a model with $Ra=10^{8},\,\eta=0.6$ 
and $g=(r_o/r)^2$. Panels (\textit{c}) and (\textit{f}) correspond to a model 
with $Ra=4\times 10^{7},\,\eta=0.8$ and $g=r/r_o$.}
 \label{fig:3D}
\end{figure}

As demonstrated in the previous section, none of the hypotheses 
classically employed accurately account for the temperature drops and the 
boundary layer asymmetry observed in spherical shells. We must therefore 
find a dynamical quantity that could be possibly identified between the two 
boundary layers.

Figure~\ref{fig:3D} shows visualisations of the thermal boundary 
layers for three selected numerical models with different radius ratios and 
gravity profiles. The isocontours displayed in panels (\textit{a}-\textit{c}) 
reveal the intricate plume structure. Long and thin sheet-like structures 
form the main network of plumes. During their migration along the 
spherical surfaces, these sheet-like plumes can collide and convolute with each 
other to give rise to mushroom-type plumes \citep[see][]{Zhou10,Chilla12}. 
During this morphological evolution, mushroom-type plumes acquire a strong 
radial vorticity component. These mushroom-type plumes are particularly 
visible at the connection points of the sheet plumes network at the 
inner thermal boundary layer (red isosurface in 
figure~\ref{fig:3D}\textit{a}-\textit{c}). 
Figure~\ref{fig:3D}(\textit{d}-\textit{f}) shows the corresponding equatorial
and radial cuts of the temperature 
fluctuation $T'=T-\vartheta$. These panels further highlight the 
plume asymmetry between the inner and the outer thermal boundary layers.
For instance, the case with $\eta=0.3$ and $g=(r_o/r)^5$ (top
panels) features an outer boundary layer approximately 4.5 times thicker 
than the inner one. Accordingly, the mushroom-like plumes that depart from the 
outer boundary layer are significantly thicker than the ones emitted from the 
inner boundary. This discrepancy tends to vanish in the thin shell case 
($\eta=0.8$, bottom panels) in which curvature and gravity variations 
play a less significant role ($\tblo/\tbli \simeq 1.02$ in that case).

\cite{Puthenveettil05} and \cite{Zhou10} performed statistical analysis 
of the geometrical properties of thermal plumes in experimental 
RB convection. By tracking a large number of plumes, their 
analysis revealed that both the plume separation and the width of the 
sheet-like plumes follow a log-normal probability density function (PDF). 

\begin{figure}
 \centering
 \includegraphics[width=\textwidth]{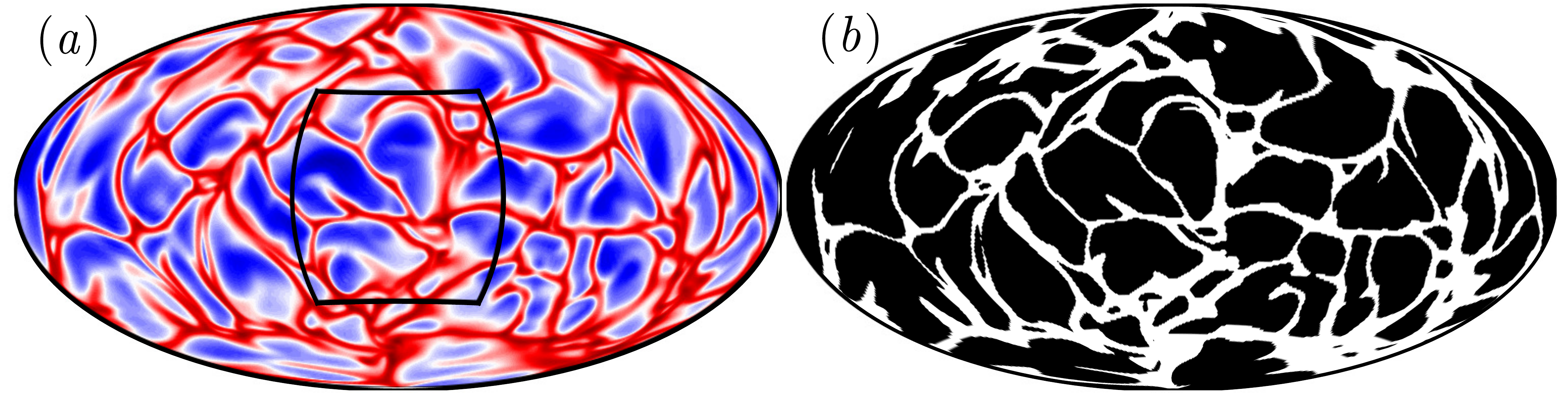}
  \includegraphics[width=\textwidth]{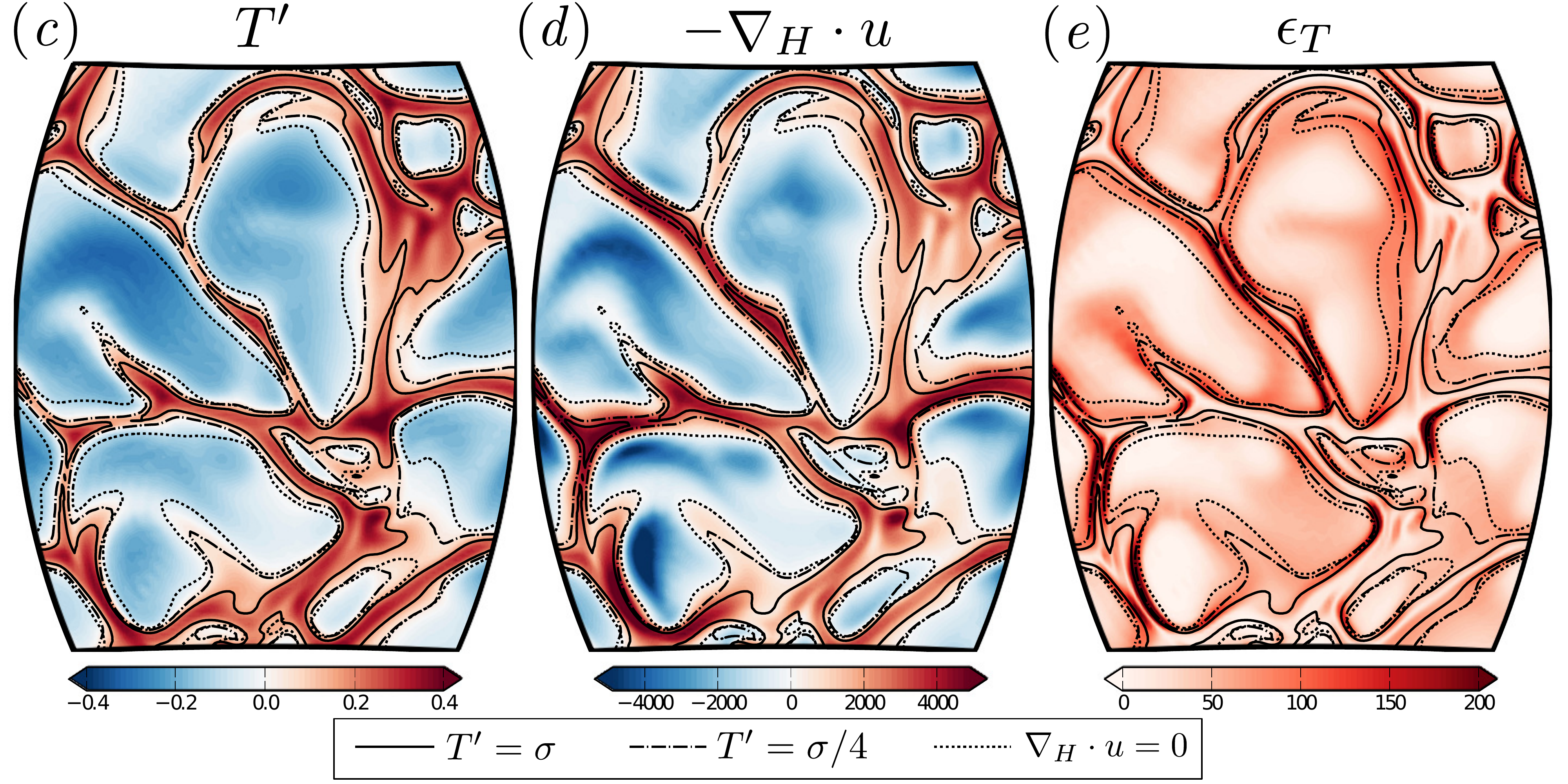}

 \caption{(\textit{a}) Temperature fluctuation at the inner 
thermal boundary layer $T'(r=r_i+\tbli)$ displayed in a Hammer projection for a 
case with $Ra=10^{6}$, $\eta=0.6$, $g=(r_o/r)^2$. (\textit{b}) Corresponding 
binarised extraction of the plumes boundaries using (\ref{eq:binary}) and 
$T' \leq \sigma/2$ to define the inter-plume area.  Zoomed-in 
contour of the temperature fluctuation $T'$ (\textit{c}), 
the horizontal divergence $\vec{\nabla}_H\cdot \vec{u}$ (\textit{d}) and the 
thermal dissipation rate $\epsilon_T$ (\textit{e}). The three black contour 
lines in panels (\textit{c}-\textit{e}) correspond to several criteria to 
extract the plume boundaries (\ref{eq:tauLim}).}
 \label{fig:binary}
\end{figure}

To further assess how the average plume properties of the inner and outer 
thermal boundary layers compare with each other in spherical geometry, we adopt 
a simpler strategy by only focussing on the statistics of the plume 
density. The plume density per surface unit at a given radius $r$ is expressed 
by
\begin{equation}
 \rho_p \sim \dfrac{N}{4\pi r^2},
\label{eq:rhop}
\end{equation}
where $N$ is the number of plumes, approximated here by the ratio of the 
spherical surface area to the mean inter-plume area $\bar{\cal 
S}$: 
\begin{equation}
N \sim \frac{4\pi r^2}{\bar{\cal S}}.
\end{equation}
This inter-plume area $\bar{\cal S}$ can be further related to the average 
plume separation  $\bar{\ell}$ via $\bar{\cal S} \sim (\pi/4)\,\bar{\ell}^2$. 

An accurate evaluation of the inter-plume area for each thermal boundary 
layer however requires to separate the plumes from the background fluid. 
Most of the criteria employed to determine the location of the plume 
boundaries are based on thresholds of certain physical quantities 
\citep[see][for a review of the different plume extraction 
techniques]{Shishkina08}. This encompasses threshold values on the temperature 
fluctuations $T'$ \citep{Zhou02}, on the vertical velocity $u_r$ 
\citep{Ching04} 
or on the thermal dissipation rate $\epsilon_T$ \citep{Shishkina05}. The choice 
of the threshold value however remains an open problem. Alternatively, 
\cite{Vipin13} show that the sign of the horizontal divergence of the velocity 
$\vec{\nabla}_H\cdot \vec{u}$ might provide a simple and threshold-free 
criterion to separate the plumes from the background fluid
\[
 \vec{\nabla}_H\cdot \vec{u} = \frac{1}{r\sin\theta}\frac{\partial}{\partial 
\theta}\left(\sin\theta\,u_\theta\right)+ 
\frac{1}{r\sin\theta}\frac{\partial u_\phi}{\partial 
\phi} = -\frac{1}{r^2}\frac{\partial}{\partial r}\left(r^2\, u_r\right).
\]
Fluid regions with $\vec{\nabla}_H\cdot \vec{u} < 0$ indeed correspond to local 
regions of positive vertical acceleration, expected inside the plumes, while 
the fluid regions with $\vec{\nabla}_H\cdot \vec{u} > 0$ characterise 
the inter-plume area. 

To analyse the statistics of ${\cal S}$, we thus consider here several criteria
based either on a threshold value of the temperature fluctuations or on the 
sign of the horizontal divergence. This means that a given inter-plume 
area at the inner (outer) thermal boundary layer is either defined 
as an enclosed region surrounded by hot (cold) sheet-like plumes carrying a 
temperature perturbation $|T'| > t$; or by an enclosed region with $ 
\vec{\nabla}_H\cdot \vec{u}>0$. To further estimate the possible 
impact of the chosen threshold value on ${\cal S}$, we vary $t$ 
between $\sigma/4$ and $\sigma$. This yields
\begin{equation}
{\cal S}(r) \equiv r^2 \oint_{{\cal T}} 
\sin\theta\, \mathrm{d}\theta\,\mathrm{d}\phi,   
\label{eq:binary}
\end{equation}
where the physical criterion ${\cal T}_i$ (${\cal T}_o$) to extract the plume 
boundaries at the inner (outer) boundary layer is given by
\begin{equation}
\begin{aligned}
 {\cal T}_i =  \left\lbrace
 \begin{aligned}
  T'(r_\lambda^i, \theta, \phi) & \leq t, \quad t \in [\sigma(r_\lambda^i), 
\sigma(r_\lambda^i/2), \sigma(r_\lambda^i/4)], \\
  \vec{\nabla}_H\cdot \vec{u} & \geq 0,
 \end{aligned}
 \right.  \\
   {\cal T}_o =  \left\lbrace
 \begin{aligned}
  T'(r_\lambda^o, \theta, \phi) & \geq t, \quad t \in [\sigma(r_\lambda^i), 
\sigma(r_\lambda^i/2), \sigma(r_\lambda^i/4)],  \\
  \vec{\nabla}_H\cdot \vec{u} & \geq 0,
 \end{aligned}
 \right. 
\end{aligned}
\label{eq:tauLim}
\end{equation}
where $r_\lambda^i=r_i+\tbli$ ($r_\lambda^o=r_o-\tblo$) for the inner (outer) 
thermal boundary layer.

Figure~\ref{fig:binary} shows an example of such a characterisation procedure 
for the inner thermal boundary layer of a numerical model with 
$Ra=10^{6}$, $\eta=0.6$, $g=(r_o/r)^2$. Panel (\textit{b}) illustrates a 
plume extraction process when using  $|T'| > \sigma/2$ to determine the 
location of the plumes: the black area correspond to the inter-plume spacing 
while the white area correspond to the complementary plume network location. 
The fainter emerging sheet-like plumes are filtered out and only the remaining 
``skeleton'' of the plume network is selected by this extraction process. 
The choice of $\sigma/2$ is however arbitrary and can influence the 
evaluation of the number of plumes. The insets displayed in panels 
(\textit{c}-\textit{e}) illustrate the sensitivity of the plume extraction 
process on the criterion employed to detect the plumes. In particular, using 
the threshold based on the largest temperature fluctuations $|T'| > \sigma$ can 
lead to the fragmentation of the detected plume lanes into several 
isolated smaller regions. As a consequence, several neighbouring inter-plume 
areas can possibly be artificially connected when using this criterion. In 
contrast, using the sign of the horizontal divergence to estimate the plumes 
location yields much broader sheet-like plumes. As visible on panel 
(\textit{e}), the plume boundaries frequently correspond to local maxima of the 
thermal dissipation rate $\epsilon_T$ \citep{Shishkina08}.

\begin{figure}
 \centering
 \includegraphics[width=10cm]{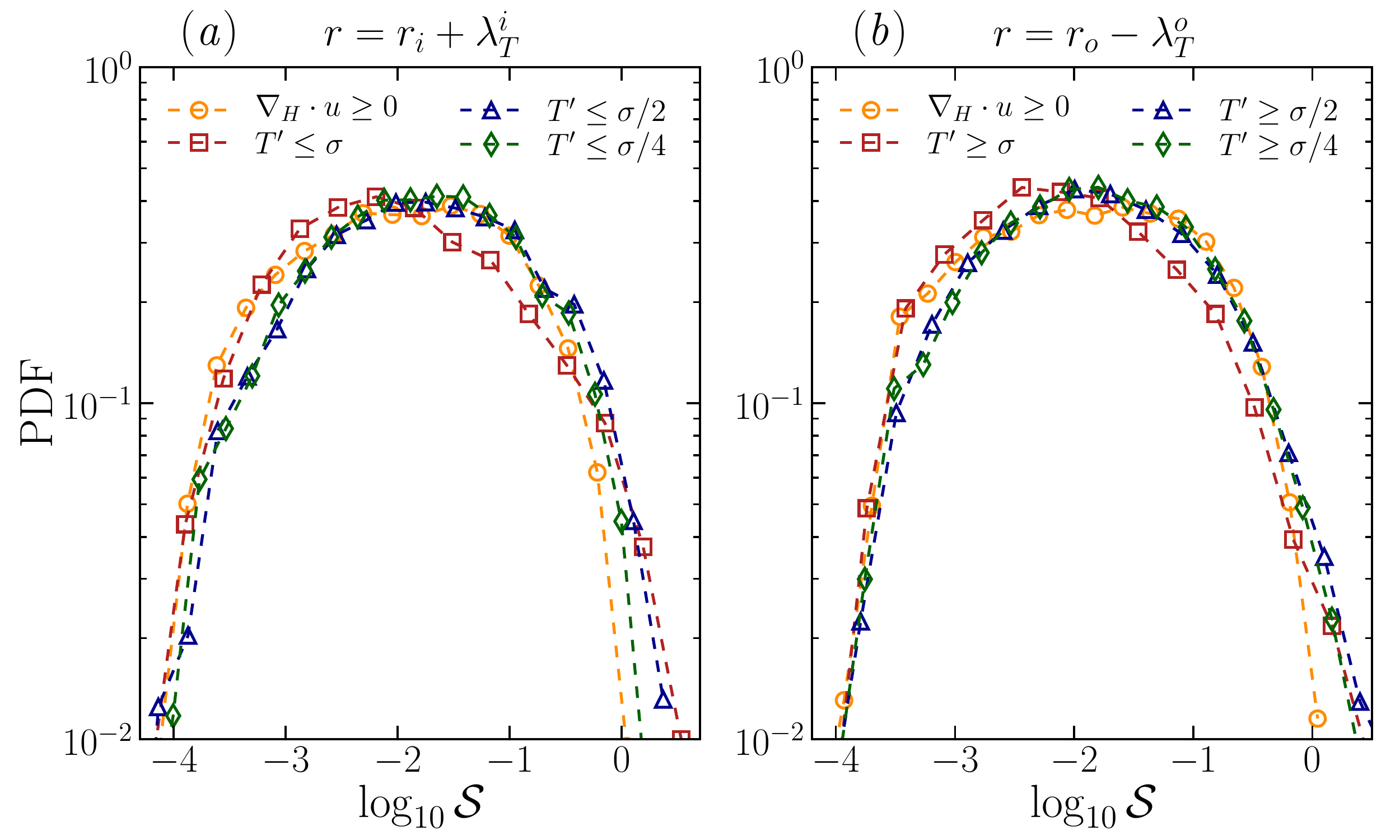}
 \caption{Probability density functions (PDFs) of the dimensionless
inter-plume area $\cal S$ at the inner (\textit{a}) and at the outer 
(\textit{b}) thermal boundary layers using different criteria to extract the 
plumes (\ref{eq:tauLim}) for a model with $Ra=4\times 10^7$, $\eta=0.8$ 
and $g=r/r_o$.}
 \label{fig:pdfs}
\end{figure}

For each criterion given in (\ref{eq:tauLim}), we then calculate the area 
of each bounded black surface visible in figure~\ref{fig:binary}(\textit{b}) 
to construct the statistical distribution of the inter-plume area for both 
thermal boundary layers. Figure~\ref{fig:pdfs} compares the resulting PDFs 
obtained by combining several snapshots for a numerical model with $Ra=4\times 
10^7$, $\eta=0.8$ and $g=r/r_o$. Besides the criterion $|T'| > 
\sigma$ which yields PDFs that are slightly shifted towards smaller inter-plume 
spacing areas, the statistical distributions are found to be relatively
insensitive to the detection criterion (\ref{eq:tauLim}). We therefore restrict 
the following comparison to the criterion $|T'| > \sigma /2$ only.


\begin{figure}
 \centering
 \includegraphics[width=\textwidth]{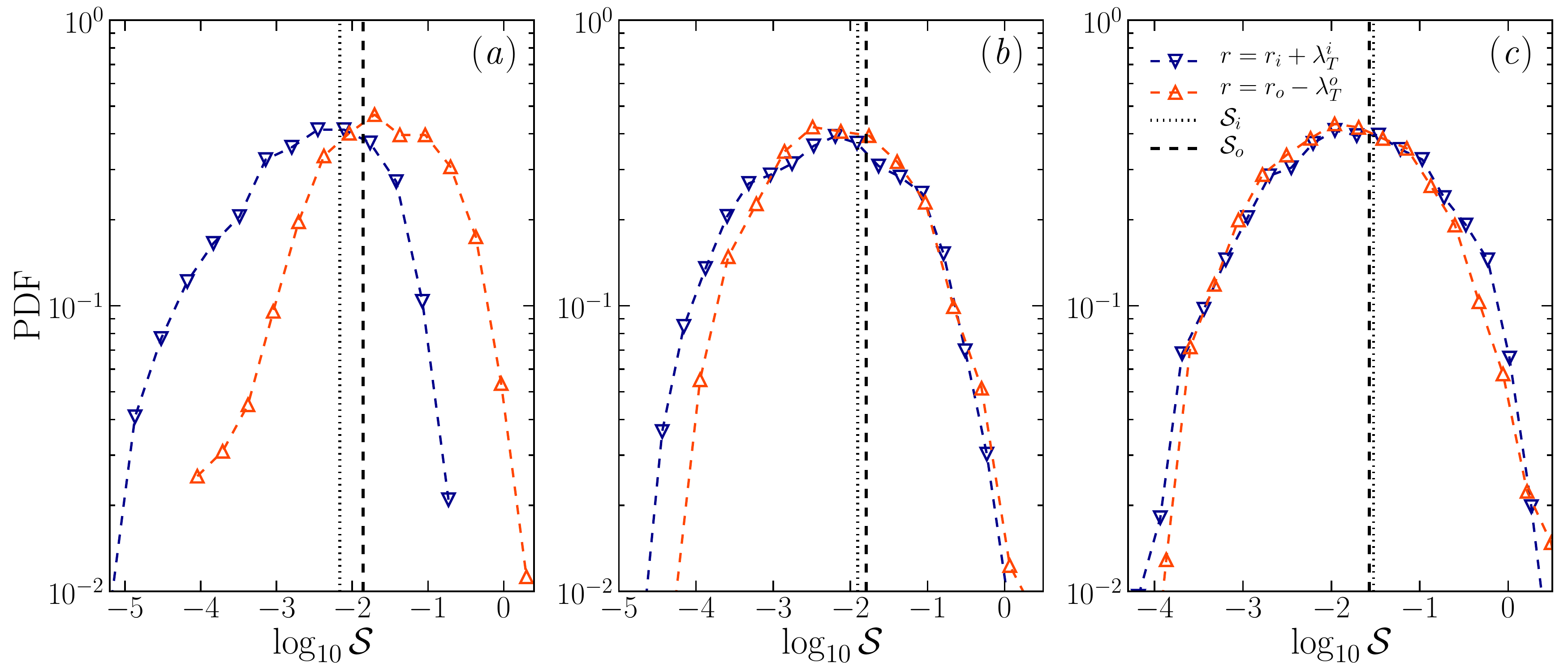}
 \caption{PDF of the dimensionless inter-plume area ${\cal 
S}$ at the outer (orange upward triangles) and at the inner 
(blue downward triangles) thermal boundary layers using $|T'| > \sigma/ 
2$ to extract plumes. Panel (\textit{a}) corresponds to a numerical model with 
$Ra=3\times 10^{6},\,\eta=0.3$ and $g=(r_o/r)^5$. Panel 
(\textit{b}) corresponds to a model with $Ra=10^{8},\,\eta=0.6$ 
and $g=(r_o/r)^2$. Panel (\textit{c}) corresponds to a model 
with $Ra=4\times 10^{7},\,\eta=0.8$ and $g=r/r_o$.
The two vertical lines correspond to the predicted values of the mean 
inter-plume area $\bar{\cal S}$  derived from (\ref{eq:plumeSpacing}) for 
both thermal boundary layers.}
 \label{fig:distVoids}
\end{figure}

Figure~\ref{fig:distVoids} shows the PDFs for the three numerical models of 
figure~\ref{fig:3D}. For the 
two cases with $\eta=0.6$ and $\eta=0.8$ (panels 
\textit{b-c}), the statistical distributions for both thermal boundary layers 
nearly overlap. This means that the inter-plume area is similar at 
both spherical shell surfaces. In contrast, for the case with $\eta=0.3$ (panel 
\textit{a}), the two PDFs are offset relative to each other. However, 
the peaks of the distributions remain relatively close, meaning that once again 
the inner and the outer thermal boundary layers share a similar average 
inter-plume area. \cite{Puthenveettil05} and \cite{Zhou10} demonstrated that 
the thermal plume statistics in turbulent RB convection follow a log-normal 
distribution \citep[see also][]{Shishkina08,Puthenveettil11}.
The large number of plumes in the cases with $\eta=0.6$ and $\eta=0.8$ 
(figure~\ref{fig:3D}\textit{b-c}) would allow a characterisation of the nature 
of the statistical distributions. However, this would be much more difficult in 
the $\eta=0.3$ case (figure~\ref{fig:3D}\textit{a}) in which the plume density 
is significantly weaker. As a consequence, no further attempt has been made to 
characterise the exact nature of the PDFs visible in figure~\ref{fig:distVoids}, 
although the universality of the log-normal statistics reported by 
\cite{Puthenveettil05} and \cite{Zhou10} likely indicates that the same 
statistical distribution should hold here too.

\begin{figure}
 \centering
 \includegraphics[width=11cm]{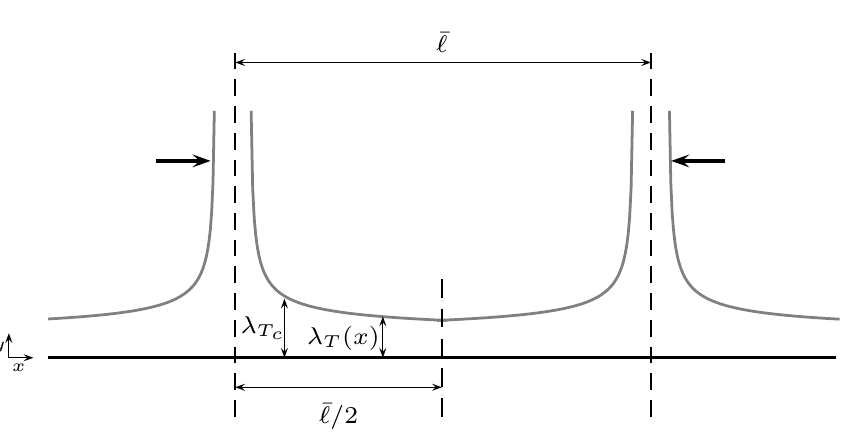}
 \caption{Schematic showing two adjacent plumes separated by a distance 
$\bar{\ell}$. The thick black arrows indicate the direction of merging of the 
two plumes.}
 \label{fig:sketch}
\end{figure}

The inter-plume area statistics therefore reveals that the inner and 
the outer thermal boundary layers exhibit a similar average plume density, 
independently of the spherical shell geometry and the gravity profile. Assuming 
$\rho_p^o \simeq \rho_p^i$ would allow us to close the system of equations 
(\ref{eq:flux}-\ref{eq:jump}) and thus finally estimate $\Delta T^i$, $\Delta 
T^o$ and $\tblo/\tbli$. This however requires us to determine an analytical 
expression of the average inter-plume area $\bar{\cal S}$ or equivalently of the 
mean plume separation $\bar{\ell}$ that depends on the boundary layer thickness 
and the temperature drop. 

Using the boundary layer equations for natural convection \citep{Rotem69}, 
\cite{Puthenveettil11} demonstrated that the  
thermal boundary layer thickness follows
\begin{equation}
 \lambda_T^{i,o}(x) \sim \frac{x}{(Ra_x^{i,o})^{1/5}}, 
\end{equation}
where $x$ is the distance along the horizontal direction and $Ra_x^{i,o} = 
\alpha g \Delta T^{i,o} x^3/\nu\kappa$ is a Rayleigh number based on the 
lengthscale $x$ and on the boundary layer temperature jumps $\Delta T^{i,o}$.
As shown on figure~\ref{fig:sketch}, using $x=\bar{\ell}/2$ 
\citep{Puthenveettil05,Puthenveettil11} then allows to establish the following 
relation for the average plume spacing
\begin{equation}
 \frac{\lambda_T}{\bar{\ell}} \sim \frac{1}{Ra_\ell^{1/5}}.
\end{equation}
which yields
\begin{equation}
 \bar{\ell}_i \sim \sqrt{\frac{\alpha g_i \Delta T^i {\tbli}^5 
}{\nu \kappa}},\quad  
 \bar{\ell}_o \sim \sqrt{\frac{\alpha g_o \Delta T^o {\tblo}^5 
}{\nu \kappa}},
 \label{eq:plumeSpacing}
\end{equation}
for both thermal boundary layers.
We note that an equivalent expression for the average plume spacing can be 
derived from a simple mechanical description of the equilibrium between 
production and coalescence of plumes in each boundary layer 
\citep[see][]{Parmentier00,King13}.

Equation~(\ref{eq:plumeSpacing}) is however expected to be only valid at 
the scaling level. The vertical lines in figure~\ref{fig:distVoids} therefore
correspond to the estimated 
average inter-plume area for both thermal boundary layers using 
(\ref{eq:plumeSpacing}) and $\bar{\cal S}_{i,o} = 0.3\,\bar{\ell}_{i,o}^2$. 
The predicted average inter-plume area is in good agreement 
with the peaks of the statistical distributions for the three cases discussed 
here. The expression~(\ref{eq:plumeSpacing}) therefore provides a reasonable 
estimate of the average plume separation 
\citep{Puthenveettil05,Puthenveettil11,Gunasegarane14}. The comparable
observed plume density at both thermal boundary layers thus yields
\begin{equation}
 \rho_p^i = \rho_p^o \,\rightarrow\, \frac{\alpha g_i \Delta T^i 
{\tbli}^5}{\nu\kappa} =\frac{\alpha g_o \Delta T^o {\tblo}^5}{\nu\kappa}.
\label{eq:plumeSpacing2}
\end{equation}
Using Eqs.~(\ref{eq:flux}-\ref{eq:jump}) then allows us to finally estimate the 
temperature jumps and the ratio of the thermal boundary layer thicknesses in 
our dimensionless units:
\begin{equation}
 \Delta T^i = \frac{1}{1+\eta^{5/3}\,\chi_g^{1/6}},\quad \Delta T^o = T_m =
\frac{\eta^{5/3}\,\chi_g^{1/6}}{1+\eta^{5/3}\,\chi_g^{1/6}}, \quad  
\frac{\tblo}{\tbli} = \frac{\chi_g^{1/6}}{\eta^{1/3}}.
\label{eq:predTemp}
\end{equation}

\begin{figure}
 \centering
 \includegraphics[width=\textwidth]{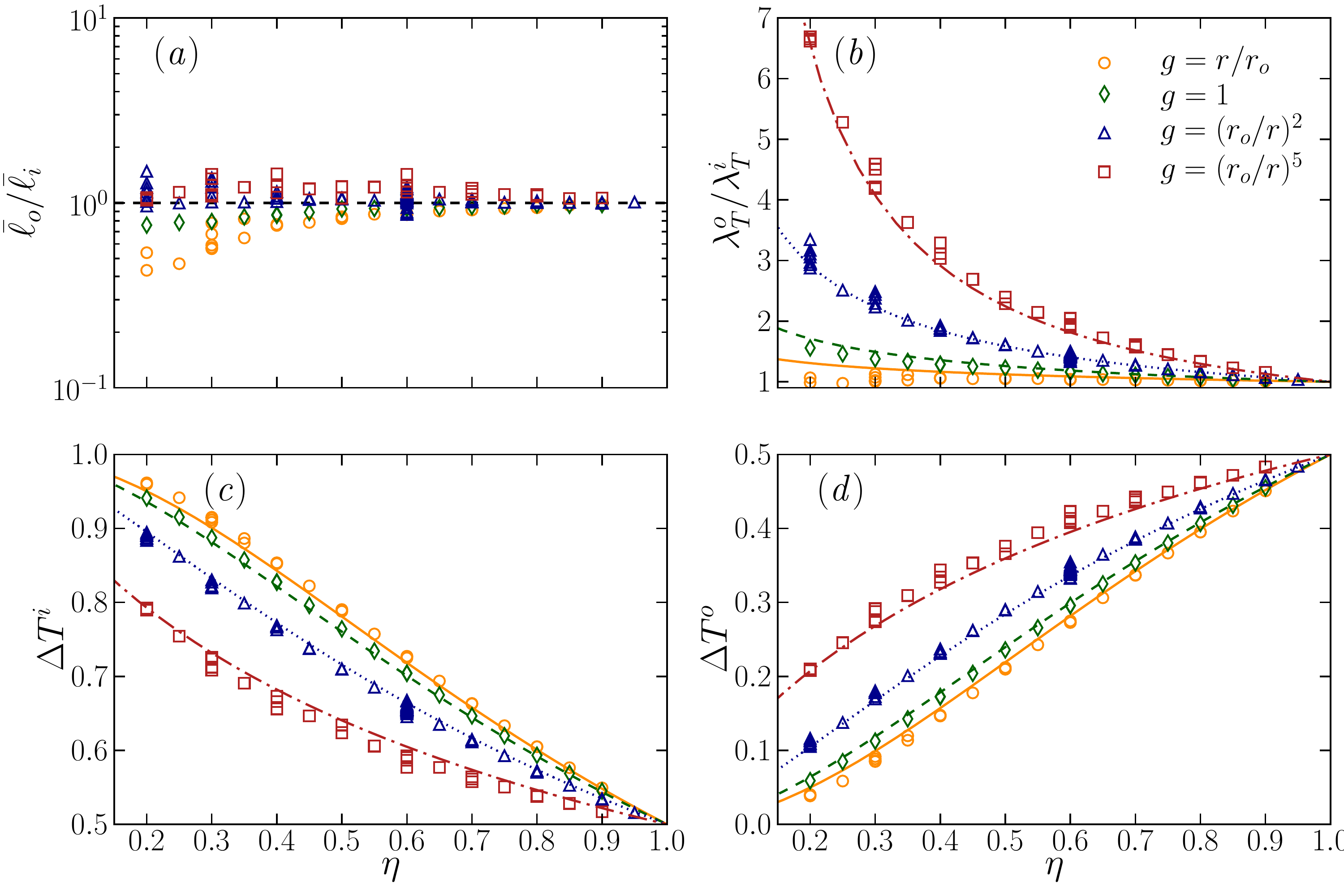}
 \caption{(\textit{a}) Ratio of the thermal plume separation estimated by 
(\ref{eq:plumeSpacing}) for various radius ratios and gravity profiles. 
The horizontal dashed line corresponds to the identity of the average plume 
separation between both thermal boundary layers, i.e. $\bar{\ell}^i = 
\bar{\ell}^o$. (\textit{b}) Ratio of thermal boundary layer thicknesses. 
(\textit{c}) Temperature drop at the inner boundary layer. (\textit{d}) 
Temperature drop at the outer boundary layer. The lines in panels 
(\textit{b}-\textit{d}) correspond to the theoretical prediction given in 
(\ref{eq:predTemp}).}
 \label{fig:predictedTh}
\end{figure}

Figure~\ref{fig:predictedTh} shows the ratios $\bar{\ell}_o/\bar{\ell}_i$, 
$\tblo/\tbli$ and the temperature jumps $\Delta T^i$ and $\Delta T^o$. In 
contrast to the previous criteria, either coming from 
the marginal stability of the boundary layer (\ref{eq:jarvis}, 
figure~\ref{fig:ra}) 
or from the identity of the temperature fluctuations at mid-shell  
(\ref{eq:predTemp}, figure~\ref{fig:wl}), the ratio of the 
average plume separation $\bar{\ell}_o/\bar{\ell}_i$ now falls much closer to 
the unity line. Some deviations are nevertheless still visible for spherical 
shells with $\eta \leq 0.4$ and $g=r/r_o$ (orange circles). The 
comparable average plume density between both boundary layers allows us 
to accurately predict 
the asymmetry of the thermal boundary layers $\tblo/\tbli$ and the 
corresponding temperature drops for the vast majority of the numerical cases 
explored here (solid lines in panels \textit{b}-\textit{d}).

\begin{figure}
 \centering
 \includegraphics[width=8cm]{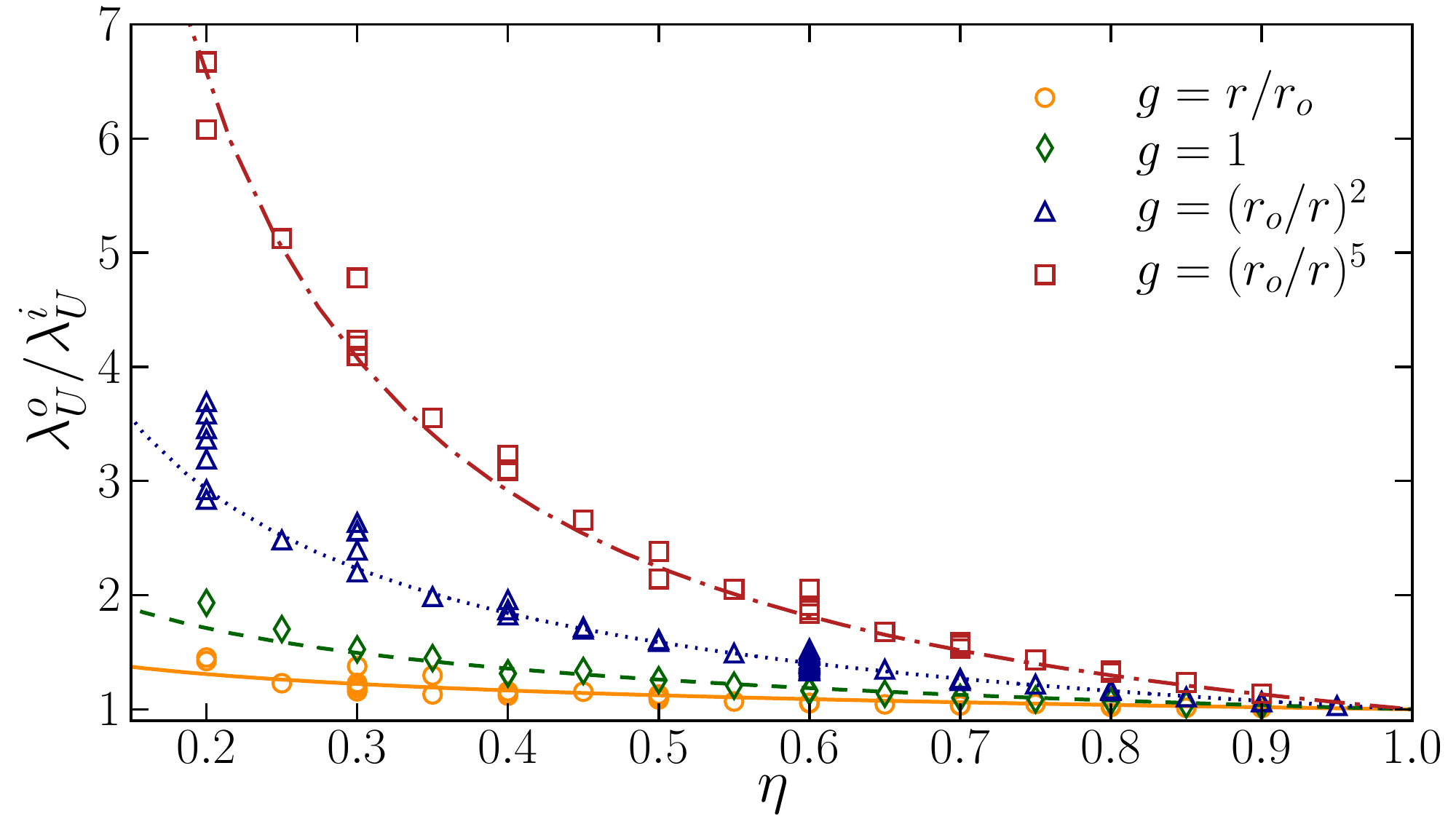}
 \caption{Ratio of the viscous boundary layer thicknesses for various aspect 
ratios and gravity profiles. The lines correspond to the theoretical 
prediction given in (\ref{eq:predvisc}).}
 \label{fig:predictedUbl}
\end{figure}

As we consider a fluid with $Pr=1$, the viscous boundary layers should 
show a comparable degree of asymmetry as the thermal boundary layers. 
(\ref{eq:predTemp}) thus implies
\begin{equation}
 \frac{\ublo}{\ubli} = \frac{\tblo}{\tbli} =\frac{\chi_g^{1/6}}{\eta^{1/3}}.
 \label{eq:predvisc}
\end{equation}
Figure~\ref{fig:predictedUbl} shows the ratio of the viscous boundary 
layer thicknesses for the different setups explored in this 
study. The observed asymmetry between the two spherical shell surfaces is in a 
good agreement with (\ref{eq:predvisc}) (solid black lines).

\subsection{Thermal boundary layer scalings}

\begin{figure}
 \centering
 \includegraphics[width=8.5cm]{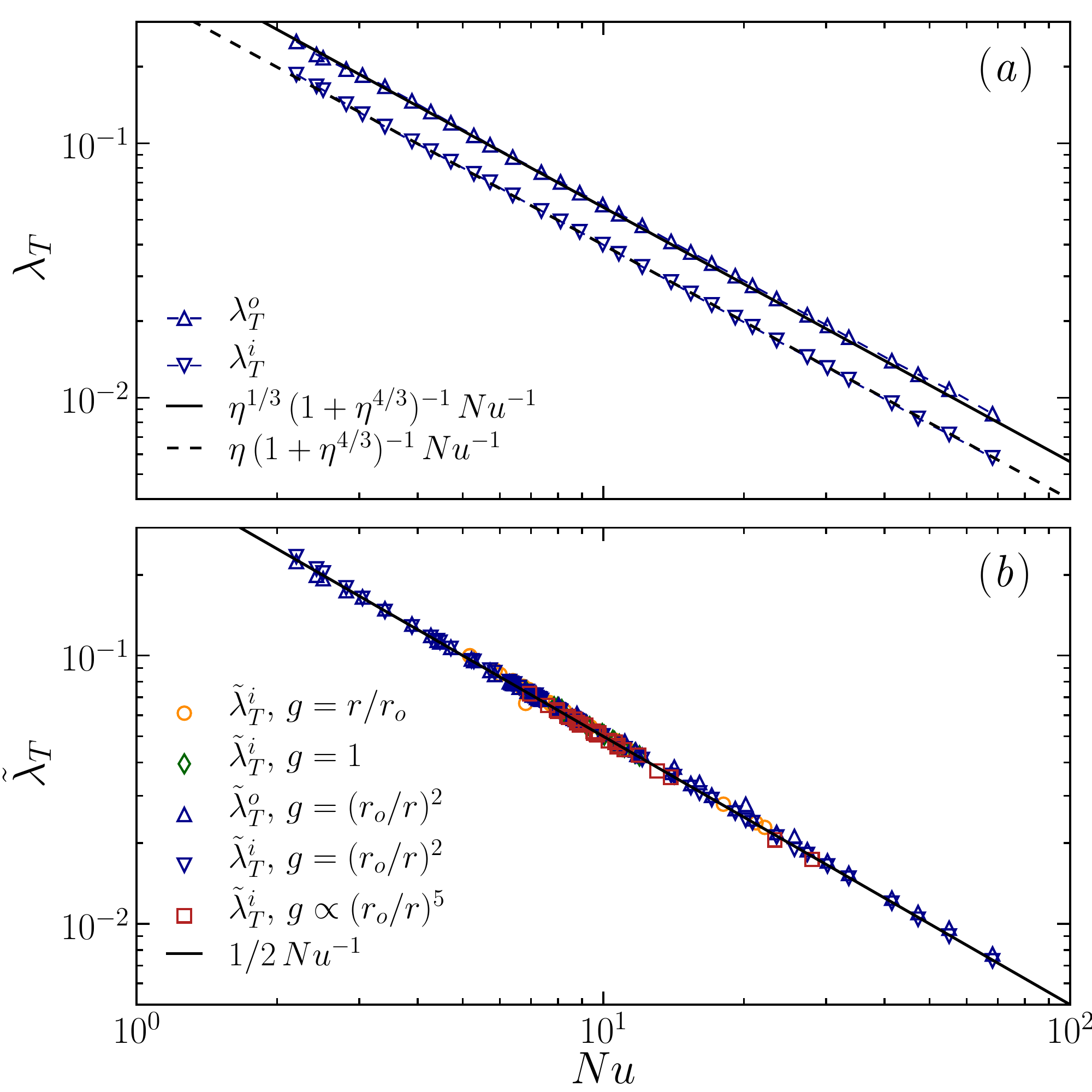}
 \caption{(\textit{a}) Thermal boundary layer thicknesses at the outer 
boundary ($\tbli$) and at the inner boundary ($\tblo$)  as a function 
of the Nusselt number for the cases of Table~\ref{tab:results}. The two lines 
correspond to the theoretical predictions from 
(\ref{eq:tblnu}). (\textit{b}) Normalised boundary layer thicknesses 
as a function of the Nusselt number for different radius ratios and gravity 
profiles. For the sake of clarity, the outer boundary layer is only displayed 
for the cases with $g=(r_o/r)^2$. The solid line corresponds to 
$\tilde{\lambda}_T = 0.5\,Nu^{-1}$ (\ref{eq:redtbl}).}
 \label{fig:thBL}
\end{figure}

Using (\ref{eq:predTemp}) and the definition of the Nusselt number 
(\ref{eq:nudef}), we can derive the following scaling relations for the 
thermal boundary layer thicknesses:
\begin{equation}
 \tbli = \frac{\eta}{1+\eta^{5/3}\,\chi_g^{1/6}}\,\frac{1}{Nu}\,, \quad \tblo = 
\frac{\eta^{2/3}\chi_g^{1/6}}{1+\eta^{5/3}\,\chi_g^{1/6}}\,\frac{1}{Nu}.
\label{eq:tblnu}
\end{equation}
Figure~\ref{fig:thBL}(\textit{a}) demonstrates that the boundary layer 
thicknesses for the numerical simulations of Table~\ref{tab:results} 
($g=(r_o/r)^2$ and $\eta=0.6$) are indeed in close agreement with the 
theoretical predictions. To further check this scaling for other spherical 
shell configurations, we introduce the following 
normalisation of the thermal boundary layer thicknesses
\[
\tilde{\lambda}_T^i=\frac{1}{2}\,\frac{1+\eta^{5/3}\,\chi_g^{1/6}}{\eta}\tbli,
\quad \tilde{\lambda}_T^o= \frac{1}{2}\,
\frac{1+\eta^{5/3}\,\chi_g^{1/6}}{\eta^{2/3}\,\chi_g^{1/6}}
\tblo.
\]
This allows us to derive a unified scaling that does not depend on the choice 
of the gravity profile or on the spherical shell geometry
\begin{equation}
\tilde{\lambda}_T = \tilde{\lambda}_T^i = \tilde{\lambda}_T^o = \frac{1}{2\,Nu}.
\label{eq:redtbl}
\end{equation}
Figure~\ref{fig:thBL}(\textit{b}) shows this normalised boundary layer 
thickness for the different spherical shell configurations explored here. 
Despite the variety of the physical setups, the normalised boundary layer 
thicknesses are in good agreement with the predicted behaviour.

\begin{figure}
 \centering
 \includegraphics[width=8.5cm]{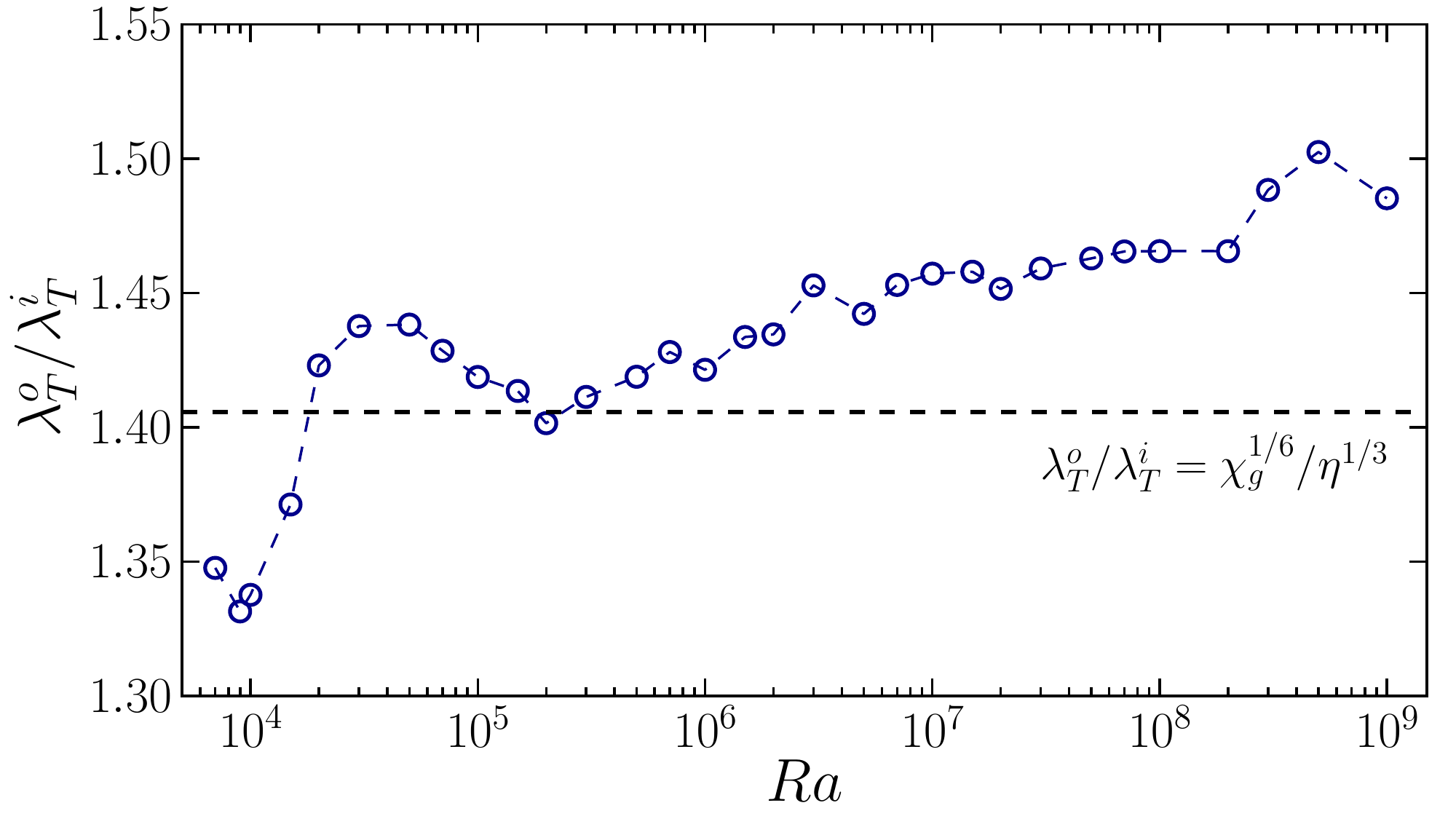}
 \caption{Ratio of the thermal boundary layer thicknesses $\tblo/\tbli$  as a 
function of the Rayleigh number for the numerical models of 
Table~\ref{tab:results} with $\eta=0.6$ and $g=(r_o/r)^2$. The horizontal 
dashed line corresponds to the predicted ratio $\tblo/\tbli$ given in 
(\ref{eq:predTemp}).}
 \label{fig:raDep}
\end{figure}

A closer inspection of figure~\ref{fig:predictedTh}(\textit{b}) and 
Table~\ref{tab:results} nevertheless reveals a remaining weak dependence of the 
ratio $\tblo/\tbli$ on the Rayleigh number. This 
is illustrated in figure~\ref{fig:raDep} 
which shows $\tblo/\tbli$  as a function of $Ra$ for the $\eta=0.6$, 
$g=(r_o/r)^2$ cases of 
Table~\ref{tab:results}. Although some complex variations are visible, the 
first-order trend is a very slow increase of $\tblo/\tbli$ with $Ra$ 
($10\%$ increase over five decades of $Ra$). No evidence 
of saturation is however visible and further deviations from the predicted 
ratio 
(horizontal dashed line) might therefore be expected at larger $Ra$. 
This variation might cast some doubts on the validity of the 
previous derivation for higher Rayleigh numbers. This may imply that either the 
plume separation is not conserved at higher $Ra$; 
or that the estimate of the average plume spacing is too simplistic to 
capture the detailed plume physics in turbulent convection.

\section{Boundary layer analysis}
\label{sec:viscBl}

The Grossmann Lohse (GL) theory relies on the assumption that the viscous and 
the thermal boundary layers are not yet turbulent. This is motivated by the 
observation of small boundary layer Reynolds numbers $Re_s=Re\,\lambda/d <200$  
in experimental convection up to $Ra \simeq 10^{14}$, which remain well 
below the expected transition to fully turbulent boundary layers  
\citep[expected at $Re_s\sim 420$, see][]{Ahlers09}. The boundary layer flow is 
therefore likely laminar and follows the 
Prandtl-Blasius (PB) laminar boundary layer theory 
\citep{Prandtl04,Blasius08,Schlichting00}. The PB theory assumes a balance 
between the viscous forces, important in the boundary 
layers, and inertia which dominates in the bulk of the fluid. For the numerical 
models with unity Prandtl number, this directly implies that the 
boundary layer thicknesses are inversely proportional to the square-root of $Re$
\begin{equation}
 \lambda_U \sim \lambda_T \sim Re^{-1/2}\,.
 \label{eq:lam_bl}
\end{equation}

\begin{figure}
 \centering
 \includegraphics[width=8.5cm]{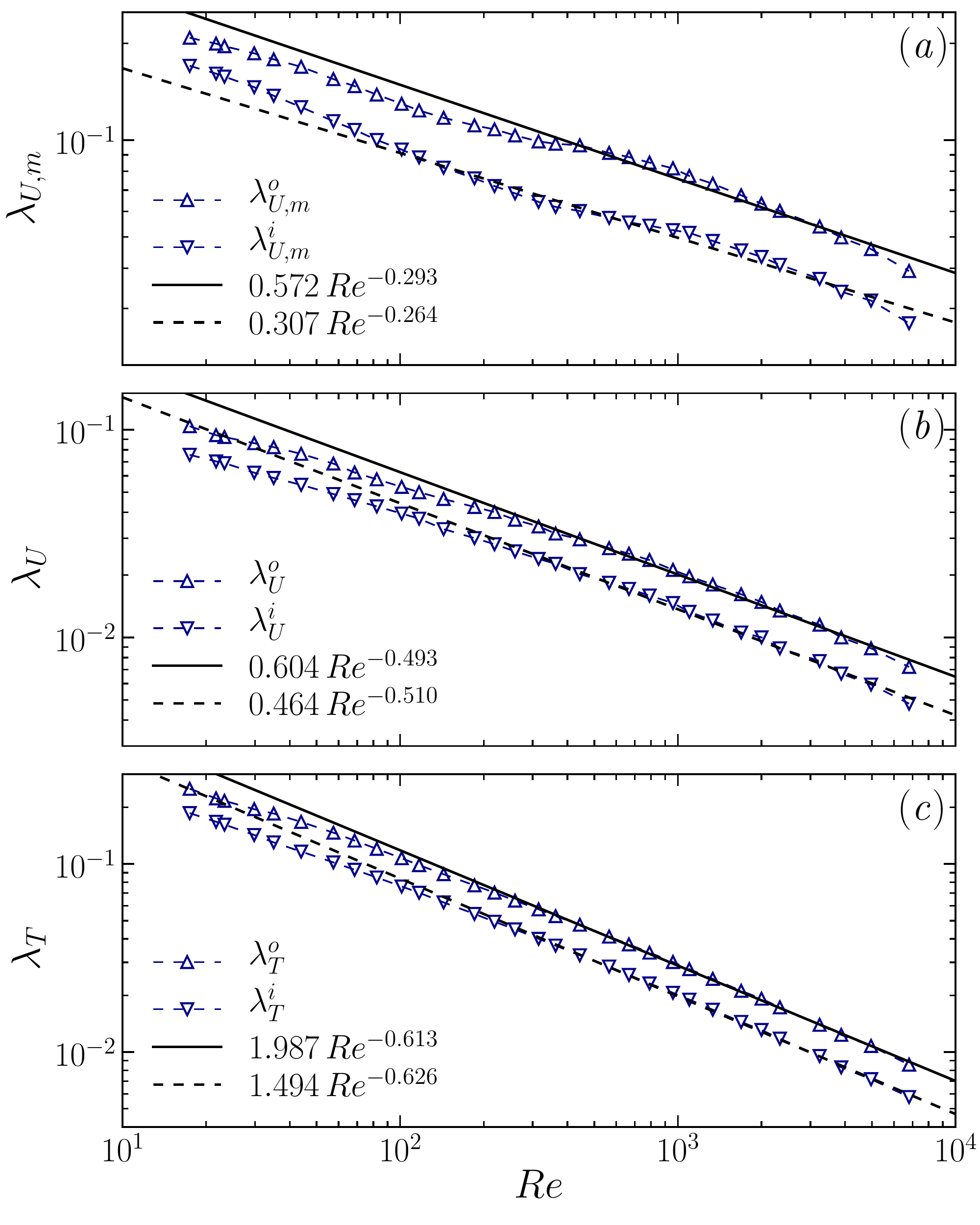}
 \caption{(\textit{a}) Viscous boundary layer thicknesses at the outer 
boundary ($\umbli$) and at the inner boundary ($\umblo$)  as a function 
of the Reynolds number for the cases of Table~\ref{tab:results} ($\eta=0.6$, 
$g=(r_o/r)^2$). (\textit{b}) 
Same for the other definition of the viscous boundary layer, i.e. $\ubli$ and 
$\ublo$. (\textit{c}) Thermal boundary layer thicknesses at the outer boundary 
($\tbli$) and at the inner boundary ($\tblo$) as a function of $Re$. The black 
lines in panels (\textit{a}, \textit{b} and \textit{c}) correspond to the 
least-square fit to the data for the numerical models with $Ra\geq 10^{6}$ 
(i.e. $Re > 250$).}
 \label{fig:bl_re} 
\end{figure}

Figure~\ref{fig:bl_re}(\textit{a}-\textit{b}) shows a test of this theoretical 
scaling for the two different definitions of the viscous boundary layer 
introduced in \S~\ref{subsec:defs}. Confirming 
previous findings by \cite{Breuer04}, the commonly-employed definition based on 
the location of the horizontal velocity maxima 
yields values that significantly differ from the theoretical prediction 
(\ref{eq:lam_bl}). The least-square fit to the data for the cases with 
$Re > 250$ indeed gives values relatively close to 
$\lambda_{U,m}\sim Re^{-1/4}$, an exponent already reported in the 
experiments by \cite{Lam02} and in the numerical models in cartesian geometry 
by \cite{Breuer04} and \cite{King13}. In addition, $\lambda_{U,m}$ is 
always significantly larger than $\lambda_T$, which is at odds with the 
expectation $\lambda_T\simeq\lambda_U$ when $Pr=1$ (see Table~\ref{tab:results} 
for detailed values).

Adopting the intersection of the two tangents to define the viscous boundary 
layers (figure~\ref{fig:bl_re}\textit{b}) leads to exponents much closer to 
the predicted value of $1/2$ in the high-$Re$ regime. The viscous boundary 
layer thicknesses obtained with this second definition are in addition
found to be relatively close to the thermal boundary layer thicknesses in the 
high Reynolds regime, i.e. $\lambda_U\simeq\lambda_T$.  Therefore, both the 
expected scaling of $\lambda_U$ with $Re$ and the similarities between thermal 
and viscous boundary layer thicknesses strongly suggest that the 
tangent-intersection method is a more appropriate way to estimate the actual 
viscous boundary layer. We therefore focus on this definition in the following.
For low Reynolds numbers ($Re < 200$), however, the viscous boundary layer 
thicknesses deviate from the PB theory and follow a shallower slope around 
$Re^{-0.4}$. This deviation implies a possible inaccurate description of the 
low $Ra$ cases by the GL theory (see below).

Figure~\ref{fig:bl_re}(\textit{c}) shows that the corresponding scaling of 
the thermal boundary layer with $Re$ follows a similar trend as the 
viscous boundary layers. The best fit to the data for the cases 
with $Re > 250$ yields exponents moderately larger than the theoretical 
prediction (\ref{eq:lam_bl}), while the low$-Re$ cases follow a shallower 
exponent. At this point, we can simply speculate that this difference might 
possibly arise because of the inherent dynamical nature of the thermal boundary 
layers.

\begin{figure}
 \centering
 \includegraphics[width=\textwidth]{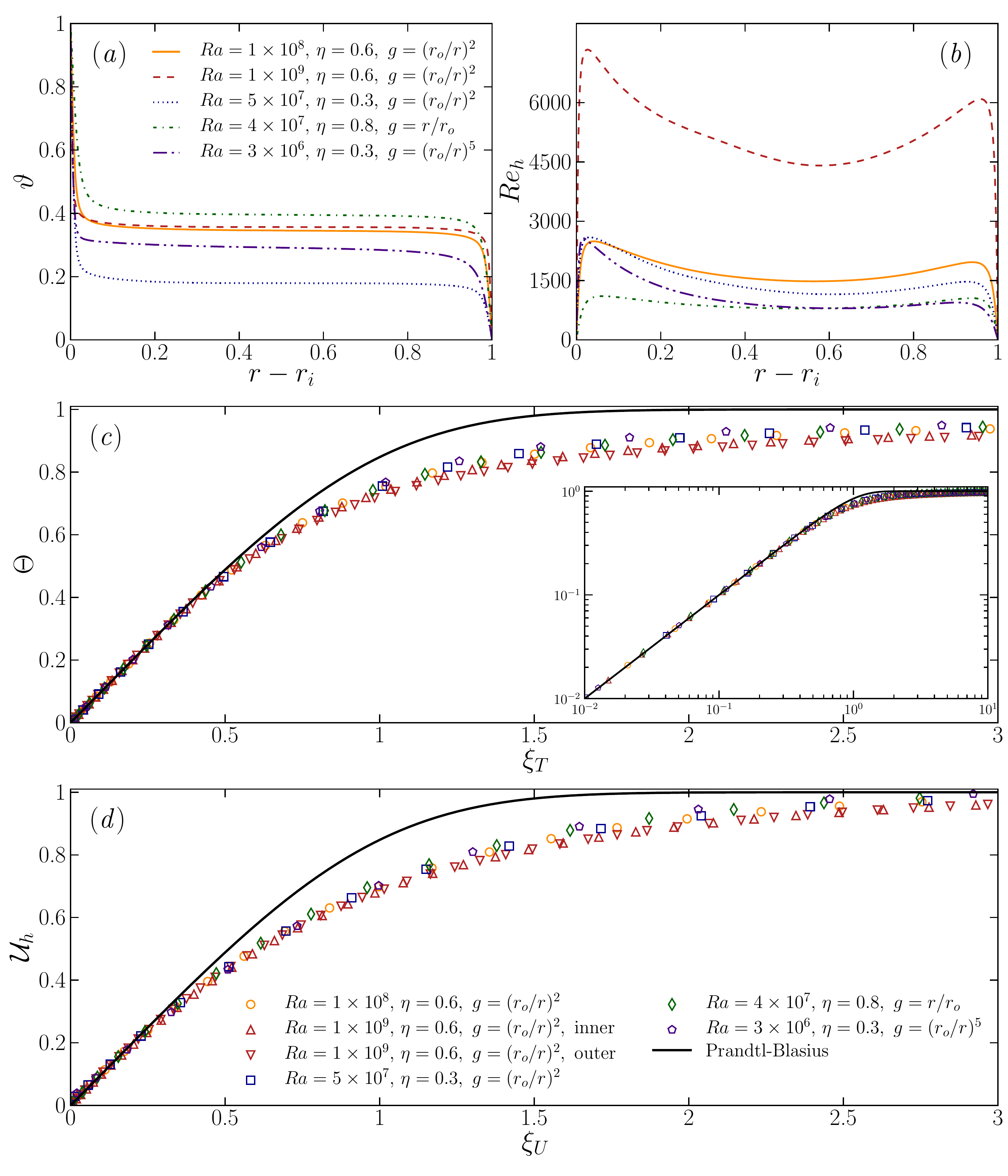}
 \caption{ (\textit{a}-\textit{b}) Radial profiles of the time and 
horizontally-averaged temperature $\vartheta$ and horizontal velocity 
$Re_h$. (\textit{c}) $\Theta$ as a function of $\xi_T$. (\textit{d})
${\cal U}_h$ as a function of $\xi_U$. The solid lines 
in panels (\textit{c-d}) corresponds to the Prandtl-Blasius solution. The inset 
in panel (\textit{c}) shows $\Theta$ in double-logarithmic scale. For the 
sake of clarity, the outer boundary layer is only displayed for one single case 
in the panels (\textit{c}-\textit{d}) ($Ra=10^9$, $\eta=0.6$, $g=(r_o/r)^2$).}
 \label{fig:selfSim}
\end{figure}

For a meaningful comparison with the boundary layer theory, we define
new scaling variables for the distance to the spherical shell 
boundaries. These variables are introduced to compensate for the changes in the 
boundary layer thicknesses that arise when $Ra$, $\eta$ or $g$ are modified. 
This then allows us to accurately characterise the shape of both the 
temperature and the flow profiles in the boundary layers and to 
compare them with the PB boundary layer profiles. To do so, we introduce the 
self-similarity variables $\xi_T$ and $\xi_U$ for both the 
inner and the outer spherical shell boundaries:
\begin{equation}
\xi_T^i = \frac{r-r_i}{\tbli}, \quad \xi_T^o = \frac{r_o-r}{\tblo}, \quad
\xi_U^i = \frac{r-r_i}{\ubli}, \quad \xi_U^o = \frac{r_o-r}{\ublo}.
\label{eq:xi}
\end{equation}
We accordingly define the following rescaled temperatures for both 
boundaries
\begin{equation}
\tilde{T}^i(r,\theta,\phi,t)=\frac{T(r,\theta,\phi,t)-\vartheta(r_m)}{T_{bot}
-\vartheta(r_m)}, \quad
\tilde{T}^o(r,\theta,\phi,t)=\frac{\vartheta(r_m)-T(r,\theta,\phi,t)}{
\vartheta(r_m)-T_ { top } },
\label{eq:thred}
\end{equation}
where $r_m=(r_i+r_o)/2$ is the mid-shell radius. The rescaled horizontal 
velocity is simply obtained by normalisation with its local 
maximum for each boundary layer:
\begin{equation}
\tilde{u}_h^i(r,\theta,\phi,t)= \dfrac{u_h(r,\theta,\phi,t)}{\max_i(Re_h)}, 
\quad
\tilde{u}_h^o(r,\theta,\phi,t)= \dfrac{u_h(r,\theta,\phi,t)}{\max_o(Re_h)}.
\label{eq:ured}
\end{equation}
To check the similarity of the profiles, we consider five 
numerical models with different $Ra$, $\eta$ and $g$. 
Figure~\ref{fig:selfSim}(\textit{a}-\textit{b}) show the typical mean 
horizontal velocity and temperature for these cases, 
while figure~\ref{fig:selfSim}(\textit{c}-\textit{d}) show the corresponding 
time and horizontally-averaged normalised quantities:
\begin{equation}  
 \Theta = 
\overline{\left\langle \tilde{T}^{i,o}_{\phantom{h}} \right\rangle_s}\,, \quad 
{\cal U}_h = \overline{\left\langle \tilde{u}_h^{i,o}\right\rangle_s}.
\label{eq:blstatic}
\end{equation}
As already shown in the previous section, the bulk temperature as well as 
the boundary layer asymmetry strongly depend on the gravity profile and the 
radius ratio of the spherical shell. Increasing $Ra$ leads to 
a steepening of the temperature profiles near the boundaries accompanied by a 
shift of the horizontal velocity maxima towards the walls.
Although $\vartheta$ and $Re_h$ 
drastically differ in the five cases considered here, introducing the 
normalised variables $\Theta$ and ${\cal U}_h$ allows to merge all 
the different configurations into one single radial profile. The 
upward and downward pointing triangles further indicate that those profiles are 
also independent of the choice of the boundary layer (at the inner or at the 
outer boundary). Finally, the solutions remain similar to each other when $Ra$ 
is varied, at least in the interval considered here (i.e. $10^8 \leq Ra \leq 
10^9$). These results are in good agreement with the profiles obtained in the 
numerical simulations by \cite{Shishkina09} that cover a similar range of 
Rayleigh numbers in cylindrical cells with $\Gamma=1$.

\begin{figure}
 \centering
 \includegraphics[width=\textwidth]{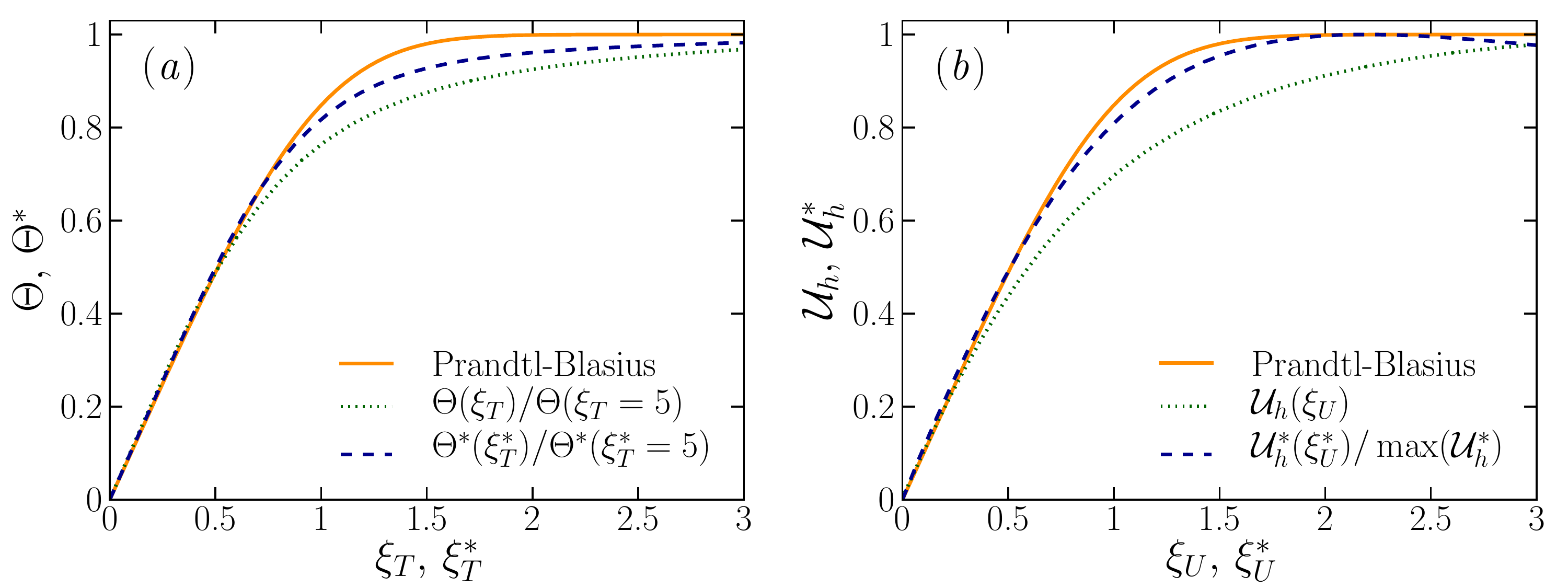}
 \caption{ (\textit{a}) Time and  horizontally-averaged normalised temperature 
profile in the fixed reference frame (dotted line, \ref{eq:blstatic}) and in 
the dynamical frame (dashed line, \ref{eq:blTdyn}) for a case with 
$Ra=10^8$, $\eta=0.6$ and $g=(r_o/r)^2$. (\textit{b}) Corresponding horizontal 
velocity profile in the fixed reference frame (dotted line, 
\ref{eq:blstatic})  and in the dynamical frame (dashed line, 
\ref{eq:blUdyn}). The solid lines in 
both panels correspond to the Prandtl-Blasius solution.}
 \label{fig:dynBls}
\end{figure}

Figure~\ref{fig:selfSim}(\textit{c}-\textit{d}) also compares
the numerical profiles with those derived from the PB boundary layer theory.
The time-averaged normalised temperature and velocity profiles slightly deviate 
from the PB profiles, confirming previous 2-D and 3-D numerical models by 
\cite{Zhou10a}, \cite{Shishkina09} and \cite{Stevens10} This 
deviation can be attributed to the intermittent 
nature of plumes that permanently detach from the boundary layers 
\citep{Sun08,Zhou10a,duPuits13}. When the boundary layer profiles are 
obtained from a time-averaging procedure at a fixed height with respect 
to the container frame (i.e. $\xi_T$ and $\xi_U$ are time-independent), they 
can actually sample both the bulk and the boundary layer dynamics as the 
measurement position can be at time either inside or outside the boundary layer 
(for instance during a plume emission). To better isolate the boundary layer 
dynamics, \cite{Zhou10a} therefore suggested  to study the physical properties 
in a time-dependent frame that accounts for the instantaneous boundary layer
fluctuations \citep[see also][]{Zhou10b,Stevens12,Shishkina15}.

We apply this dynamical rescaling method to our numerical models by defining 
local and instantaneous boundary layer thicknesses
\[
 \xi_T^*(\theta,\phi,t) = \frac{r_o-r}{\lambda_T^o(\theta,\phi,t)}, \quad
 \xi_U^*(\theta,\phi,t) = \frac{r_o-r}{\lambda_U^o(\theta,\phi,t)}\,.
\]
As the inner and the outer boundary layers exhibit the same 
behaviour (figure~\ref{fig:selfSim}), we restrict the following 
discussion to the outer boundary layer. Following \cite{Zhou10b} and 
\cite{Shi12}, the horizontal velocity and temperature profiles are given 
by
\begin{equation}
{\cal U}_h^*(\xi_U^*)= \overline{\left\langle 
\tilde{u}_h\left(r,\theta,\phi,t\,\big|\,r=r_o-\xi_U^*\lambda_U^o(\theta,\phi,
t) \right)
\right\rangle_s}\,,
\label{eq:blTdyn}
\end{equation}
\begin{equation}
\Theta^*(\xi_T^*)= \overline{\left\langle 
\tilde{T}\left(r,\theta,\phi,t\,\big|\,r=r_o-\xi_T^*\lambda_T^o(\theta,\phi,
t)\right)\right\rangle_s}\,.
\label{eq:blUdyn}
\end{equation}
Practically, this dynamical rescaling strategy has been achieved by measuring 
the local and instantaneous boundary layer for each grid coordinates 
$(\theta,\phi)$ for several snapshots. Following \cite{Zhou10a} and 
\cite{Stevens12}, the temperature profiles have been further normalised to some 
position outside the thermal boundary layer (here $\xi_T=5$ or $\xi_T^*=5$) to 
ease the comparison with the classical definition of the boundary layer in the 
fixed reference frame (\ref{eq:blstatic}). Figure~\ref{fig:dynBls} shows an 
example of this dynamical rescaling method applied to a case with $Ra=10^8$, 
$\eta=0.6$ and $g=(r_o/r)^2$. The temperature and horizontal velocity 
profiles in the spatially and 
temporally varying local frame are now in much closer agreement with the PB 
laminar 
profiles than those obtained in the fixed reference frame.

Because of the numerical cost of the whole procedure, the 
dynamical rescaling has only been tested on a limited number of cases. 
Applying the same method to the numerical model with $Ra=10^9$ 
($\eta=0.6$, $g=(r_o/r)^2$) yields nearly indistinguishable profiles
from those displayed in figure~\ref{fig:dynBls}. This further indicates 
that boundary layers in spherical shells are laminar in the $Ra$ 
range explored here and can be well described by the PB theory, provided 
boundary layers are analysed in a time-dependent frame, which fluctuates with 
the local and instantaneous boundary layer thicknesses.

\section{Dissipation analysis}
\label{sec:dissip}

\subsection{Bulk and boundary layer contributions to viscous and thermal 
dissipation rates}

The prerequisite of a laminar boundary layer seems fulfilled in our numerical 
models and we can thus try to apply the GL formalism to our dataset.
The idea of the GL theory is to separate the 
viscous and thermal dissipation rates into two contributions, one coming from 
the fluid bulk (indicated by the superscript $bu$ in the following) and one 
coming from the boundary layers ($bl$), such that
\begin{equation}
 \epsilon_T = \epsilon_T^{bu}+\epsilon_T^{bl}, \quad \epsilon_U = 
\epsilon_U^{bu}+\epsilon_U^{bl},
\label{eq:dissiprates}
\end{equation}
where the contributions from the bulk are defined by
\[
 \epsilon_T^{bu} = \frac{4\pi}{V}\int_{r_i+\tbli}^{r_o-\tblo}
\overline{\left\langle\left(\nabla T\right)^2\right\rangle_s}\,
r^2{\rm d}r, \quad  \epsilon_U^{bu} = \frac{4\pi}{V}\int_{r_i+\ubli}^{r_o-\ublo}
\overline{\left\langle\left(\vec{\nabla}\times\vec{u}\right)^2\right\rangle_s}\,
r^2{\rm d}r,
\]
and the boundary layer contributions are given by
\[
  \begin{aligned}
 \epsilon_T^{bl} &= 
\frac{4\pi}{V}\int_{r_i}^{r_i+\tbli}\overline{%
\left\langle\left(\nabla T\right)^2\right\rangle_s}\,r^2{\rm 
d}r + \frac{4\pi}{V}\int_{r_o-\tblo}^{r_o}
\overline{\left\langle\left(\nabla T\right)^2\right\rangle_s}\,
r^2{\rm d}r\,, \\
 \epsilon_U^{bl} &= 
\frac{4\pi}{V}\int_{r_i}^{r_i+\ubli}\overline{%
\left\langle\left(\vec{\nabla}\times\vec{u}\right)^2\right\rangle_s}\,r^2{\rm 
d}r+ \frac{4\pi}{V}\int_{r_o-\ublo}^{r_o}
\overline{\left\langle\left(\vec{\nabla}\times\vec{u}\right)^2\right\rangle_s}\, 
r^2{\rm d}r \,.
\end{aligned}
\]
The RB flows are then classified in the $Ra-Pr$ parameter space according to the 
dominant contributions to the viscous and thermal dissipation 
rates. This defines four regimes depending on $Ra$ and $Pr$: regime I when  
$\epsilon_U\simeq \epsilon_U^{bl}$ and $\epsilon_T\simeq \epsilon_T^{bl}$;  
regime II when 
$\epsilon_U\simeq \epsilon_U^{bu}$ and $\epsilon_T\simeq \epsilon_T^{bl}$; 
regime III when 
$\epsilon_U\simeq \epsilon_U^{bl}$ and $\epsilon_T\simeq \epsilon_T^{bu}$; and 
finally regime IV when
$\epsilon_U\simeq \epsilon_U^{bu}$ and $\epsilon_T\simeq \epsilon_T^{bu}$.

For a unity Prandtl number, the GL theory predicts that the flows should be 
dominated by dissipations in the boundary layer regions at low $Ra$  (regime I). 
A transition to another regime where dissipations in the fluid bulk dominate  
(regime IV) is expected to happen roughly around $Ra\simeq 10^8-10^{10}$  
\citep[][]{Grossmann00,Ahlers09,Stevens13}.

\begin{figure}
\centering 
\includegraphics[width=8.5cm]{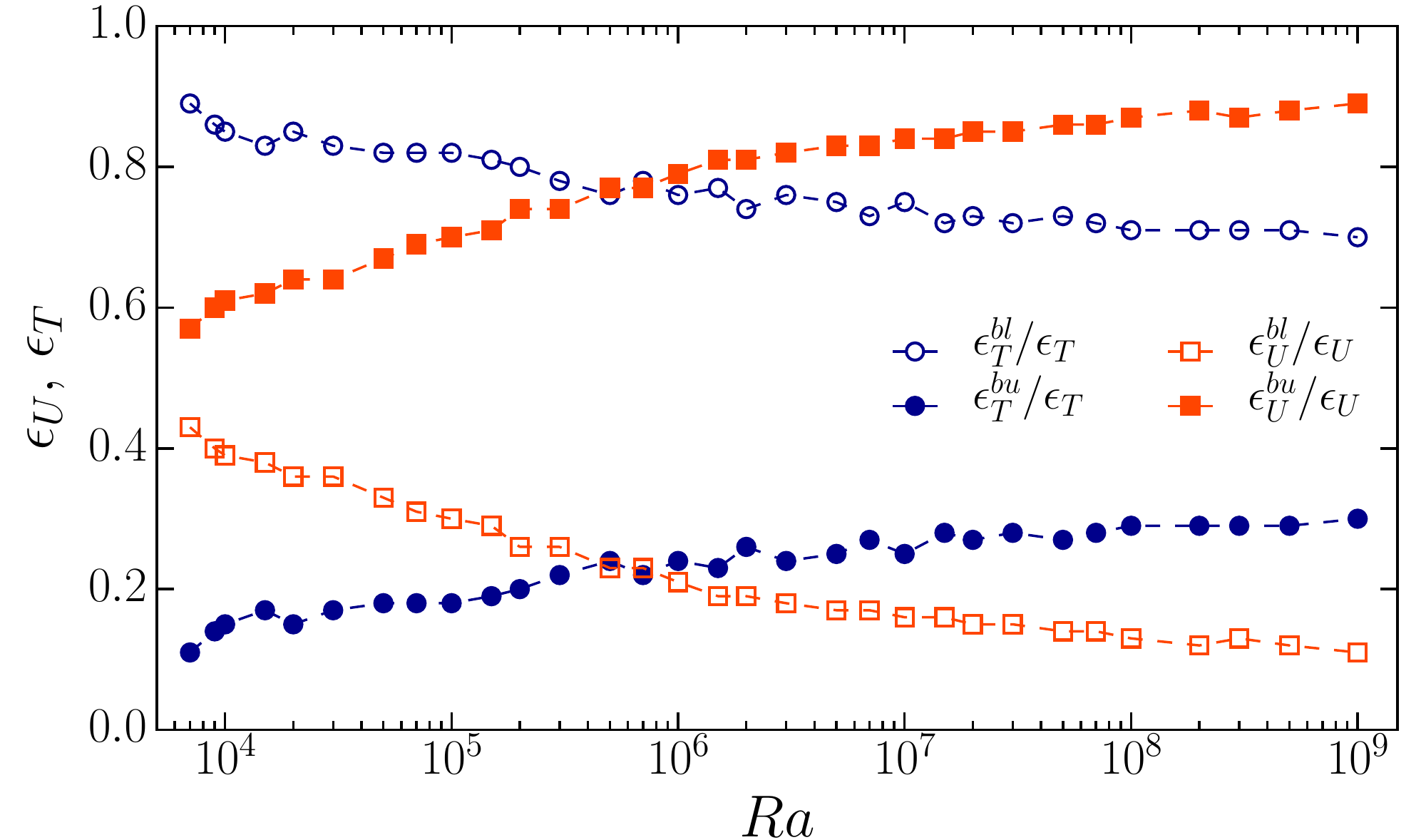}
 \caption{Measured contributions of the boundary layer (open symbols) and the 
fluid bulk (filled symbols) to the total viscous dissipation rate $\epsilon_U$ 
(orange squares) and the total thermal dissipation rate $\epsilon_T$ (blue 
circles).}
\label{fig:dissip}
\end{figure}

Figure~\ref{fig:dissip} shows the relative contributions of the bulk and 
boundary layers to the viscous and thermal dissipation rates. The viscous 
dissipation $\epsilon_U$ is 
always dominated by the bulk contribution: starting from roughly 60\% at 
$Ra=10^4$, it nearly reaches 90\% at $Ra=10^9$. In contrast, the 
thermal dissipation rate is always dominated by the boundary layer regions, 
such that $\epsilon_T^{bu}$ slowly increases from 10\% at $Ra=10^4$ to 30\% at 
$Ra=10^9$. According to the GL classification, all our numerical simulations 
thus belong to the regime II of the $Ra-Pr$ parameter space, in which 
$\epsilon_U^{bu}>\epsilon_U^{bl}$ and $\epsilon_T^{bl}>\epsilon_T^{bu}$. 
This seems at odds with the GL theory, which predicts that our DNS 
should be located either in the regime I or in the regime IV of the parameter 
space for the range of $Ra$ explored here ($10^3\leq Ra\leq 10^9$).

The dominance of the boundary layer contribution in the thermal dissipation 
rate was already reported by \cite{Verzicco03a} for the same range of $Ra$ 
values. This phenomenon may be attributed to the dynamical nature of the plumes 
which permanently detach from the boundary layers and penetrate in the bulk of 
the fluid. These thermal plumes have the same typical size as the boundary 
layer thickness and can thus be thought as ``detached 
boundary layers''. \cite{Grossmann04} have therefore suggested to modify 
their scaling theory to incorporate these detached boundary layers in the 
thermal dissipation rate. They propose to decompose $\epsilon_T$ into one 
contribution coming from the plumes ($\epsilon_T^{pl}$) and one coming from the
turbulent background ($\epsilon_T^{bg}$)
\begin{equation}
 \epsilon_T = \epsilon_T^{pl}+\epsilon_T^{bg},
 \label{eq:dissipGL04}
\end{equation}
Such a decomposition is however extremely difficult to conduct in spherical 
shells in which the very large aspect ratio of the convective layer yields a 
complex and time-dependent multi-cellular large scale circulation (LSC) pattern
\citep[see for instance][for the influence of  large $\Gamma$ on the 
LSC]{Bailon10}. In the following, we therefore first keep the initial 
decomposition (\ref{eq:dissiprates}) before coming back to the inherent problem 
of accurately separating the different contributions to the dissipation rate 
in \S~\ref{subsec:glob}.

\subsection{Individual scaling laws for the dissipation rates}

\begin{figure}
\centering 
\includegraphics[width=\textwidth]{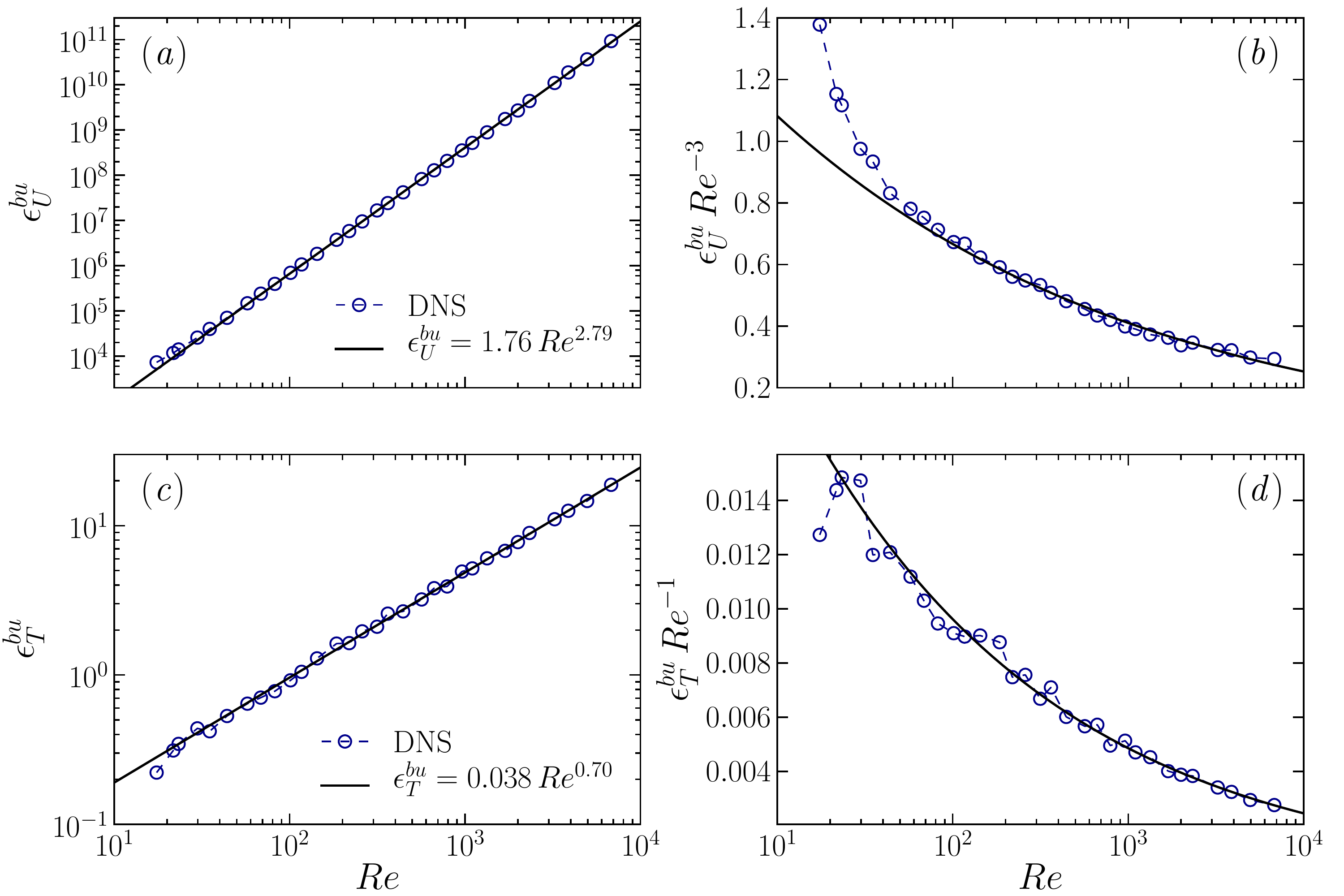}
 \caption{(\textit{a}) Viscous dissipation in the bulk of the fluid as a 
function of $Re$. (\textit{b}) Corresponding compensated $\epsilon_U^{bu}$ 
scaling. (\textit{c}) Thermal dissipation in the bulk of the fluid as a 
function of $Re$. (\textit{d}) Corresponding compensated $\epsilon_T^{bu}$ 
scaling. The solid black lines in the four panels correspond to the 
least-square fit to the data for the numerical models with
$Ra\geq 10^{5}$.}
 \label{fig:bulkdiss}
\end{figure}

Based on the hypothesis of homogeneous and isotropic turbulence, the GL theory 
assumes that the thermal and viscous dissipation rates in the bulk of the fluid 
scale like
\begin{equation}
 \epsilon_U^{bu}\sim Re^3,\quad \epsilon_T^{bu}\sim Re,
 \label{eq:predbulk}
\end{equation}
in our dimensionless units. 
Figure~\ref{fig:bulkdiss} shows the bulk dissipation rates as a function of 
$Re$ for the numerical models of Table~\ref{tab:results}. The best fit to the 
data (solid lines) for the cases with $Ra\geq 10^{5}$ yields 
$\epsilon_U^{bu}\sim Re^{2.79}$ and $\epsilon_T^{bu}\sim Re^{0.7}$, only 
roughly similar to the prediction (\ref{eq:predbulk}). These 
deviations from the theoretical exponents are further confirmed by the
compensated scalings $\epsilon_U^{bu}\,Re^{-3}$ and $\epsilon_T^{bu}\,Re^{-1}$ 
shown in panels (\textit{b}) and (\textit{d}), which
show a coherent remaining dependence on $Re$. Even at high Reynolds numbers, 
there is no evidence of convergence towards the exact expected scalings from 
the GL theory. This is particularly obvious for $\epsilon_T^{bu}$ which remains 
in close agreement with $\epsilon_T^{bu}\sim Re^{0.7}$ for the whole range of 
$Re$ values explored here (solid line in figure~\ref{fig:bulkdiss}\textit{d}). 
The dependence of  $\epsilon_U^{bu}$ on $Re$ shows a gradual steepening of the 
slope when $Re$ increases, which implies that $\epsilon_U^{bu}(Re)$ cannot be 
accurately represented by a simple power law. Similar deviations from 
(\ref{eq:predbulk}) have already been reported in the Taylor-Couette flow 
experiments by \cite{Lathrop92} and in the numerical simulations of RB 
homogeneous turbulence by \cite{Calzavarini05}.

A similar procedure can be applied to the dissipation rates in the 
boundary layers. In spherical shells, the volume fraction occupied by the 
boundary layers can be approximated by the following Taylor expansion to the 
first order
\[
 f_V(\lambda) = \frac{4\pi}{V}\left(\int_{r_i}^{r_i+ \lambda^i} r^2 {\rm d} r +
\int_{r_o-\lambda^o}^{r_o} r^2 {\rm d} r \right)
\simeq 3\,\frac{\lambda^i \eta^2+\lambda^o}{1+\eta+\eta^2},
\]
in the limit of thin boundary layers $\lambda \ll r_o-r_i$. 
The viscous dissipation rate in the boundary layers can 
then be estimated by
\[
 \epsilon_U^{bl} \sim \frac{U_{rms}^2}{\lambda_U^2}\,f_V(\lambda_U) \sim 
\frac{Re^2}{\lambda_U},
\]
in our dimensionless units. As demonstrated in \S~\ref{sec:viscBl}, the 
boundary layers are laminar and are in reasonable agreement with the PB 
boundary layer theory. This implies that $\lambda_U \sim 
Re^{-1/2}$ and thus yields
\begin{equation}
 \epsilon_U^{bl} \sim Re^{5/2}.
\end{equation}

\begin{figure}
\centering 
\includegraphics[width=\textwidth]{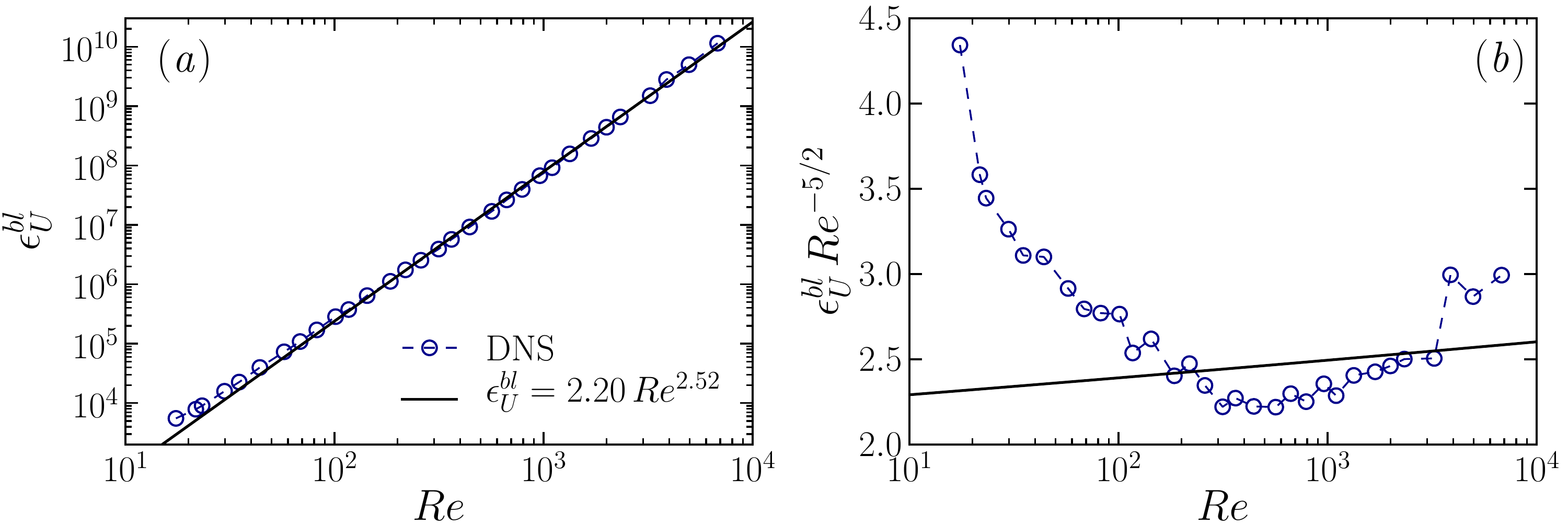}
 \caption{(\textit{a}) Viscous dissipation in the boundary layers as a 
function of $Re$. (\textit{b}) Corresponding compensated $\epsilon_U^{bl}$ 
scaling. The solid black lines in the four panels correspond to the 
least-square fit to the data for the numerical models with $Ra\geq 
10^{5}$.}
 \label{fig:blUdiss}
\end{figure}

Figure~\ref{fig:blUdiss} shows the viscous dissipation rate in the boundary 
layers as a function of $Re$ for the numerical models of 
Table~\ref{tab:results}. The least-square fit to the data 
(solid lines) yields $\epsilon_U^{bl}\sim Re^{2.52}$, in close agreement 
with the expected theoretical exponent. The compensated scaling displayed in 
panel (\textit{b}) reveals  a remaining weak secondary dependence of 
$\epsilon_U^{bl}$ on 
$Re$, which is not accurately captured by the power law. The local slope of 
$\epsilon_U^{bl}(Re)$ initially decreases with $Re$ (when $Re 
\lesssim 200$) before slowly increasing at higher Reynolds numbers 
($Re\gtrsim 200$). This behaviour 
suggests that, although simple power laws are at first glance in very good 
agreement with the GL theory, they may not account for 
the detailed variations of $\epsilon_U^{bl}(Re)$.

\begin{figure}
\centering 
\includegraphics[width=\textwidth]{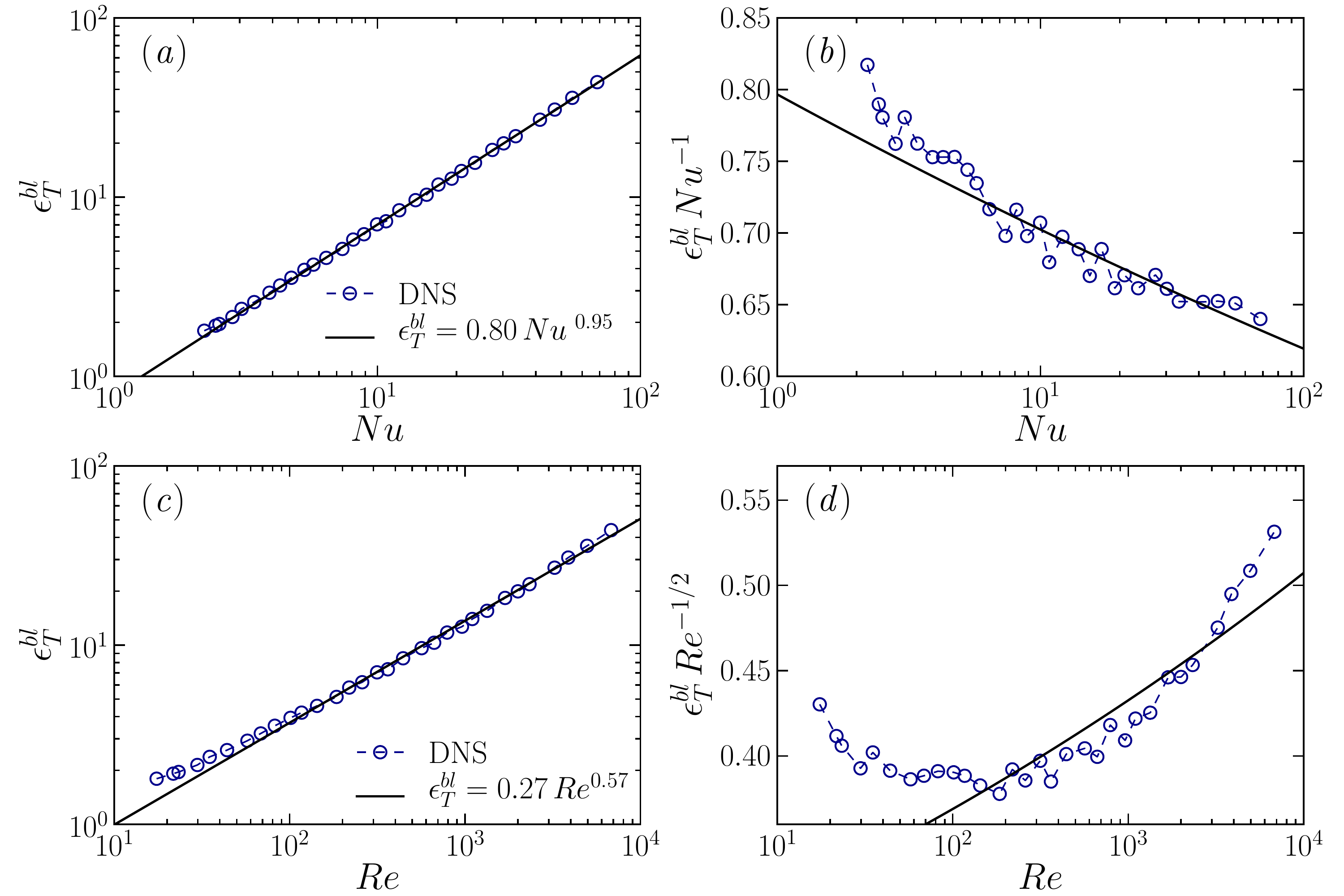}
 \caption{(\textit{a}) Thermal dissipation in the boundary layers as a 
function of $Nu$. (\textit{b}) Corresponding compensated 
$\epsilon_T^{bl}$ scaling. (\textit{c}) Thermal dissipation in the boundary 
layers as a function of $Re$. (\textit{d}) Corresponding compensated 
$\epsilon_T^{bl}$ scaling. The solid black lines in the four panels correspond 
to the least-square fit to the data for the numerical models with $Ra\geq 
10^{5}$.}
 \label{fig:blTdiss}
\end{figure}

The boundary layer contribution to the thermal dissipation rate is estimated in 
a similar way:
\[
 \epsilon_T^{bl} \sim \frac{\Delta T^2}{\lambda_T^2}\,f_V(\lambda_T)\sim 
\frac{1}{\lambda_T},
\]
which yields
\begin{equation}
 \epsilon_T^{bl} \sim Nu.
 \label{eq:etblnu}
\end{equation}
The laminar nature of the boundary layers also implies  $\lambda_T 
\sim Re^{-1/2}$ and thus 
\begin{equation}
 \epsilon_T^{bl} \sim Re^{1/2}.
 \label{eq:etblre}
\end{equation}
Figure~\ref{fig:blTdiss} shows $\epsilon_T^{bl}$ as a function of $Nu$ and $Re$ 
for the cases of Table~\ref{tab:results}. The least-square fits yield 
$\epsilon_T^{bl}\sim Nu^{0.95}$ and $\epsilon_T^{bl}\sim Re^{0.57}$ (solid 
lines), close to the expected exponents. However, the compensated 
scalings displayed in panels (\textit{b}) and (\textit{d}) reveal that the 
linear fit to $\epsilon_T^{bl}(Nu)$ remains in good agreement with the data, 
while $\epsilon_T^{bl}(Re)$ is not accurately described by such a simple fit. 
The solutions increasingly deviate from the power law at high 
Reynolds numbers with the local slope of 
$\epsilon_T^{bl}(Re)$ that gradually steepens with $Re$.

\subsection{Individual versus global scalings}
\label{subsec:glob}

Despite the overall fair agreement with the GL predictions, a close 
inspection of the dependence of the four dissipation rates on the Reynolds 
number reveals some remaining dependence on $Re$, which cannot be 
perfectly described by simple power laws. This is particularly obvious in the 
boundary layer contributions $\epsilon_U^{bl}(Re)$ and $\epsilon_T^{bl}(Re)$.
In addition, both thermal dissipation rates deviate 
stronger from the theoretical exponents than their viscous counterparts 
\citep[see also][]{Verzicco03a}. One obvious problem is the 
inherent difficult separation of bulk and boundary layer contributions already
discussed above. The dynamical plumes constantly departing from the boundary 
layers obviously complicate matters.
To check whether the general idea of a boundary layer and a bulk contribution
that both scale with the predicted exponents is at 
least compatible with the \emph{total} dissipation rates, we directly fit
\[
  \begin{aligned}
 \widehat{\epsilon_U} &=\widehat{\epsilon_U^{bu}}+\widehat{\epsilon_U^{bl}}  = 
& a\,Re^{3}+b\,Re^{5/2}\, , \\
  \widehat{\epsilon_T} &=\widehat{\epsilon_T^{bu}}+\widehat{\epsilon_T^{bl}} = 
& c\,Re+d\,Re^{1/2}\, .
\end{aligned}
\]
This leaves only the four prefactors ($a, b, c, d$) as free fitting parameters 
and yields
\begin{equation}
  \begin{aligned}
 \widehat{\epsilon_U} &=\widehat{\epsilon_U^{bu}}+\widehat{\epsilon_U^{bl}}  = 
& 0.248\,Re^{3}+7.084\,Re^{5/2}\, , \\
  \widehat{\epsilon_T} &=\widehat{\epsilon_T^{bu}}+\widehat{\epsilon_T^{bl}} = 
& 0.004\,Re+0.453\,Re^{1/2}\, .
\end{aligned}
\label{eq:scal_glob}
\end{equation}
We then compare this direct least-square fit of the total dissipation rates to 
the sum of the individual scalings obtained in the previous section
\begin{equation}
 \begin{aligned}
 \widehat{\epsilon_U} & =\widehat{\epsilon_U^{bu}}+\widehat{\epsilon_U^{bl}}
= &1.756\,Re^{2.79}+2.197\,Re^{2.52}\,, \\
  \widehat{\epsilon_T} & =\widehat{\epsilon_T^{bu}}+\widehat{\epsilon_T^{bl}}  
= &0.038\,Re^{0.7}+0.268\,Re^{0.57}\,.
 \end{aligned}
 \label{eq:scal_indiv}
\end{equation}

\begin{figure}
\centering 
\includegraphics[width=\textwidth]{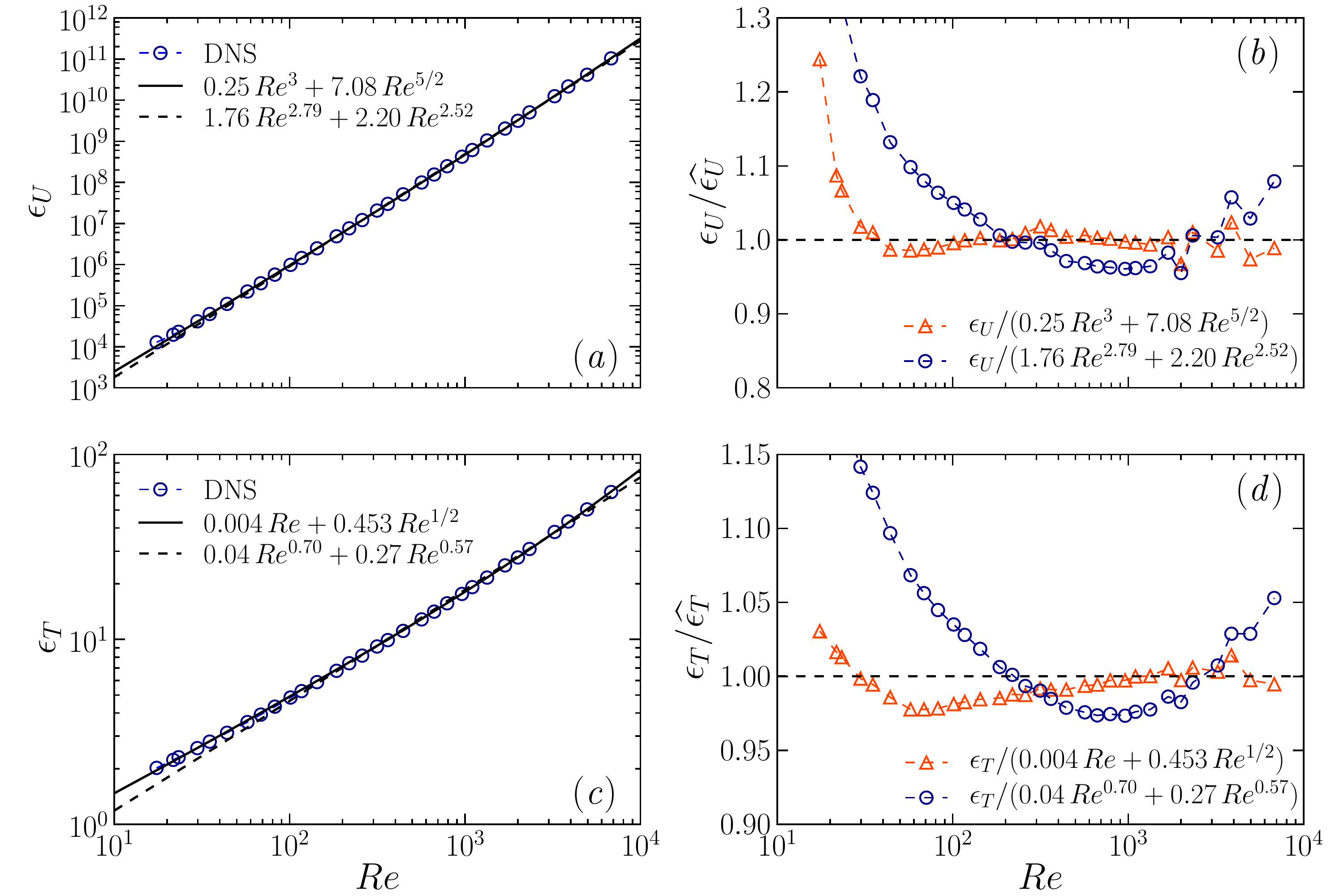}
 \caption{(\textit{a}) $\epsilon_U$ as a 
function of $Re$. (\textit{b}) $\epsilon_U$ normalised by  
the predictions coming from (\ref{eq:scal_glob}) (orange 
triangles) and (\ref{eq:scal_indiv}) (blue circles) as a function of $Re$. 
(\textit{c}) $\epsilon_T$ as 
a function of $Re$. (\textit{d}) $\epsilon_T$ normalised by 
the predictions coming from (\ref{eq:scal_glob}) (orange triangles) and 
(\ref{eq:scal_indiv}) (blue circles) as a function of $Re$. The dashed 
black lines in panels (\textit{c}-\textit{d}) correspond to the equality between 
the asymptotic scalings and the data.}
 \label{fig:globDiss}
\end{figure}

The accuracy of the two scalings (\ref{eq:scal_glob}) and (\ref{eq:scal_indiv}) 
are compared in figure~\ref{fig:globDiss}, which shows the total viscous and 
thermal dissipation rates as a function of $Re$ for the numerical models of 
Table~\ref{tab:results}. While the two scalings are nearly indistinguishable on 
the left panels (\textit{a}) and (\textit{c}), the corresponding normalised
scalings shown in panels (\textit{b}) and (\textit{d}) reveal some important 
differences. The scalings based on the sum of the power laws 
derived in the last section (\ref{eq:scal_indiv}) are in relatively poor 
agreement with the data ($5-10\%$ error for $Re > 10^2$) with no obvious 
asymptotic behaviour. On the other hand, the global scalings
(\ref{eq:scal_glob}) fall much closer to the actual values for the 
range $10^2<Re<10^4$ and approach an asymptote  for $Re > 10^2$. The deviations 
observed for the highest $Re$ cases have probably a numerical origin: the 
averaging timespan used to estimate the dissipation rates are likely too short 
in the most demanding cases to perfectly average out all the fluctuations.

The total thermal and viscous dissipation rates in our 
spherical shell simulations are thus better described by the sum of two 
power laws that follow the GL theory than by the sum of the asymptotic 
laws derived from the individual contributions, which suffers from an unclear 
separation of the boundary layer and bulk dynamics.

\begin{figure}
\centering 
\includegraphics[width=8.5cm]{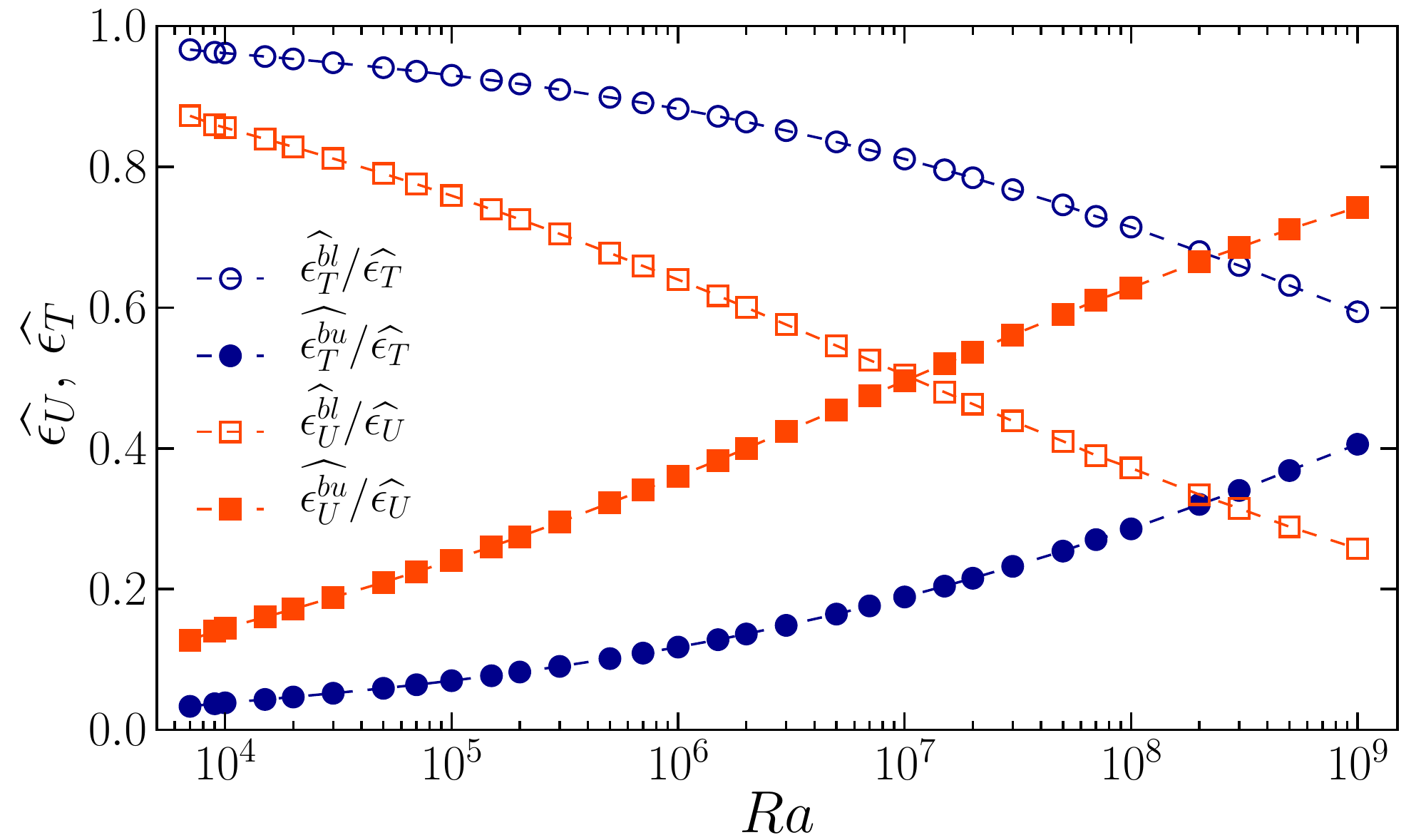}
 \caption{Estimated relative contributions of the boundary layer (open symbols) 
and the fluid bulk (filled symbols) to $\widehat{\epsilon_U}$ (orange 
squares) and $\widehat{\epsilon_T}$ (blue circles) using 
(\ref{eq:scal_glob}).}
\label{fig:newDissip}
\end{figure}

This result also sheds a new light on the placement of our 
numerical simulations in the GL regime diagram (figure~\ref{fig:dissip}). 
Equation~(\ref{eq:scal_glob}) directly provides the estimated relative 
contributions of the bulk and boundary layers to the viscous and thermal 
dissipation rates. Figure~\ref{fig:newDissip} shows these different 
contributions for the numerical models of Table~\ref{tab:results} and reveals a 
completely different balance than in figure~\ref{fig:dissip}. At low Rayleigh 
numbers, the estimated boundary layer 
contributions now dominate both the viscous and thermal dissipation rates. The 
viscous dissipation rate in the fluid bulk gradually increases with 
$Ra$ and dominates beyond $Ra > 10^7$. The bulk contribution to the thermal 
dissipation rate exhibits a similar 
trend, gradually increasing from roughly 5\% at $Ra=10^4$ to more than 40\% at 
$Ra=10^9$. While the thermal dissipation rate in the fluid bulk never dominates 
in the regime explored here, our scaling predict that it will do so for $Ra 
\gtrsim 3\times 10^9$.
These values would then locate the numerical models with $Ra \leq 
10^7$ in the GL regime I of the $Ra-Pr$ parameter space. The cases with $10^7 < 
Ra < 3\times 10^9$ would then belong to regime II.  The transition to 
regime IV  where the bulk contributions dominate the dissipation rates would 
then happen around $Ra\simeq 3\times 10^9$.

While it is not clear that the contributions inferred from the 
total dissipation rates actually reflect the exact bulk and boundary layer 
contributions to $\epsilon_U$ and $\epsilon_T$, this separation nevertheless 
allows us to reconcile the classification of our spherical shell convection 
models in the $Ra-Pr$ parameter space with the prediction of the GL theory for a 
fluid with $Pr=1$. Future work that will better characterise and separate the 
bulk and boundary layer dynamics might help to reconcile the individual 
scalings (figure~\ref{fig:dissip}) with the global ones 
(figure~\ref{fig:newDissip}), especially at low $Ra$. In particular, 
considering dissipation layers as defined by \cite{Petschel13} instead of 
classical boundary layers might possibly help to better separate the bulk and 
boundary layer contributions.

\section{Nusselt and Reynolds numbers scalings}
\label{sec:nurera}

\begin{figure}
\centering 
\includegraphics[width=\textwidth]{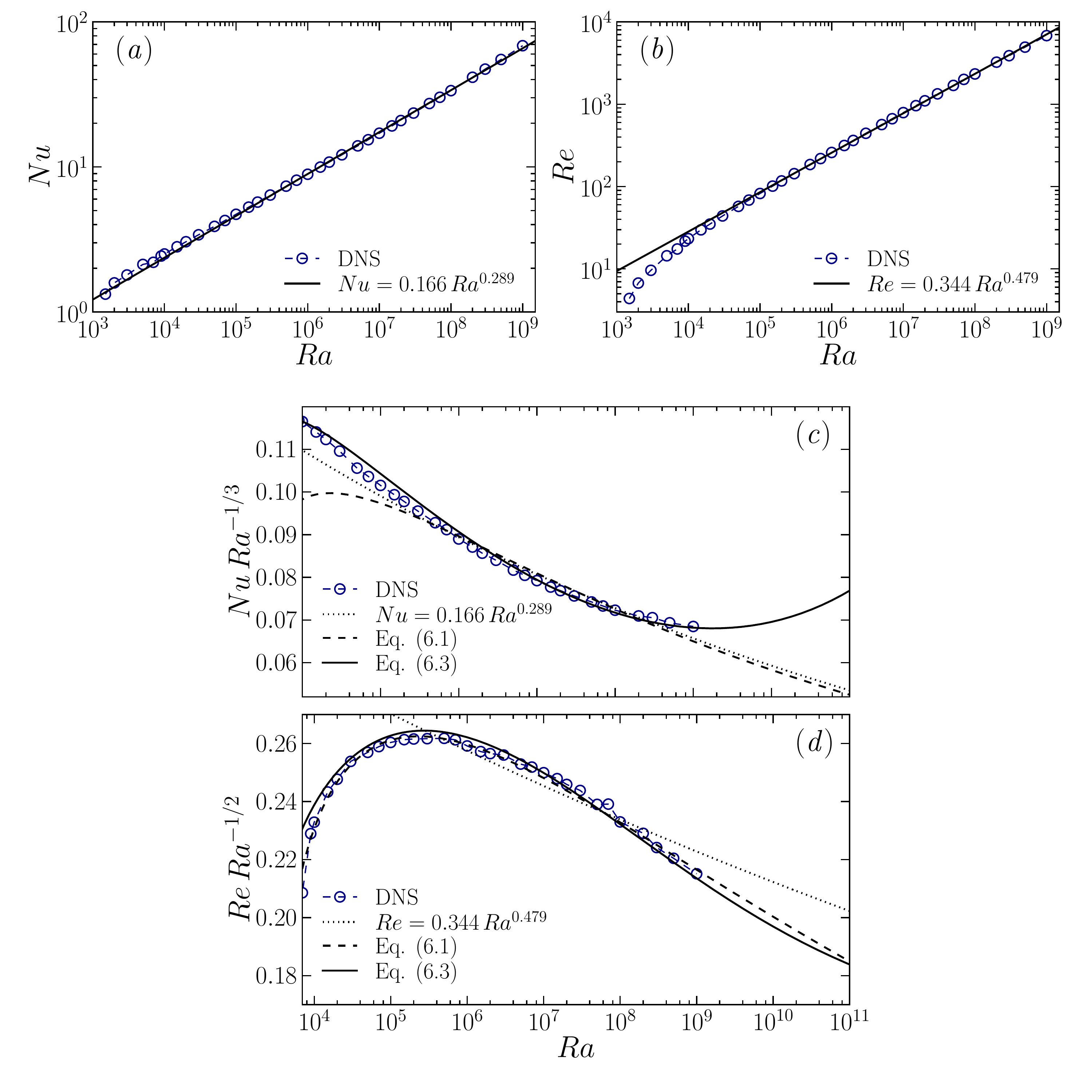}
 \caption{(\textit{a}) Nusselt number versus Rayleigh number.  (\textit{b}) 
Reynolds number versus Rayleigh number. (\textit{c}) Compensated Nusselt number 
versus Rayleigh number. (\textit{d}) Compensated Reynolds number versus Rayleigh 
number. The power laws given in panels (\textit{a-b}) have been derived from a 
best fit to the cases of Table~\ref{tab:results} with $Ra \geq 10^5$. In panels 
(\textit{c-d}), the dashed lines correspond to the numerical solution of 
(\ref{eq:syseq}) and the solid lines to the numerical solution of 
(\ref{eq:syseqGL}).}
 \label{fig:nurera}
\end{figure}

 Figure~\ref{fig:nurera} shows $Nu$ and $Re$ as a 
function of $Ra$ for the cases of Table~\ref{tab:results}. A simple best fit 
to the data for the cases with $Ra\geq 10^5$ yields $Nu\sim Ra^{0.289}$ and 
$Re\sim Ra^{0.479}$, relatively close to $Nu\sim Ra^{2/7}$ and $Re\sim 
Ra^{1/2}$.  While reducing the scaling behaviours of 
$Nu$ and $Re$ to such simple power laws is a common practice in studies of 
convection in spherical shells with infinite Prandtl number 
\citep[e.g.][]{Wolstencroft09,Deschamps10}, this description might be too
simplistic to account for the complex dependence of $Nu$ and $Re$ upon
$Ra$. To illustrate this issue, the panels (\textit{c}) and (\textit{d}) of 
figure~\ref{fig:nurera} show the compensated scalings of $Nu$ and $Re$. The 
power 
laws fail to capture the complex behaviour of $Nu(Ra)$ and $Re(Ra)$ and show 
an increasing deviation from the data at high $Ra$. For instance, the $0.289$ 
scaling exponent obtained for the Nusselt number is too steep for $Ra \leq 
10^7$ and too shallow for higher Rayleigh numbers.

\begin{figure}
\centering 
\includegraphics[width=\textwidth]{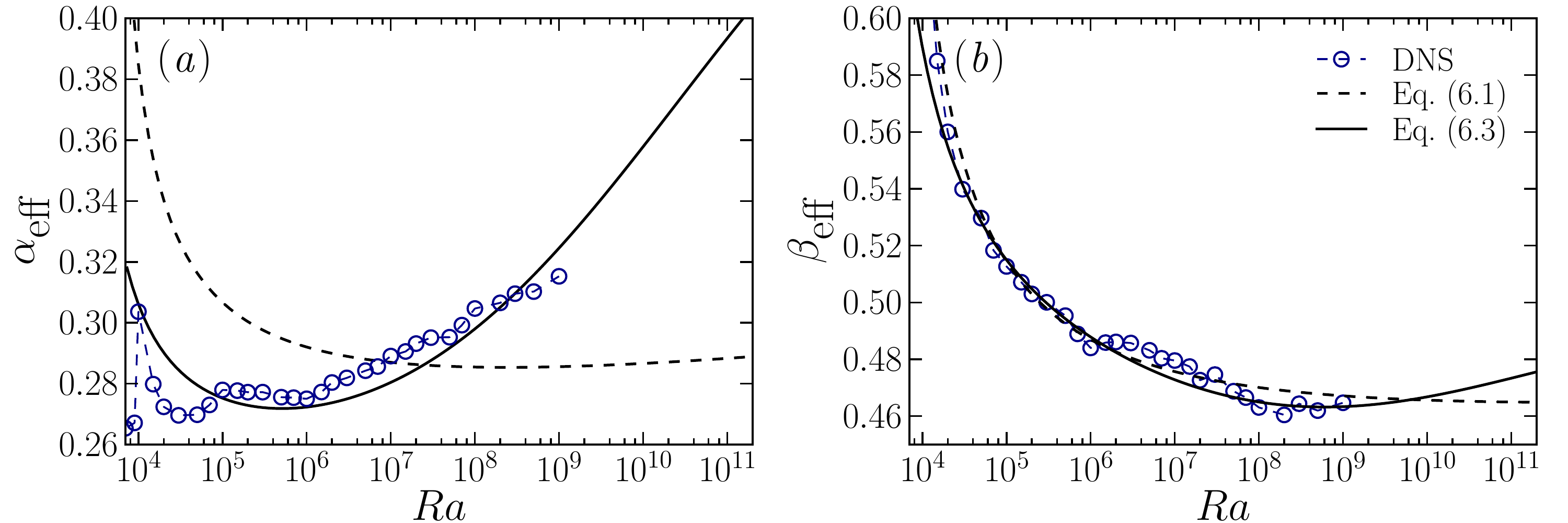}
 \caption{(\textit{a}) $\alpha_{\text{eff}}$ versus $Ra$. (\textit{b}) 
$\beta_{\text{eff}}$ 
versus $Ra$. The dashed black line correspond to the solution 
of (\ref{eq:syseq}), while the solid black line correspond to the solution 
of (\ref{eq:syseqGL}).}
 \label{fig:effExpo}
\end{figure}

The GL theory predicts a gradual change of the slopes of $Nu(Ra)$ and 
$Re(Ra)$ since the flows cross different dynamical regimes when $Ra$ 
increases. Using the asymptotic laws obtained for the different contributions 
to the dissipation rates (\ref{eq:scal_indiv}) and the 
dissipation relations (\ref{eq:viscDiss}) and (\ref{eq:tDiss}), we can 
derive the following equations that relate $Nu$ and $Re$ to $Ra$:
\begin{equation}
\begin{aligned}
 \epsilon_U & = \frac{3}{1+\eta+\eta^2}\,\frac{Ra}{Pr^2}\,(Nu-1)  & = & 
1.756\,Re^{2.79}+2.197\,Re^{2.52}\,,  \\
  \epsilon_T &  = 
\frac{3\eta}{1+\eta+\eta^2}\,Nu &= & 0.038\,Re^{0.7}+0.268\,Re^{0.57}\,.
 \end{aligned}
 \label{eq:syseq}
\end{equation}
This system of equations can be numerically integrated to derive the scaling 
laws for  $Nu$ and $Re$. Figure~\ref{fig:nurera}(\textit{c}-\textit{d}) 
illustrates the comparison between these integrated values and the actual 
data, while figure~\ref{fig:effExpo} shows the local effective exponents 
$\alpha_{\text{eff}}$ and 
$\beta_{\text{eff}}$ of the $Nu(Ra)$ and $Re(Ra)$ laws as a function of $Ra$:
\begin{equation}
 \alpha_{\text{eff}} = \frac{\partial \ln Nu}{\partial \ln Ra}; \quad
\beta_{\text{eff}} = \frac{\partial \ln Re}{\partial \ln Ra}\, .
\label{eq:alphaeff}
\end{equation}
While $Re(Ra)$ is nicely described by the solution of (\ref{eq:syseq}), 
some persistent deviations in the $Nu(Ra)$ scaling are 
noticeable. In particular, $\alpha_{\text{eff}}(Ra)$ increases much faster 
than expected from the scaling law (dashed lines): $\alpha_{\text{eff}}(10^9) 
\simeq 0.32$ while the predicted slope remains close to $0.285$. 
The difficulties to accurately separate the 
bulk and boundary layer dynamics when deriving the scaling laws for 
the different contributions to the dissipation rates are once again likely 
responsible of this misfit.

As demonstrated in the previous section, replacing the sum of the individual 
dissipation contributions by the global scalings provides a much better fit to 
$\epsilon_U$ and $\epsilon_T$ (figure~\ref{fig:globDiss}). We can thus construct
another set of equations based on the scaling laws for the total dissipation 
rates (\ref{eq:scal_glob}):
\begin{equation}
\begin{aligned}
\epsilon_U & =
 \frac{3}{1+\eta+\eta^2}\,\frac{Ra}{Pr^2}\,(Nu-1)  & =  &
0.248\,Re^{3}+7.084\,Re^{5/2}\,,  \\
  \epsilon_T & = 
\frac{3\eta}{1+\eta+\eta^2}\,Nu & = & 0.004\,Re+0.453\,Re^{1/2}\,.
 \end{aligned}
 \label{eq:syseqGL}
\end{equation}
This system of equation is once again numerically integrated to derive the 
scaling laws for $Nu(Ra)$ and $Re(Ra)$. The solid  black lines displayed in 
figure~\ref{fig:nurera}(\textit{c}-\textit{d}) and figure~\ref{fig:effExpo} 
show that these scaling laws fall now much closer to the 
data. They accurately reproduce both $Nu(Ra)$ and $Re(Ra)$ for the whole range 
of Rayleigh numbers and the gradual change in the slopes $\alpha_{\text{eff}}$ 
and $\beta_{\text{eff}}$ are also correctly captured. The improvement of the fit 
to the data when using 
(\ref{eq:syseqGL}) instead of (\ref{eq:syseq}) is the direct 
consequence of the better description of the total dissipation rates by the 
global scalings (\ref{eq:scal_glob}) rather than by the sum of the 
individual scalings (\ref{eq:scal_indiv}).

Figures~\ref{fig:nurera}(\textit{c}-\textit{d}) and \ref{fig:effExpo} also 
show an extrapolation of the scaling laws, solution of (\ref{eq:syseqGL}), 
up to $Ra=10^{11}$. Interestingly, the scaling laws predict that 
$\alpha_{\text{eff}}$ 
would become steeper than $1/3$ for $Ra > 5 \times 10^{9}$. This 
transition point to an enhanced heat transport efficiency would then 
occur at much lower $Ra$ than in RB convection in cartesian or cylindrical 
cells. For instance, the experiments by \cite{Roche10} showed an enhanced 
scaling of $Nu \sim Ra^{0.38}$ for $Ra > 7\times 10^{11}$.
The predicted effective exponent $\alpha_{\text{eff}}$ however 
seems to increase slightly faster than suggested by our numerical data. The 
possible transition to an enhanced heat transport regime at lower $Ra$ than in 
planar geometry thus remains an open issue. Furthermore, the extrapolation of 
the obtained scaling laws to high 
Rayleigh numbers is debatable since the underlying decomposition 
(\ref{eq:scal_glob}) relies on the assumption of laminar boundary layers, which 
will not hold beyond the transition point. Future RB models in spherical 
shells that will possibly reach $Ra\simeq 10^{10}$ could certainly help to 
confirm this trend and check the robustness of the
best-fit coefficients obtained in (\ref{eq:scal_glob}) \citep{Stevens13}.

\section{Conclusion and outlooks}
\label{sec:conclu}

We have studied Rayleigh-B\'enard (RB) convection in spherical shells for 
Rayleigh 
numbers up to $10^9$ and Prandtl number unity. Because of both curvature and 
radial variations of buoyancy, convection  in spherical shells exhibits 
asymmetric boundary layers. To better characterise this asymmetry, we have 
conducted a systematic parameter study, varying both the radius ratio and the 
radial distribution of gravity. Two theories were developed in the 
past to determine this boundary layer asymmetry. The first one by 
\cite{Jarvis93} and 
\cite{Vangelov94} hypothesises that both boundary layers adjust their thickness 
to maintain the same critical boundary layer Rayleigh number; while the second 
one by \cite{Wu91} assumes that the thermal fluctuations at mid-depth are 
statistically symmetrically distributed. Both theories however yield scaling 
laws in poor agreement with our numerical simulations. On the contrary, we 
found that the average plume density, or equivalently the average inter-plume 
spacing, is comparable for both boundary layers. An estimation of the 
average plume density at both spherical bounding surfaces 
has allowed us to 
accurately predict the boundary layer asymmetry and the mean 
bulk temperature for the wide range of spherical shell configurations explored 
here ($\eta=r_i/r_o$ spanning the range $0.2  \leq \eta \leq 0.95$ and gravity 
profiles $g\in[r/r_o,\,1,\,(r_o/r)^2,\,(r_o/r)^5]$).

Because of the lack of experiments and numerical models of non-rotating 
convection in spherical shells at finite Prandtl numbers, the scaling 
properties of the Nusselt and the Reynolds numbers are poorly characterised in 
this geometry. To further address this question, we have conducted numerical 
models in spherical shells with $\eta=0.6$ up to $Ra=10^9$. 
We have adopted a gravity profile of the form $g=(r_o/r)^2$, which has allowed
us to conduct a full dissipation analysis. One of the aims of this study 
was to check the applicability of the scaling theory by 
\cite{Grossmann00,Grossmann04} (GL) to convection in spherical shells. One of 
the prerequisites of this theory is the assumption of Prandtl-Blasius-type (PB) 
boundary layers. We have thus studied the temperature and horizontal velocity 
boundary layer profiles. In agreement with the previous findings by 
\cite{Zhou10a}, the boundary layer profiles have been found to be in
fair agreement with the PB profiles, provided  the numerical simulations are 
analysed in a dynamical frame that incorporates the time and spatial variations 
of the boundary layers. Following the GL central idea, we have then decomposed
the viscous and thermal dissipation rates into contributions coming from the 
fluid bulk and from the boundary layer regions. The detailed analysis of the 
individual contributions to the viscous and thermal dissipation rates reveals 
some noticeable discrepancies to the GL theory ($\epsilon_U^{bu} \sim 
Re^{2.79}$, $\epsilon_U^{bl} \sim Re^{2.52}$, $\epsilon_T^{bu} \sim Re^{0.7}$ 
and $\epsilon_T^{bl} \sim Re^{0.57}$). The total dissipation rates, however, 
can nevertheless be nicely described by the sum of bulk and boundary layer 
contributions that follow the predicted GL exponents ($\epsilon_U \sim 
a\,Re^3+b\,Re^{5/2}$ and $\epsilon_T \sim a\,Re+b\,Re^{1/2}$). This strongly 
suggests that the inaccurate separation of the boundary layer and bulk dynamics 
is the reason for the inferior fitting of the individual contributions.
 These scaling laws have finally been employed to study the scaling 
properties of the Nusselt and the Reynolds numbers and provide laws that 
accurately fit the data. Although these laws exhibit a similar behaviour than 
experiments and numerical simulations of RB convection in 
cartesian or cylindrical coordinates; some distinction to classical RB cells 
have also been reported. Our scaling laws predict a continuous increase of the 
local effective slope of $Nu(Ra)$ from $0.28$ at $Ra=10^6$ to $0.32$ 
at $Ra=10^9$ and suggest a possible enhanced heat transfer scaling with 
an effective exponent steeper than $1/3$ for $Ra > 5 \times 10^{9}$. 
Similar transitions have been observed in some experiments, though at 
significantly higher Rayleigh numbers \citep[$Ra \sim 10^{11}-10^{12}$, 
see][]{Roche10}.

To explore whether the spherical shell geometry is responsible for this 
difference, additional numerical simulations at higher Rayleigh numbers are 
required. Ongoing improvements of pseudo-spectral codes for modelling convection 
in three dimensional spherical shells might help to reach spatial resolutions of 
the order ($N_r\times \ell_{max}=2048\times 2048$) in the coming years 
\citep[e.g.][]{Schaeffer13}. Assuming that the minimum admissible mesh size $h$ 
has to be smaller than the global Kolmogorov scale 
\citep{Grotzbach83,Shishkina10} yields
\[
 h \leq \eta_K = \frac{\nu^{3/4}}{\epsilon_U^{1/4}} = 
\left(\frac{1+\eta+\eta^2}{3}\right)^{1/4}\,\frac{Pr^{1/2}}{Ra^{1/4}\,(Nu-1)^{
1/4}}\,.
\]
An extrapolation of the spatial resolutions employed in this study 
(Table~\ref{tab:results}) then implies that typical resolutions ($N_r\times 
\ell_{max}=2048\times  2048$) might be sufficient to reach $Ra \simeq 10^{10}$ 
for the configuration we considered here ($\eta=0.6$, $g=(r_o/r)^2$). This 
additional decade in $Ra$ might already be sufficient to ascertain the derived 
asymptotic scalings.

\begin{acknowledgments}
We thank Andreas Tilgner for fruitful discussions. All the 
computations have been carried out on the GWDG computer facilities in 
G\"ottingen and on the IBM iDataPlex HPC System Hydra at the MPG Rechenzentrum 
Garching. TG is supported by the Special Priority Program 1488 (PlanetMag, 
\url{www.planetmag.de}) of the German Science Foundation.
\end{acknowledgments}

\appendix

\section{Table of results for the numerical models with different geometry and 
gravity profiles}

\begin{center}
\begin{longtable}{ccccccccc}
\hline
\multicolumn{9}{c}{} \\
$\eta$ & $Ra$ & $Nu$ & $Re$ & $\lambda_T^{i}/\lambda_T^{o}$ & $\lambda_U^{i}/\lambda_U^{o}$ & $\epsilon_T^{bu} (\%)$ & $\epsilon_U^{bu} (\%)$ & $N_r\times \ell_{max}$ \\
\hline
\multicolumn{9}{c}{} \\
\multicolumn{9}{c}{$g=r/r_o$} \\
0.2 & $1\times10^{8}$ & 8.27 & 723.1 & $0.024/0.023$ & $0.015/0.021$ & 0.27 & 0.73 & $97 \times 170$ \\
0.2 & $3\times10^{8}$ & 11.23 & 1252.4 & $0.017/0.018$ & $0.012/0.017$ & 0.29 & 0.76 & $129 \times 341$ \\
0.25 & $2\times10^{7}$ & 6.92 & 407.8 & $0.035/0.034$ & $0.022/0.026$ & 0.27 & 0.72 & $65 \times 128$ \\
0.3 & $3\times10^{6}$ & 5.18 & 188.1 & $0.054/0.054$ & $0.030/0.035$ & 0.24 & 0.69 & $65 \times 128$ \\
0.3 & $5\times10^{6}$ & 5.87 & 242.1 & $0.048/0.048$ & $0.027/0.031$ & 0.26 & 0.71 & $65 \times 128$ \\
0.3 & $7\times10^{6}$ & 6.40 & 287.2 & $0.044/0.045$ & $0.025/0.030$ & 0.26 & 0.72 & $73 \times 133$ \\
0.3 & $3\times10^{7}$ & 9.38 & 595.5 & $0.029/0.032$ & $0.019/0.023$ & 0.28 & 0.76 & $73 \times 133$ \\
0.3 & $3\times10^{8}$ & 18.08 & 1824.8 & $0.015/0.017$ & $0.011/0.015$ & 0.29 & 0.81 & $97 \times 341$ \\
0.35 & $5\times10^{6}$ & 6.74 & 274.1 & $0.047/0.048$ & $0.027/0.031$ & 0.26 & 0.74 & $65 \times 128$ \\
0.35 & $3\times10^{8}$ & 21.23 & 2016.0 & $0.015/0.016$ & $0.011/0.014$ & 0.30 & 0.82 & $129 \times 341$ \\
0.4 & $1\times10^{6}$ & 5.17 & 139.7 & $0.067/0.071$ & $0.036/0.041$ & 0.23 & 0.68 & $65 \times 128$ \\
0.4 & $3\times10^{6}$ & 6.72 & 235.3 & $0.052/0.055$ & $0.029/0.033$ & 0.23 & 0.75 & $65 \times 128$ \\
0.4 & $5\times10^{6}$ & 7.65 & 302.4 & $0.045/0.048$ & $0.026/0.030$ & 0.26 & 0.76 & $73 \times 133$ \\
0.45 & $2\times10^{6}$ & 6.71 & 215.6 & $0.056/0.059$ & $0.032/0.036$ & 0.24 & 0.73 & $65 \times 128$ \\
0.5 & $1\times10^{6}$ & 6.01 & 162.4 & $0.066/0.070$ & $0.037/0.040$ & 0.23 & 0.72 & $65 \times 128$ \\
0.5 & $2\times10^{6}$ & 7.22 & 229.4 & $0.055/0.058$ & $0.032/0.036$ & 0.24 & 0.75 & $65 \times 128$ \\
0.5 & $5\times10^{6}$ & 9.24 & 359.9 & $0.043/0.045$ & $0.027/0.029$ & 0.25 & 0.78 & $81 \times 133$ \\
0.55 & $2\times10^{6}$ & 7.71 & 240.2 & $0.054/0.057$ & $0.031/0.033$ & 0.24 & 0.76 & $65 \times 128$ \\
0.6 & $1\times10^{6}$ & 6.80 & 179.2 & $0.065/0.067$ & $0.036/0.038$ & 0.22 & 0.74 & $65 \times 128$ \\
0.6 & $5\times10^{6}$ & 10.55 & 392.3 & $0.042/0.043$ & $0.026/0.027$ & 0.24 & 0.80 & $81 \times 133$ \\
0.6 & $5\times10^{6}$ & 10.56 & 392.2 & $0.042/0.043$ & $0.026/0.027$ & 0.24 & 0.80 & $81 \times 133$ \\
0.65 & $1\times10^{6}$ & 7.11 & 186.6 & $0.064/0.066$ & $0.035/0.037$ & 0.23 & 0.76 & $73 \times 170$ \\
0.7 & $7\times10^{5}$ & 6.72 & 162.0 & $0.070/0.072$ & $0.037/0.039$ & 0.21 & 0.74 & $73 \times 170$ \\
0.7 & $1\times10^{6}$ & 7.41 & 193.1 & $0.063/0.065$ & $0.035/0.036$ & 0.23 & 0.76 & $73 \times 170$ \\
0.75 & $1\times10^{6}$ & 7.64 & 198.9 & $0.063/0.064$ & $0.034/0.036$ & 0.22 & 0.76 & $97 \times 213$ \\
0.8 & $3\times10^{6}$ & 10.60 & 347.8 & $0.046/0.047$ & $0.028/0.028$ & 0.23 & 0.80 & $97 \times 426$ \\
0.8 & $4\times10^{7}$ & 22.16 & 1198.3 & $0.022/0.022$ & $0.016/0.016$ & 0.26 & 0.85 & $129 \times 1024$ \\
0.85 & $7\times10^{5}$ & 7.26 & 175.5 & $0.068/0.069$ & $0.037/0.038$ & 0.20 & 0.76 & $97 \times 341$ \\
0.9 & $5\times10^{5}$ & 6.73 & 151.4 & $0.074/0.075$ & $0.040/0.040$ & 0.22 & 0.74 & $97 \times 426$ \\
\multicolumn{9}{c}{} \\
\multicolumn{9}{c}{$g=1$} \\
0.2 & $1\times10^{8}$ & 11.95 & 1082.7 & $0.016/0.025$ & $0.011/0.021$ & 0.29 & 0.78 & $97 \times 170$ \\
0.25 & $2\times10^{7}$ & 9.35 & 572.0 & $0.025/0.036$ & $0.016/0.027$ & 0.25 & 0.78 & $81 \times 133$ \\
0.3 & $7\times10^{6}$ & 8.15 & 377.8 & $0.033/0.046$ & $0.020/0.031$ & 0.26 & 0.78 & $73 \times 133$ \\
0.35 & $6\times10^{6}$ & 8.79 & 383.9 & $0.035/0.046$ & $0.021/0.030$ & 0.26 & 0.79 & $73 \times 133$ \\
0.4 & $3\times10^{6}$ & 7.98 & 288.7 & $0.042/0.054$ & $0.025/0.032$ & 0.24 & 0.79 & $65 \times 128$ \\
0.4 & $5\times10^{6}$ & 9.19 & 371.4 & $0.036/0.047$ & $0.022/0.029$ & 0.27 & 0.80 & $65 \times 128$ \\
0.45 & $5\times10^{6}$ & 9.95 & 395.4 & $0.036/0.046$ & $0.022/0.029$ & 0.27 & 0.80 & $73 \times 128$ \\
0.5 & $5\times10^{6}$ & 10.57 & 421.7 & $0.036/0.045$ & $0.023/0.029$ & 0.26 & 0.80 & $81 \times 133$ \\
0.55 & $5\times10^{6}$ & 11.09 & 434.8 & $0.037/0.044$ & $0.023/0.028$ & 0.26 & 0.81 & $81 \times 133$ \\
0.6 & $3\times10^{6}$ & 10.04 & 343.3 & $0.042/0.049$ & $0.026/0.030$ & 0.24 & 0.80 & $81 \times 133$ \\
0.6 & $5\times10^{6}$ & 11.68 & 442.7 & $0.036/0.042$ & $0.023/0.027$ & 0.26 & 0.81 & $81 \times 133$ \\
0.65 & $3\times10^{6}$ & 10.48 & 355.3 & $0.042/0.048$ & $0.026/0.029$ & 0.24 & 0.80 & $81 \times 133$ \\
0.7 & $1\times10^{6}$ & 7.85 & 210.0 & $0.058/0.064$ & $0.033/0.036$ & 0.22 & 0.77 & $73 \times 213$ \\
0.75 & $3\times10^{6}$ & 10.82 & 363.0 & $0.043/0.047$ & $0.026/0.029$ & 0.24 & 0.80 & $97 \times 341$ \\
0.8 & $3\times10^{6}$ & 10.98 & 367.4 & $0.043/0.046$ & $0.027/0.028$ & 0.24 & 0.80 & $97 \times 426$ \\
0.85 & $1\times10^{6}$ & 8.22 & 217.5 & $0.059/0.062$ & $0.033/0.035$ & 0.21 & 0.76 & $97 \times 426$ \\
0.9 & $5\times10^{5}$ & 6.83 & 155.2 & $0.072/0.074$ & $0.039/0.040$ & 0.21 & 0.74 & $97 \times 426$ \\
\multicolumn{9}{c}{} \\
\multicolumn{9}{c}{$g=(r_o/r)^2$} \\
0.2 & $5\times10^{5}$ & 5.85 & 171.9 & $0.031/0.089$ & $0.017/0.047$ & 0.25 & 0.80 & $49 \times 85$ \\
0.2 & $7\times10^{5}$ & 6.45 & 206.2 & $0.028/0.082$ & $0.016/0.046$ & 0.24 & 0.80 & $49 \times 85$ \\
0.2 & $1\times10^{6}$ & 7.17 & 248.8 & $0.026/0.076$ & $0.014/0.044$ & 0.24 & 0.80 & $61 \times 106$ \\
0.2 & $1.5\times10^{6}$ & 8.03 & 308.1 & $0.022/0.068$ & $0.012/0.041$ & 0.28 & 0.81 & $81 \times 170$ \\
0.2 & $2\times10^{6}$ & 8.78 & 351.9 & $0.020/0.064$ & $0.011/0.039$ & 0.26 & 0.82 & $81 \times 170$ \\
0.2 & $1\times10^{7}$ & 14.19 & 794.2 & $0.013/0.040$ & $0.008/0.027$ & 0.30 & 0.84 & $97 \times 256$ \\
0.2 & $3\times10^{7}$ & 20.25 & 1369.5 & $0.009/0.029$ & $0.006/0.021$ & 0.31 & 0.84 & $97 \times 341$ \\
0.25 & $5\times10^{5}$ & 6.29 & 180.2 & $0.035/0.087$ & $0.019/0.047$ & 0.26 & 0.79 & $49 \times 85$ \\
0.3 & $5\times10^{5}$ & 6.60 & 185.7 & $0.038/0.085$ & $0.021/0.046$ & 0.22 & 0.80 & $65 \times 128$ \\
0.3 & $1\times10^{6}$ & 8.08 & 263.2 & $0.031/0.071$ & $0.017/0.041$ & 0.23 & 0.81 & $65 \times 128$ \\
0.3 & $3\times10^{6}$ & 11.11 & 458.1 & $0.022/0.053$ & $0.013/0.034$ & 0.29 & 0.83 & $73 \times 133$ \\
0.3 & $1\times10^{7}$ & 16.07 & 803.7 & $0.015/0.037$ & $0.010/0.025$ & 0.28 & 0.86 & $97 \times 256$ \\
0.3 & $5\times10^{7}$ & 25.68 & 1810.7 & $0.010/0.024$ & $0.007/0.018$ & 0.30 & 0.86 & $161 \times 512$ \\
0.35 & $5\times10^{5}$ & 6.87 & 187.4 & $0.041/0.083$ & $0.023/0.045$ & 0.23 & 0.79 & $65 \times 128$ \\
0.4 & $5\times10^{5}$ & 7.15 & 189.0 & $0.044/0.081$ & $0.024/0.043$ & 0.24 & 0.78 & $65 \times 128$ \\
0.4 & $8\times10^{5}$ & 8.07 & 237.7 & $0.038/0.072$ & $0.022/0.041$ & 0.22 & 0.80 & $65 \times 128$ \\
0.4 & $1\times10^{6}$ & 8.59 & 266.6 & $0.036/0.068$ & $0.021/0.039$ & 0.26 & 0.81 & $65 \times 128$ \\
0.4 & $3\times10^{6}$ & 11.72 & 459.1 & $0.026/0.051$ & $0.016/0.032$ & 0.27 & 0.83 & $97 \times 170$ \\
0.45 & $5\times10^{5}$ & 7.26 & 192.7 & $0.046/0.080$ & $0.026/0.044$ & 0.24 & 0.77 & $65 \times 128$ \\
0.45 & $7\times10^{5}$ & 7.98 & 226.9 & $0.042/0.073$ & $0.024/0.041$ & 0.23 & 0.78 & $65 \times 128$ \\
0.5 & $5\times10^{5}$ & 7.29 & 190.6 & $0.049/0.079$ & $0.028/0.044$ & 0.23 & 0.76 & $61 \times 106$ \\
0.5 & $7\times10^{5}$ & 8.01 & 225.7 & $0.045/0.072$ & $0.026/0.041$ & 0.23 & 0.78 & $65 \times 128$ \\
0.5 & $1\times10^{6}$ & 8.83 & 270.1 & $0.041/0.066$ & $0.024/0.039$ & 0.23 & 0.79 & $81 \times 170$ \\
0.55 & $5\times10^{5}$ & 7.31 & 187.3 & $0.052/0.078$ & $0.029/0.043$ & 0.24 & 0.78 & $61 \times 106$ \\
0.65 & $5\times10^{5}$ & 7.36 & 182.2 & $0.057/0.076$ & $0.031/0.042$ & 0.21 & 0.77 & $61 \times 106$ \\
0.7 & $1.5\times10^{5}$ & 5.22 & 97.7 & $0.083/0.106$ & $0.042/0.053$ & 0.19 & 0.72 & $73 \times 170$ \\
0.7 & $3\times10^{5}$ & 6.35 & 138.2 & $0.068/0.087$ & $0.036/0.046$ & 0.20 & 0.75 & $73 \times 170$ \\
0.7 & $5\times10^{5}$ & 7.32 & 178.3 & $0.059/0.075$ & $0.032/0.041$ & 0.21 & 0.76 & $81 \times 266$ \\
0.7 & $1\times10^{6}$ & 8.83 & 251.0 & $0.049/0.063$ & $0.029/0.036$ & 0.22 & 0.79 & $73 \times 213$ \\
0.75 & $5\times10^{5}$ & 7.26 & 174.9 & $0.062/0.075$ & $0.033/0.041$ & 0.22 & 0.75 & $81 \times 266$ \\
0.8 & $5\times10^{5}$ & 7.20 & 171.2 & $0.064/0.074$ & $0.035/0.040$ & 0.20 & 0.75 & $81 \times 266$ \\
0.8 & $1\times10^{6}$ & 8.71 & 240.8 & $0.053/0.062$ & $0.030/0.035$ & 0.22 & 0.78 & $97 \times 341$ \\
0.8 & $3\times10^{6}$ & 11.81 & 409.2 & $0.039/0.046$ & $0.024/0.028$ & 0.23 & 0.81 & $97 \times 426$ \\
0.85 & $1\times10^{6}$ & 8.62 & 235.5 & $0.055/0.061$ & $0.031/0.034$ & 0.22 & 0.78 & $97 \times 426$ \\
0.9 & $1\times10^{5}$ & 4.47 & 72.2 & $0.108/0.116$ & $0.053/0.056$ & 0.17 & 0.68 & $97 \times 266$ \\
0.9 & $5\times10^{5}$ & 7.04 & 163.8 & $0.069/0.074$ & $0.037/0.040$ & 0.21 & 0.75 & $97 \times 426$ \\
0.9 & $1\times10^{6}$ & 8.54 & 230.2 & $0.057/0.061$ & $0.032/0.034$ & 0.22 & 0.78 & $97 \times 512$ \\
0.95 & $1\times10^{5}$ & 4.41 & 70.4 & $0.112/0.116$ & $0.054/0.056$ & 0.17 & 0.68 & $97 \times 512$ \\
\multicolumn{9}{c}{} \\
\multicolumn{9}{c}{$g=(r_o/r)^5$} \\
0.2 & $3\times10^{4}$ & 9.15 & 206.2 & $0.017/0.117$ & $0.009/0.055$ & 0.24 & 0.85 & $65 \times 128$ \\
0.2 & $5\times10^{4}$ & 10.59 & 265.7 & $0.015/0.100$ & $0.007/0.048$ & 0.23 & 0.85 & $65 \times 128$ \\
0.2 & $7\times10^{4}$ & 11.90 & 313.9 & $0.013/0.089$ & $0.007/0.043$ & 0.28 & 0.85 & $65 \times 133$ \\
0.25 & $6\times10^{4}$ & 9.52 & 238.3 & $0.020/0.105$ & $0.011/0.055$ & 0.26 & 0.82 & $65 \times 128$ \\
0.3 & $7\times10^{4}$ & 8.69 & 212.7 & $0.025/0.107$ & $0.013/0.056$ & 0.24 & 0.83 & $65 \times 128$ \\
0.3 & $1\times10^{5}$ & 9.77 & 251.9 & $0.023/0.095$ & $0.012/0.050$ & 0.23 & 0.83 & $65 \times 133$ \\
0.3 & $3\times10^{5}$ & 13.94 & 440.1 & $0.015/0.070$ & $0.009/0.040$ & 0.24 & 0.84 & $65 \times 133$ \\
0.3 & $3\times10^{6}$ & 27.99 & 1327.6 & $0.008/0.035$ & $0.005/0.023$ & 0.29 & 0.86 & $161 \times 426$ \\
0.35 & $2\times10^{5}$ & 10.75 & 307.3 & $0.023/0.083$ & $0.013/0.045$ & 0.23 & 0.83 & $65 \times 133$ \\
0.4 & $1\times10^{5}$ & 7.94 & 189.5 & $0.034/0.104$ & $0.018/0.055$ & 0.23 & 0.80 & $65 \times 133$ \\
0.4 & $3\times10^{5}$ & 11.08 & 327.1 & $0.024/0.076$ & $0.014/0.043$ & 0.23 & 0.83 & $65 \times 133$ \\
0.4 & $5\times10^{5}$ & 13.04 & 422.0 & $0.020/0.067$ & $0.012/0.039$ & 0.27 & 0.83 & $65 \times 133$ \\
0.45 & $3\times10^{5}$ & 10.25 & 290.0 & $0.029/0.077$ & $0.016/0.042$ & 0.26 & 0.81 & $65 \times 133$ \\
0.5 & $1\times10^{5}$ & 6.95 & 153.4 & $0.046/0.106$ & $0.025/0.053$ & 0.22 & 0.76 & $65 \times 133$ \\
0.5 & $3\times10^{5}$ & 9.53 & 265.8 & $0.033/0.079$ & $0.019/0.044$ & 0.23 & 0.80 & $65 \times 133$ \\
0.55 & $3\times10^{5}$ & 8.88 & 239.3 & $0.038/0.081$ & $0.022/0.045$ & 0.25 & 0.78 & $65 \times 133$ \\
0.6 & $3\times10^{5}$ & 8.41 & 216.6 & $0.043/0.081$ & $0.024/0.044$ & 0.23 & 0.78 & $65 \times 133$ \\
0.6 & $5\times10^{5}$ & 9.69 & 278.4 & $0.037/0.071$ & $0.021/0.040$ & 0.24 & 0.80 & $73 \times 170$ \\
0.6 & $7\times10^{5}$ & 10.69 & 329.1 & $0.033/0.065$ & $0.020/0.038$ & 0.24 & 0.81 & $73 \times 170$ \\
0.6 & $1\times10^{7}$ & 23.28 & 1199.9 & $0.015/0.031$ & $0.011/0.022$ & 0.28 & 0.84 & $129 \times 341$ \\
0.65 & $3\times10^{5}$ & 7.97 & 199.8 & $0.047/0.082$ & $0.026/0.044$ & 0.24 & 0.77 & $65 \times 170$ \\
0.7 & $3\times10^{5}$ & 7.60 & 183.9 & $0.052/0.082$ & $0.029/0.044$ & 0.22 & 0.77 & $65 \times 170$ \\
0.7 & $5\times10^{5}$ & 8.76 & 236.1 & $0.045/0.072$ & $0.026/0.040$ & 0.22 & 0.78 & $65 \times 170$ \\
0.7 & $7\times10^{5}$ & 9.66 & 279.4 & $0.041/0.066$ & $0.024/0.038$ & 0.22 & 0.78 & $65 \times 170$ \\
0.75 & $5\times10^{5}$ & 8.34 & 218.8 & $0.050/0.072$ & $0.028/0.040$ & 0.21 & 0.78 & $65 \times 256$ \\
0.8 & $7\times10^{5}$ & 8.80 & 240.0 & $0.049/0.066$ & $0.028/0.038$ & 0.22 & 0.79 & $65 \times 256$ \\
0.8 & $7\times10^{5}$ & 8.78 & 239.7 & $0.049/0.066$ & $0.028/0.037$ & 0.23 & 0.78 & $97 \times 426$ \\
0.85 & $7\times10^{5}$ & 8.41 & 223.2 & $0.054/0.066$ & $0.030/0.037$ & 0.22 & 0.78 & $97 \times 426$ \\
0.9 & $1\times10^{6}$ & 8.95 & 248.9 & $0.052/0.060$ & $0.030/0.034$ & 0.21 & 
0.78 & $97 \times 512$ \\
\hline  
\multicolumn{9}{c}{} \\
\caption{Summary table of $Pr=1$ numerical simulations with various 
radius ratio $\eta$ and gravity profiles $g(r)$.}
\label{tab:etas}
\end{longtable}
\end{center}

\bibliographystyle{jfm}

\end{document}